\documentclass[aip,jcp,reprint]{revtex4-1}
\usepackage{amsmath,amssymb}
\usepackage{graphicx}
\usepackage{dcolumn}
\usepackage{bm}
\usepackage{multirow}
\usepackage{float}
\usepackage{subfigure}
\usepackage{color}
\usepackage{afterpage}
\usepackage{tabularx}

\usepackage[utf8]{inputenc}
\usepackage[T1]{fontenc}
\usepackage{mathptmx}
\renewcommand{\thefigure}{\arabic{figure}}

\def\expfunc#1{\mbox{$\displaystyle\mathbf{e}^{#1}$}} 

\def\delfunc#1{\mbox{$\displaystyle\delta\left(#1\right)$}} 

\def\ket#1{\mbox{$\displaystyle\vert\,#1\,\rangle$}}

\def\bra#1{\mbox{$\displaystyle\langle\,#1\,\vert$}}

\def\braket#1#2{\mbox{$\displaystyle\langle\,#1\,\vert\,#2\,\rangle$}}

\def\roundbracket#1{\mbox{$\displaystyle\left(#1\right)$}}
\def\squarebracket#1{\mbox{$\displaystyle\left[#1\right]$}}
\def\vertbracket#1{\mbox{$\displaystyle\left|#1\right|$}}
\def\bvec#1{\mbox{$\displaystyle\textbf{#1}$}}
\def\cF{{\cal F}}

\begin{document}

\preprint{AIP/123-QED}

\title{Photoinduced Anomalous Coulomb Blockade and the Role of Triplet States in Electron Transport through an Irradiated Molecular Transistor II: Effects of Electron-Phonon Coupling and Vibrational Relaxation}
\author{Bo Fu}
\affiliation{ 
Department of Physics and Astronomy, Northwestern University, 2145 Sheridan Road, Evanston, IL 60201, United States
}%
\author{Liang-Yan Hsu}
\email{lyhsu@gate.sinica.edu.tw}
\affiliation{Institute of Atomic and Molecular Sciences, 
Academia Sinica, Taipei 10617, Taiwan}

\date{\today}

\begin{abstract}
We generalize our previous theory [Nano Lett. \textbf{18}, 5015-5023 (2018)] to investigate the influence of electron-phonon (e-p) coupling and vibrational relaxation on photoinduced \textit{anomalous Coulomb blockade}, which originates from the triplet states and the energy level alignment. 
We derive the master equation for an irradiated molecular transistor and obtain the relevant rates via the Redfield theory instead of the phenomenological Fermi golden rule approach.
To explore the interplay between e-p coupling and vibrational relaxation, we analyze the charge stability diagrams and the current-voltage characteristics (both gate voltage and source-drain bias voltage) under different e-p coupling strengths in two extreme limits of vibrational relaxation (equilibrated and unequilibrated phonon regimes). 
From the perspective of energy level alignment, we choose four representative situations and derive the analytical formulas of the photoinduced current in the equilibrated regime. The analytical solution reveals a new type of photocurrent due to e-p coupling that does not require the perfect energy level alignment between charged states and triplet states.
In general, our study indicates that photoinduced current and anomalous Coulomb blockade caused by the triplet states are supposed to be experimentally observed.
\end{abstract}

\maketitle

\section{Introduction}
Molecular electronics attracts broad interest in the field of chemical physics because it bridges fundamental sciences and frontier nanotechnology.\cite{Aviram1974,Nitzan2003,Joachim2005,Venkataraman2006,Ratner2013,Huang2015,Hsu2017,Xin2019,Liu2019} During the past two decades, electron transport through a variety of molecules has been extensively investigated, and a lot of intriguing physical phenomena has been experimentally observed. \cite{Reed1997,Cui2001,JiwoongPark2002,Xu2003,Quek2009,Parks2010,Vazquez2012,Burzuri2014,Yoshida2015,Garner2018,Zhou2018,Lumbroso2018,Bai2019}. The experimental breakthrough together with theoretical advances has made molecular electronics thriving and robust. However, for a practical device, the capability of tuning electric current by external fields is crucial and necessary.
Thus, the manipulation of electron transport, e.g., by means of a back gate\cite{Heershe2006,Steele2009,Osorio2010,Perrin2013}, electrochemical gate\cite{Xiao2005,Huang2015}, or light\cite{Jia2016,Fung2017,JiwoongPark2002}, is a primary focus for the whole field of molecular electronics.

Electric current through a molecular junction can be controlled by an optical field via either chemical or physical processes. Chemically, the high- and low-conductance states based on different molecular conformations have been able to be switched by a laser field at specific wavelengths\cite{Katsonis2006,dulic2003,Roldan2013,Jia2016}. Physically, photoelectric current can be generated via the mechanisms of photon-assisted tunneling\cite{Schiffrin2012,Paasch2014,Yoshida2015,Cocker2016,Rybka2016,Yoshioka2016,Fung2017} or electronic excitation\cite{Yasutomi2004,Banerjee2010,Battacharyya2011,Fainberg2012,Kornbluth2013,Zhou2018,Najarian2018,Najarian2019} in a molecular junction. From the theoretical aspects, the Floquet-based methods have successfully demonstrated several fascinating phenomena, including photon-assisted tunneling\cite{Tien1963,Mujica2002,Mujica2004,Mujica2007,Tikhonov2002,Hsu2012,LiangYan2014,LiangYan2015}, coherent destruction of tunneling\cite{Grossman1991,Camalet2003}, coherent revival of tunneling\cite{Hsu2015}, and quantum ratchet effect\cite{Lehmann2002}.
For a single-photon process, kinetic approaches, e.g., quantum master equation approaches\cite{Muralidharan2006,Segal2006,May2008,Fainberg2007,Zelinskyy2011,Fainberg2012} or non-equilibrium green's function (NEGF) approaches\cite{Galperin2005,Galperin2006,Galperin2006b,Galperin2008,Chen2017,Galerpin2017}, offer a clear description of electron current induced by electronic excitation and optical excitation.
However, in these theoretical studies, molecular triplet states are not considered in the modelling because of the slow rate of an intersystem crossing process. In fact, our previous study\cite{Fu2018} has clearly shown that the presence of the triplet states is significant for the light-driven transport properties regardless of whether the rate of the intersystem crossing is fast or not.

The triplet states play a crucial role in light-driven electron transport. In the previous study\cite{Fu2018}, we introduce the concept of \textit{renormalized state energy} and find that the low bias photoinduced current can be maximally achieved when "the renormalized state energy of the charged states is aligned with the energy of the triplet states", in which the triplet dark states indeed dominate the transport properties. Furthermore, our study shows that an optical field enables a field-off Coulomb diamond to be decomposed into three smaller diamonds, and the presence of the triplet states is revealed by the central diamond in the field-on conductance spectra. The concepts of renormalized state energy, energy-level alignment and photoinduced anomalous Coulomb blockade are general and allow us to explore the importance of dark states in light-driven quantum transport. However, the previous study does not consider the effects of electron-phonon (e-p) coupling and vibrational relaxation, and these two effects may interrupt the experimental observation of anomalous Coulomb blockade or photoinduced current attributed to the energy-level alignment of the triplet states and the charged states. In order to eliminate these concerns, a general theory of light-driven electron transport involving the triplet states, e-p coupling and vibrational relaxation is required.

Electron-vibration interactions exhibit rich physical phenomena in nanoscale electron transport and inspire extensive theoretical studies.\cite{Segal2000,Mitra2004,Wegewijs2005,Galperin2005a,Ryndyk2006,Hsu2010,Galperin2007,Koch2005a,Koch2005,Koch2006,May2008,May2008a,May2008b,Wang2010,Wang2010a,Simine2014,Agarwalla2015}
Among these studies, we would like to emphasize the pioneering works by Koch\cite{Koch2004,Koch2005a,Koch2005,Koch2006} and May\cite{May2008,May2008a,May2008b,Wang2010,Wang2011,Wang2010a,WangJElectroanaChem2011}.
Koch et. al. first predicted \textit{Franck-Condon blockade}\cite{Koch2005}
using a Pauli master equation approach based on the Fermi golden rule rates, where the effect of the vibrational relaxation is introduced phenomenologically via the single-mode relaxation time approximation. May et. al. discussed the photoinduced removal of the Franck-Condon blockade using a generalized rate equation approach.\cite{May2008,May2008a,May2008b} The generalized rate equation approach is derived based on the electron-vibrational states using the projection operator method which includes the 2nd order rates of charge transfer transitions, vibrational relaxations, optical transitions and the molecular deexcitation due to the electrode. 
Based on the studies above, we present a Pauli master equation approach involving relevant transitions together with a pedagogical derivation without using the projection operator method.

This article is divided into five parts. In section~\ref{Model_and_Theory}, we introduce our model (including a molecule, two leads, thermal bath, spin-orbit coupling (SOC), and light-matter interactions) and the corresponding Hamiltonians. To adequately treat the vibronic coupling, we apply the polaron transformation and analyze the model Hamiltonians in the polaron frame. In section~\ref{Master_Equation_Approach}, in the framework of the Redfield theory, we derive the rates of charge transfer transition, vibration relaxation and optical transition together with phenomenologically including the rate of the intersystem crossing. Finally, we arrive at the master equation and the current formula in our model. In section~\ref{Parameters}, we discuss the rationality of the parameters in the work. In section~\ref{Results}, first, we define the equilibrated and unequilibrated phonon regimes. Based on the two regimes, we explore the charge stability diagrams and the current-voltage curves of an irradiated molecular junction. Besides, we study the influence of vibronic coupling and bath relaxation on photoinduced Coulomb diamonds and anomalous Coulomb blockade. Moreover, according to the energy level alignment, we analyze the origin of the maximal photoinduced current and the unexpected photoresistive behavior.

\section{Model and Theory}
\label{Model_and_Theory}

\subsection{Hamiltonian}
\begin{figure}
\center{\includegraphics[width=1.0\linewidth]{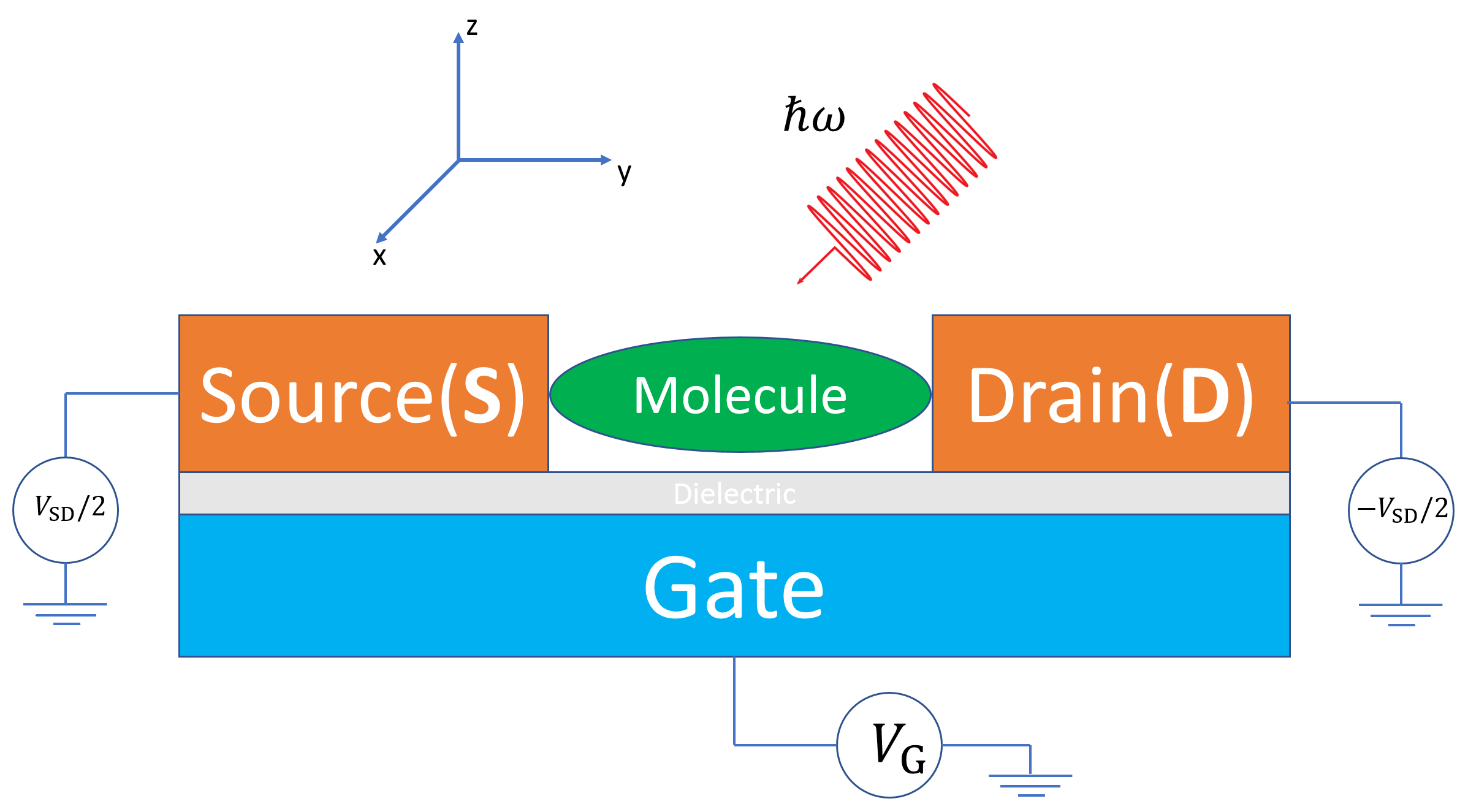}}
\caption{An illustration of the three-terminal molecular device employed in this paper.}
\label{fig:device}
\end{figure}

As illustrated in Figure \ref{fig:device}, we study the light-driven electron transport through a three-terminal molecular device, where a molecule is connected to a source (S) electrode and a drain (D) electrode under a gate electric field. We particularly focus on electric current due to the mechanism of incoherent sequential tunneling in the weak coupling limit, which can be reached by using a molecule with saturated alkyl linkers\cite{JiwoongPark2002}. The total Hamiltonian of the device is written as
\begin{align}\label{eq:Htot}
    \hat H_{\mathrm{tot}}=\hat H_0+\hat V,
\end{align}
where 
\begin{align}\label{eq:H0}
    \hat H_0=\hat H_{\mathrm{mol}}+\hat H_{\mathrm{lead}}+\hat H_{\mathrm{th}}.
\end{align}
$\hat H_{\mathrm{mol}}$ represents the isolated molecular system, $\hat H_{\mathrm{lead}}$ refers to the isolated leads (electrodes), $\hat H_{\mathrm{th}}$ describes the thermal bath, and the coupling Hamiltonian is
\begin{align}\label{eq:coupling}
    \hat V =  \hat H_{\mathrm{SOC}} + \hat H_{\mathrm{field}} + \hat H_{\mathrm{m-th}}  +\hat H_{\mathrm{m-l}},
\end{align}
which consists of the SOC $\hat H_{\mathrm{SOC}}$,  the coupling to the radiation field $\hat H_{\mathrm{field}}$, the coupling to the thermal bath $\hat H_{\mathrm{m-th}}$ and the molecule-lead coupling between the molecule and the electrodes $\hat H_{\mathrm{m-l}}$.

The two electrodes including a source lead and a drain lead are described by a non-interacting electron gas model as
\begin{align}
    \label{eq:lead}
\hat H_{\mathrm{lead}}=\sum_{\alpha}\hat H_{\mathrm{lead,\alpha}}=\sum_{\alpha\bvec k\sigma}\varepsilon_{\alpha\bvec k\sigma}\hat a^{\dagger}_{\alpha\bvec k\sigma}\hat a_{\alpha\bvec k\sigma}, 
\end{align}
where $\hat a^{\dagger}_{\alpha\bvec k\sigma}$ ($\hat a_{\alpha\bvec k\sigma}$) creates (annihilates) an electron with momentum $\bvec k$, spin $\sigma$ and energy $\varepsilon_{\alpha\bvec k\sigma}$ in the electrode $\alpha$ (S for source lead, and D for source lead).
The electronic relaxation in the electrodes is assumed to be fast compared to all the dynamical processes on the molecular system. Therefore, the electrons in the electrodes are always at the thermal equilibrium. The thermal equilibrium density operator of the isolated electrodes reads
\begin{align}
    \hat\rho_{\mathrm{lead}}=\prod_{\alpha=S,D}\frac{\expfunc{-\beta\roundbracket{\hat H_{\mathrm{lead,\alpha}}-\mu_{\alpha}\hat N_{\mathrm{lead},\alpha}}}}{\mathrm{Tr}_{\mathrm{lead}}\{\expfunc{-\beta\roundbracket{\hat H_{\mathrm{lead,\alpha}}-\mu_{\alpha}\hat N_{\mathrm{lead},\alpha}}}\}},
\end{align}
where $\beta=1/k_BT$ with the Boltzmann constant $k_B$, $\hat N_{\mathrm{lead,\alpha}}=\sum_{\bvec k\sigma}\hat a^{\dagger}_{\alpha\bvec k\sigma}\hat a_{\alpha\bvec k\sigma}$ is the number operator of electrode $\alpha$, and $\mu_{\alpha}$ denotes the chemical potential of electrons in the electrode $\alpha$. A source-drain bias $V_{\mathrm{SD}}$ applied symmetrically across the molecular junction is then taken into account by setting $\mu_{\mathrm{S}}=\mu_0+V_{\mathrm{SD}}/2$ and $\mu_{\mathrm{D}}=\mu_0-V_{\mathrm{SD}}/2$, where $\mu_0$ is the chemical potential of the two electrodes at zero bias.

The inactive intramolecular vibrations with respect to the electronic transitions, together with the environment, are treated as the thermal bath and then modelled as a set of harmonic oscillators in $\hat H_{\mathrm{th}}$,
\begin{align}
    \hat H_{\mathrm{th}}=\sum_{\alpha}\roundbracket{\frac{\hat p^2_{\alpha}}{2m_{\alpha}}+\frac{1}{2}m_{\alpha}\omega^2_{\alpha}\hat x^2_{\alpha}},
\end{align}
where $\hat p_{\alpha}$, $\hat x_{\alpha}$, $m_{\alpha}$ and $\omega_{\alpha}$ are respectively the momentum operator, the position operator, the mass and the vibrational frequency of the thermal bath mode $\alpha$. The thermal bath Hamiltonian is diagonalized by introducing $\hat q_{\alpha}= \roundbracket{\hat b^{\dagger}_{\alpha}+\hat b_{\alpha}}\ell_{\alpha}/\sqrt{2}$ and $\hat p_{\alpha}=i\roundbracket{\hat b^{\dagger}_{\alpha}-\hat b_{\alpha}}\hbar/(\sqrt{2}\ell_{\alpha})$ with $\ell_{\alpha}=\sqrt{\hbar/(m_{_{\alpha}}\omega_{_{\alpha}})}$, which results in
\begin{align}
    \hat H_{\mathrm{th}}=\sum_{\alpha}\roundbracket{\hat b^{\dagger}_{\alpha}\hat b_{\alpha}+\frac{1}{2}}\hbar\omega_{\alpha},
\end{align}
where $\hat b^{\dagger}_{\alpha}$ ($\hat b_{\alpha}$) creates (annihilates) a vibrational quantum of mode $\alpha$ with vibrational frequency $\omega_{\alpha}$. Similar to the electrons in the electrodes, the relaxation process of the phonon bath is also assumed to be faster than all relevant processes in the molecular system. Therefore, the thermal bath is always kept at its thermal equilibrium characterized by the density operator
\begin{align}
    \hat\rho_{\mathrm{th}}=\expfunc{-\beta\hat H_{\mathrm{th}}}/\mathrm{Tr}_{\mathrm{th}}\{\expfunc{-\beta\hat H_{\mathrm{th}}}\}.
\end{align}

The molecular Hamiltonian consists of an electronic part and a vibronic part, i.e.
\begin{align}
    \hat H_{\mathrm{mol}} = \hat H_{\mathrm{m-el}}+\hat H_{\mathrm{m-vib}}\,.
\end{align}
The electronic Hamiltonian of the isolated molecule is expressed in terms of many-electron states that span the Fock space $\cF$, which reads 
\begin{align}
    \label{eq:molham}
\hat H_{\mathrm{m-el}}&=\sum_{\vert N,a\rangle\in\cF}\roundbracket{E_{N,a}-NeV_{\mathrm{G}}}\ket{N,a}\bra{N,a}\,.
\end{align}
The $\ket{N,a}$ labels a many-electron state with $N$ referring to the number of excess electrons with respect to the neutral state of the molecule and $a$ labelling the electronic state for its spin multiplicity and energy level, which is consistent with the convention in the community of photophysics. The energy level of the state $\ket{N,a}$ that includes the energy shift caused by the image charge effect is denoted by $E_{N,a}$. The effect of the gate voltage $V_G$ is considered by $NeV_{\mathrm{G}}$. Although the Fock space is constructed by a complete set of many-electron states, we employ a truncated Fock space which includes a finite number of states that participate in the relevant electronic transitions. Our truncated Fock space is spanned by cation doublet charge states ($\ket{-1,D^{\sigma=\pm 1/2}_0}$, anion doublet and $\ket{1,D^{\sigma=\pm 1/2}_0}$), neutral singlet ground state ($\ket{0,S_0}$,  neutral singlet excited state $\ket{0,S_1}$) and three neutral triplet states ($\ket{0,T^{m=0,\pm 1}_1}$), where the degeneracy of doublet and triplet states are considered.

We take a molecular vibration mode as the reaction coordinate and 
write the vibronic part of the molecular Hamiltonian as
\begin{align}
\label{eq:e-p-classical}
\hat H_{\mathrm{m-vib}}=\sum_{\vert N,a}\biggr[\frac{\hat p^2}{2m_{\mathrm{vib}}}+\frac{1}{2}m_{\mathrm{vib}}\omega^2_{\mathrm{vib}}\roundbracket{\hat q - q_{N,a}}^2\biggr]\vert N,a\rangle \langle N,a\vert,
\end{align}
where $m_{\mathrm{vib}}$, $\omega_{\mathrm{vib}}$, $\hat p$ and $\hat q$ are respectively the effective mass, the frequency, the momentum operator and the position operator of the active vibrational motion, and $q_{N,a}$ is the equilibrium position of the potential energy surface (PES) associated with the electronic state $\ket{N,a}$. We introduce $\hat q= \roundbracket{\hat b^{\dagger}+\hat b}\ell/\sqrt{2}$ and $\hat p=i\roundbracket{\hat b^{\dagger}-\hat b}\hbar/(\sqrt{2}\ell)$ with $\ell=\sqrt{\hbar/(m_{\mathrm{vib}}\omega_{\mathrm{vib}})}$, and then transform the Hamiltonian $\hat H_{\mathrm{m-vib}}$ into

\begin{align}
\label{eq:e-p}
\hat H_{\mathrm{m-vib}}=&\hbar\omega_{\mathrm{vib}}\sum_{\vert N,a\rangle\in\cF}\biggr[\hat b^{\dagger}\hat b+\frac{1}{2} \nonumber \\
&-\lambda_{N,a}\roundbracket{\hat b^{\dagger}+\hat b}+\lambda_{N,a}^2\biggr]\ket{N,a}\bra{N,a},
\end{align}
where $\hat b^{\dagger}$ ($\hat b$) creates (annihilates) a vibronic quantum with frequency $\omega_{\mathrm{vib}}$, and the dimensionless e-p coupling parameter $\lambda_{N,a}$ is $q_{N,a}\sqrt{m_{\mathrm{vib}}\omega_{\mathrm{vib}}/2\hbar}$.

The coupling Hamiltonian $\hat V$ describes the couplings between the molecular system and the other external degrees of freedom, which are respectively responsible for the singlet-triplet transition, the radiation-induced transition, the vibrational relaxation and the charge transfer transition. 
The SOC (the origin of the singlet-triplet transitions) is represented as
\begin{align}\label{eq:soc}
\hat H_{\mathrm{SOC}}&=\sum_{m=0,\pm 1}\biggr(V^{\mathrm{SOC}}_{0S_0,0T^m_1}\ket{0,S_0}\bra{0,T^m_1}\nonumber\\
&+V^{\mathrm{SOC}}_{0T^m_1,0S_1}\ket{0,T^m_1}\bra{0,S_1}\biggr),
\end{align}
where the transition amplitudes $V^{\mathrm{SOC}}_{0S_0,0T^m_1}$ and $V^{\mathrm{SOC}}_{0T^m_1,0S_1}$ could be introduced by a relativistic correction to the kinetic terms.  $\hat H_{\mathrm{SOC}}$ is responsible for the phenomenological rates of  singlet-triplet transitions  in the Pauli master equation. The explicit expression of the SOC and its origin\cite{Kohler2009,Marian2012,Powell2015,Baryshnikov2017} are not discussed here.

The spin-allowed transitions between the neutral singlet states $\ket{0,S_0}$ and $\ket{0,S_1}$, due to the coupling to a radiation field, are described by $\hat H_{\mathrm{field}}$ based on the semi-classical radiation theory and the dipole approximation\cite{Schatz1993} as 
\begin{align}
    \label{eq:field}
H_{\mathrm{field}}&=-\bvec E\roundbracket{t}\cdot\roundbracket{\bvec d_{01}\ket{0,S_0}\bra{0,S_1}
+\mathrm{h.c.}
}, 
\end{align}
where h.c. stands for Hermitian conjugate, $\bvec d_{01}$ is the transition dipole moment between $\ket{0,S_0}$ and $\ket{0,S_1}$, and $\bvec E\roundbracket{t}$ the electric field component of the radiation field. The time-dependent electric field takes the form of 
\begin{align}\label{eq:cw}
\bvec E\roundbracket{t}=\bvec E_0\roundbracket{\expfunc{i\omega t}+\expfunc{-i\omega t}}
\end{align}
for a continuous wave laser source. 

The vibrational relaxation process within a PES associated with the electronic state $\ket{N,a}$ originates from the coupling of the molecular vibrations to the thermal bath, i.e., $\hat H_{\mathrm{m-th}}$, which takes the following bilinear form\cite{Caldeira1983,Garg1985,Leggett1987,Jean1992,May2008}
\begin{align}\label{eq:m-ph-classical}
    \hat H_{\mathrm{m-th}}=\sum_{N,a}\sum_{\alpha}c_\alpha\roundbracket{\hat q-q_{N,a}}\hat q_{\alpha}\vert N,a\rangle \langle N,a\vert,
\end{align}
where $c_{\alpha}$ characterizes the coupling the active molecular vibration to the thermal bath mode $\alpha$.
For convenience of the derivation of the master equation, we write Eq. (\ref{eq:m-ph-classical}) in terms of a system oporator and a bath operator as
\begin{align}\label{eq:m-ph-classical2}
    \hat H_{\mathrm{m-th}}= \hat S_{\mathrm{m-th}}\hat B_{\mathrm{m-th}},
\end{align}
where the system operator and the bath operator are respectively given by
\begin{align}\label{eq:m-ph-S}
    \hat S_{\mathrm{m-th}}=\sum_{N,a}\roundbracket{\hat q-q_{N,a}}\vert N,a\rangle \langle N,a\vert
\end{align}
and
\begin{align}\label{eq:m-ph-B}
    \hat B_{\mathrm{m-th}}=\sum_{\alpha}c_\alpha\hat q_{\alpha}\,.
\end{align}


Finally, we model the molecule-lead coupling in the following form, 
\begin{align}
    \label{eq:mol-lead}
    \hat H_{\mathrm{m-l}}
    &=
    \sum_{\alpha\bvec k\sigma,Nab}(V^*_{\alpha\bvec k\sigma,NaN-1b}\hat a^{\dagger}_{\alpha\bvec k\sigma}\ket{N-1,b}\bra{N,a}\nonumber \\
    &+\mathrm{h.c.})\,,
\end{align}
which leads to the exchange of electrons between the molecule and the two electrodes, and this coupling is the origin of the current through the molecular junction. The coupling parameter $V^*_{\alpha\bvec k\sigma,NaN-1b}$ ($V_{\alpha\bvec k\sigma,NaN-1b}$) is a scattering amplitude which describes the tunneling of an electron from the molecule (electrode $\alpha$) to the electrode $\alpha$ (molecule) together with a transition of the molecule from $\ket{N,a}$ to $\ket{N-1,b}$ (from $\ket{N-1,b}$ to $\ket{N,a}$). As detailed in Ref. \!\![\citenum{Pedersen2005}], the relationship between $\hat H_{\mathrm{m-l}}$ and its form based on a single particle picture reveals a route to evaluate the coupling parameter $V_{\alpha\bvec k\sigma,NaN-1b}$. The evaluation of $V_{\alpha\bvec k\sigma,NaN-1b}$ has been discussed in the \textbf{Supporting Information} of Ref. [\!\!\citenum{Fu2018}] and is briefly reviewed in Appendix \ref{sec:appendix1}.

The molecule-lead coupling $\hat H_{\mathrm{m-l}}$ can also be written in terms of bath operators and system operators as\cite{Galperin2005,Fainberg2007,May2008}
\begin{align}\label{eq:Hm-l2}
    \hat H_{\mathrm{m-l}}=\sum_{\alpha}\roundbracket{\hat B^+_{\alpha}\hat S^-_{\alpha}+\hat S^+_{\alpha}\hat B^-_{\alpha}}\,,
\end{align}
where the definitions of system operators and bath operators are respectively given by
\begin{align}
    \hat B^+_{\alpha}=\roundbracket{\hat B^-_{\alpha}}^{\dagger}=\sum_{\bvec k\sigma}M_{\alpha\bvec k\sigma}\hat a_{\alpha\bvec k\sigma},
\end{align}
and
\begin{align}
    \hat S^+_{\alpha}=\roundbracket{\hat S^-_{\alpha}}^{\dagger}=\sum_{N,a,b}T^{\alpha}_{Na,N-1b}\ket{N,a}\bra{N-1,b},
\end{align}
with the help of the decomposition of the coupling parameter\cite{Peskin2017}
\begin{align}\label{eq:factorization1}
    V_{\alpha\bvec k\sigma,NaN-1b}=T^{\alpha}_{NaN-1b}M_{\alpha\bvec k\sigma}\,.
\end{align}
$T^{\alpha}_{NaN-1b}$ refers to the probability amplitude of a molecular transition due to electrode $\alpha$, and 
$M_{\alpha\bvec k\sigma}$ determines the spectral density of electrode $\alpha$.

\subsection{Polaron Transformation}
In order to simplify the subsequent derivation of master equations, we employ a polaron transformation\cite{Nitzan2006textbook,Mahan2008,May2011}.
We choose a representation of density matrix operator that is defined with respect to the electron-vibrational states $\ket{N,a,\mu}$. This electron-vibrational state is expressed as a tensor product of a many-electron state and an associated displaced harmonic oscillator state, i.e.,
\begin{align}
    \ket{N,a,\nu}=\ket{N,a}\otimes\ket{\nu\roundbracket{\lambda_{N,a}}},
\end{align}
where the vibrational quantum number $\nu=0,1,2,3\dots$ and the corresponding energy level $E_{Na\nu}=E_{N,a}+(\nu+1/2)\hbar\omega_{\mathrm{vib}}$. Note that the vibrational operators $\hat b$ and $\hat b^{\dagger}$ in Eq.~(\ref{eq:e-p}) are defined for $\{\ket{\nu\roundbracket{0}}\}$, which is related to the states of a displaced harmonic oscillator $\{\ket{\nu\roundbracket{\lambda}}\}$ through $\ket{\nu\roundbracket{\lambda}}=\expfunc{\lambda\roundbracket{\hat b^{\dagger}-\hat b}}\ket{\nu\roundbracket{0}}$.\cite{Glauber1969} 

After a polaron transformation\cite{Koch2005,Mahan2008,Hsu2010},
the new molecular Hamiltonian $\hat{H}'_{\mathrm{mol}}=\hat U\hat H_{\mathrm{mol}}\hat U^{-1}$ is given in terms of electron-vibrational states $\ket{N,a,\mu}$ as 
\begin{align}
    &\hat{H}'_{\mathrm{mol}}
    =  \nonumber \\ 
    &\sum_{N,a,\nu}\roundbracket{E_{N,a}-NeV_G+(\nu+\frac{1}{2})\hbar\omega_{\mathrm{vib}}}\ket{N,a,\nu}\bra{N,a,\nu}\,,
\end{align}
where the unitary operator $\hat U$ is 
\begin{align}
    \hat U=\expfunc{\sum_{N,a}\lambda_{N,a}\ket{N,a}\bra{N,a}\roundbracket{\hat b^{\dagger}-\hat b}}\,.
\end{align}
The coupling of the molecule to the thermal bath in Eq. (\ref{eq:m-ph-classical2}) is transformed similarly as
\begin{align}\label{eq:m-ph-classical-transformed}
    \hat H'_{\mathrm{m-th}}= \hat S'_{\mathrm{m-th}}\hat B'_{\mathrm{m-th}}
\end{align}
with $\hat B'_{\mathrm{m-th}}=\hat B_{\mathrm{m-th}}$ and
\begin{align}\label{eq:m-ph-S-transformed}
    \hat S'_{\mathrm{m-th}}=\sum_{N,a}\hat q\vert N,a\rangle \langle N,a\vert\,.
\end{align}

Furthermore, $\hat U$ allows us to transform the coupling Hamiltonians $\hat H_{\mathrm{field}}$, $\hat H_{\mathrm{m-l}}$ and $\hat H_{\mathrm{SOC}}$ into $\hat H'_{\mathrm{field}}$, $\hat H'_{\mathrm{m-l}}$ and $\hat H'_{\mathrm{SOC}}$ in the polaron frame
by using the following relationship,
\begin{align}\label{eq:polarontrasform}
    &\hat U\ket{N_1,a_1}\bra{N_2,a_2}\hat U^{-1}\nonumber \\
    =&\expfunc{\roundbracket{\hat b^{\dagger}-\hat b}\lambda_{N_1,a_1}}\ket{N_1,a_1}\bra{N_2,a_2}\expfunc{-\roundbracket{\hat b^{\dagger}-\hat b}\lambda_{N_2,a_2}}\,.
\end{align}
The transformed molecule-lead coupling Hamiltonian then reads 
\begin{align}\label{eq:HamMolLeadTransformed}
    \hat H'_{\mathrm{m-l}}=\sum_{\alpha}\roundbracket{\hat B'^+_{\alpha}\hat S'^-_{\alpha}+\hat S'^+_{\alpha}\hat B'^-_{\alpha}}
\end{align}
with  $B'^{\pm}_{\alpha}=B^{\pm}_{\alpha}$ and
\begin{align}\label{eq:sys-op-transformed}
&\hat S'^{+}_ {\alpha}=(\hat S'^{-}_ {\alpha}) ^{\dagger}\nonumber \\
=&\sum_{Nab}\sum_{\nu_b\nu_b}\biggr(M_{\nu_a\nu_b}\roundbracket{\lambda_{N-1,b}-\lambda_{N,a}}\nonumber \\
&\times T^{\alpha}_{NaN-1b}\vert N,a,\nu_a\rangle\langle N-1,b,\nu_b\vert +\mathrm{h.c.}\biggr)\,,
\end{align}
where $M_{\nu_a\nu_b}\roundbracket{\lambda_{N-1,b}-\lambda_{N,a}}$ absorbs the exponential factor introduced in Eq.~(\ref{eq:polarontrasform}) and refers to the vibrational overlap $\braket{\nu_a\roundbracket{\lambda_{N,a}}}{\nu_b\roundbracket{\lambda_{N-1,b}}}$. 
$M_{\nu_a\nu_b}\roundbracket{\lambda}$ for real $\lambda$ is then given by\cite{Glauber1969,Koch2004}
\begin{align}
    M_{\nu_a\nu_b}\roundbracket{\lambda}=&
    \bra{\nu_a\roundbracket{0}}\expfunc{\lambda\roundbracket{\hat a^{\dagger}-\hat a}}\ket{\nu_b\roundbracket{0}}\nonumber \\
    =&\squarebracket{\mathrm{sgn}\roundbracket{\nu_b-\nu_a}}^{\nu_a-\nu_b}\lambda^{\vert \nu_a-\nu_b\vert}\expfunc{-\lambda^2/2}\roundbracket{\frac{\nu_{\mathrm{min}}!}{\nu_{\mathrm{max}}!}}^{1/2}\nonumber \\
    &\times\mathrm{L}^{\vert\nu_a-\nu_b\vert}_{\nu_{\mathrm{min}}}\roundbracket{\lambda^2},
\end{align}
where $\mathrm{sgn}\roundbracket{x}$ returns the sign of $x$, $\nu_{\mathrm{min}}=\mathrm{min}\{\nu_a,\nu_b\}$, $\nu_{\mathrm{max}}=\mathrm{max}\{\nu_a,\nu_b\}$ and $\mathrm{L}^n_m\roundbracket{x}$ denotes the generalized Laguerre polynomial.

Similarly, one can derive the radiative coupling Hamiltonian $\hat H_{\mathrm{field}}$ and the SOC Hamiltonian $\hat H_{\mathrm{SOC}}$ in the polaron frame as
\begin{align}
    \hat H'_{\mathrm{field}}=&-\textbf{E}(t)\cdot\textbf{d}_{01}\nonumber \\
    &\times \sum_{\nu_0,\nu_1}\biggr(M_{\nu_0\nu_1}\roundbracket{\lambda_{0,S_1}-\lambda_{0,S_0}}\ket{0,S_0,\nu_0}\bra{0,S_1,\nu_1}\nonumber \\
    &+M_{\nu_1\nu_0}\roundbracket{\lambda_{0,S_0}-\lambda_{0,S_1}}\ket{0,S_1,\nu_1}\bra{0,S_0,\nu_0}\biggr) \,,
\end{align}
and
\begin{align}\label{eq:SOCTransformed}
&\hat H'_{\mathrm{SOC}}\nonumber \\
&=\sum_{m,\nu,\nu'}\biggr(V^{\mathrm{SOC}}_{0S_0,0T^m_1}M_{\nu\nu'}\roundbracket{\lambda_{0,T^m_1}-\lambda_{0,S_0}}\ket{0,S_0,\nu}\bra{0,T^m_1,\nu'}\nonumber\\
&+V^{\mathrm{SOC}}_{0T^m_1,0S_1}M_{\nu\nu'}\roundbracket{\lambda_{0,S_1}-\lambda_{0,T^m_1}}\ket{0,T^m_1,\nu}\bra{0,S_1\nu'}\biggr)\,.
\end{align}



\section{Master Equation Approach}
\label{Master_Equation_Approach}

In this section, we will show that the electron transport characteristics of an irradiated molecular junction can be described using Pauli master equations (rate equations). The Pauli master equations built upon electron-vibrational states have been extensively discussed\cite{Koch2004,Koch2005,HsuJCP2010}. In most studies, the rate equations of incoherent sequential tunneling are phenomenologically constructed based on the Fermi golden rule in the framework of a T-matrix approach.\cite{Averin1990,Averin1992,Esteve1992,Schoeller1994,Averin1994,Matveev2002,Koch2004,Koch2005,Walldorf2017} In our work, we derive the rate equations for irradiated incoherent sequential tunneling explicitly from the reduced density matrix approach\cite{Blum1981,CohenTannoudji1992,Nitzan2006textbook,May2011} via the Redfield theory\cite{Redfield1965}, incorporating the optical transitions, vibrational relaxations and singlet-triplet transitions. 

First of all, the reduced density operator of the molecular system is defined as $\hat \rho\roundbracket{t}=\mathrm{Tr}_{\mathrm{lead+th}}\{\hat \rho_{\mathrm{tot}}\}$, where $\hat \rho_{\mathrm{tot}}$ is the density operator of the total system and the trace $\mathrm{Tr}_{\mathrm{lead+th}}\{\cdots\}$ averages over the states of the lead electrons and the thermal bath modes. The population of a state $\ket{N,a,\nu}$ is then defined as $P_{N,a,\nu}\roundbracket{t}=\bra{N,a,\nu}\hat \rho\roundbracket{t}\ket{N,a,\nu}$

We partition the total Hamiltonian $\hat H_{\mathrm{tot}}$ as in Eq. (\ref{eq:Htot}) and obtain the Liouville equation of the density operator $\hat \rho_{\mathrm{tot}}$ in the interaction picture as,
\begin{align}\label{eq:eom-1storder}
    \frac{d}{dt}\hat \rho^I_{\mathrm{tot}}\roundbracket{t}=-\frac{i}{\hbar}\squarebracket{\hat V^I\roundbracket{t},\hat \rho^I_{\mathrm{tot}}\roundbracket{t}}
\end{align}
where an operator in the interaction picture $\hat A^I\roundbracket{t}$ relates to its Schrodinger picture counterpart $\hat A\roundbracket{t}$ through $\hat A^I\roundbracket{t}=\expfunc{\frac{i}{\hbar}\hat H_0t}\hat A\roundbracket{t}\expfunc{-\frac{i}{\hbar}\hat H_0t}$. 

Due to different coupling terms, the evolution of the density operator can be separated into
\begin{align}\label{eq:eom-3terms}
    \frac{d}{dt}\hat \rho^I_{\mathrm{tot}}\roundbracket{t}=&
    -\frac{i}{\hbar}\squarebracket{\hat H^I_{\mathrm{m-l}}\roundbracket{t},\hat \rho^I_{\mathrm{tot}}\roundbracket{t}}
    -\frac{i}{\hbar}\squarebracket{\hat H^I_{\mathrm{m-th}}\roundbracket{t},\hat \rho^I_{\mathrm{tot}}\roundbracket{t}}\nonumber \\
    &-\frac{i}{\hbar}\squarebracket{\hat H^I_{\mathrm{field}}\roundbracket{t},\hat \rho^I_{\mathrm{tot}}\roundbracket{t}}
    -\frac{i}{\hbar}\squarebracket{\hat H^I_{\mathrm{SOC}}\roundbracket{t},\hat \rho^I_{\mathrm{tot}}\roundbracket{t}}\,,
\end{align}
where the four terms on the right hand side 
are respectively denoted as 
$\frac{d \hat \rho^I_{\mathrm{tot}}\roundbracket{t}}{dt}\vert_{\mathrm{m-l}}$, 
$\frac{d \hat \rho^I_{\mathrm{tot}}\roundbracket{t}}{dt}\vert_{\mathrm{m-th}}$, 
$\frac{d \hat \rho^I_{\mathrm{tot}}\roundbracket{t}}{dt}\vert_{\mathrm{field}}$ 
and 
$\frac{d \hat \rho^I_{\mathrm{tot}}\roundbracket{t}}{dt}\vert_{\mathrm{SOC}}$ in the following.

Based on Eq. (\ref{eq:eom-3terms}), we will derive the rate equations of $\{P_{Na\nu}\}$ separately according to $\hat H^I_{\mathrm{m-l}}$, $\hat H^I_{\mathrm{m-th}}$, $\hat H^I_{\mathrm{field}}$ and $\hat H^I_{\mathrm{SOC}}$.
Note that the rate equations due to $\hat H_{\mathrm{SOC}}$ are not derived explicitly. Instead, we treat the rates due to $\hat H_{\mathrm{SOC}}$ phenomenologically, because the relevant processes are insignificant compared to the other electronic transitions. 

The molecule-lead rate equations $\frac{d P_{Na\nu}}{dt}\vert_{\mathrm{m-l}}$ and the vibrational relaxation rate equation $\frac{d P_{Na\nu}}{dt}\vert_{\mathrm{m-th}}$ will be derived through a 2nd order expansion of the Liouville equation using Redfield theory, whereas the derivation of the radiative rate equations $\frac{d P_{Na\nu}}{dt}\vert_{\mathrm{field}}$ does not require a perturbation expansion since the radiative coupling is semiclassically treated in $\hat H_{\mathrm{field}}$.  

The rest of this section is organized as follows. In section  \ref{sec:RedfieldQME}, we briefly review Redfield theory. In section \ref{sec:QME-m-l}, \ref{sec:vib} and \ref{sec:opt}, we respectively discuss the derivations of charge transfer transitions, vibrational relaxations and optical transitions. Section \ref{sec:pme} finalizes the derivation of the Pauli master equation via phenomenologically introducing the other relevant processes. Finally, section \ref{sec:current} presents the current formula provided by the master equation approach.

\subsection{Redfield Theory}\label{sec:RedfieldQME}
We start from a component of the quantum master equation in Eq. (\ref{eq:eom-3terms}), i.e.,
\begin{align}\label{eq:RedQMEeq1}
    \left.\frac{d}{d t} \hat \rho^I_{\mathrm{tot}}\roundbracket{t}\right\vert_{\mathrm{coup}}= -\frac{i}{\hbar}\squarebracket{\hat H^I_{\mathrm{coup}}\roundbracket{t},\hat \rho^I_{\mathrm{tot}}\roundbracket{t}}\,,
\end{align}
where "coup" refers to the contributions from $\hat H_{\mathrm{m-l}}$ or $\hat H_{\mathrm{m-th}}$ by "m-l" or "m-th".

We then substitute the solution of Eq. (\ref{eq:eom-1storder}), i.e., $\hat \rho^I_{\mathrm{tot}}\roundbracket{t}=\hat\rho^I_{\mathrm{tot}}\roundbracket{t_0}-\frac{i}{\hbar}\int_{t_0}^{t}du[\hat V^I\roundbracket{u}\,,\hat \rho^I_{\mathrm{tot}}\roundbracket{u}]$,  into Eq.  (\ref{eq:RedQMEeq1}), apply $\mathrm{Tr}_{\mathrm{lead+th}}\{\cdots\}$ on both sides, and employ the Born-Markov approximation. Finally we arrive at the following Redfield master equation, 
\begin{align}\label{eq:eombr1}
    &\left.\frac{d}{d t} \hat \rho^I\roundbracket{t}\right\vert_{\mathrm{coup}}= \nonumber \\
    &-\frac{1}{\hbar^2}\int^{\infty}_0du\,\mathrm{Tr}_{\mathrm{lead+th}}\{\squarebracket{\hat H^I_{\mathrm{coup}}\roundbracket{t},\squarebracket{\hat H^I_{\mathrm{coup}}\roundbracket{t-u},\hat \rho^I\roundbracket{t}\hat\rho_{\mathrm{lead}}\hat\rho_{\mathrm{th}}}}\}\,.
\end{align}

When the molecule is weakly coupled to the electrodes and the thermal bath,
the electrons in the electrodes and the thermal bath modes are supposed to be relaxed significantly faster than all the other relevant processes, and the two baths are therefore assumed to be in their thermal equilibrium characterized by $\hat \rho_{\mathrm{lead}}$ and $\hat \rho_{\mathrm{th}}$ at all times. As a result, the Born-Markov approximation required by Redfield theory is valid in our problem. The \textit{Born approximation} $\hat \rho_{\mathrm{tot}}\roundbracket{t}\approx\hat \rho\roundbracket{t}\hat \rho_{\mathrm{lead}}\hat \rho_{\mathrm{th}}$ is first invoked to 
decouple the dynamics of the molecular system from that of the lead electrons and the thermal bath.
The non-local memory effect is then disregarded by replacing $\hat \rho^I_{\mathrm{tot}}\roundbracket{u}$ with $\hat \rho^I_{\mathrm{tot}}\roundbracket{t}$, which is the \textit{first Markov approximation}. The \textit{second Markov approximation} follows by increasing the upper limit of the integral to infinity after a variable transformation from $u$ to $u=t-u$.

In Eq.~(\ref{eq:eombr1}), it should be noted that the first order term $\mathrm{Tr}_{\mathrm{coup}}\{\squarebracket{\hat H_{\mathrm{m-l}}\roundbracket{t}\,,\hat \rho^I_{\mathrm{tot}}\roundbracket{t_0}}\}$ has been eliminated by invoking an uncoupled initial state $\hat\rho^I_{\mathrm{tot}}\roundbracket{t_0}=\hat \rho^I\roundbracket{t_0}\hat\rho_{\mathrm{lead}}\hat\rho_{\mathrm{th}}$. Besides, the 2nd order terms including the mix of  $\hat H_{\mathrm{coup}}$ and other coupling Hamiltonians are excluded because they do not survive after the trace operation $\mathrm{Tr}_{\mathrm{lead+th}}\{\cdots\}$.
\subsection{Charge Transfer Transition}\label{sec:QME-m-l}

In order to derive the charge transfer transitions, we first write down the transformed molecule-lead coupling Hamiltonian in the interaction picture as 
\begin{align}\label{eq:HamMolLeadTransformedInteractionPicture}
    \hat H'^I_{\mathrm{m-l}}=\sum_{\alpha}\roundbracket{\hat B^{+I}_{\alpha}\hat S'^{-I}_{\alpha}+\hat S'^{+I}_{\alpha}\hat B^{-I}_{\alpha}}
\end{align}
with 
\begin{align}\label{eq:sys-op-transformedInteractionPicture}
&\hat S'^{+I}_ {\alpha}=(\hat S'^{-I}_ {\alpha})^{\dagger}
=
\expfunc{\frac{i}{\hbar}\hat H_{\mathrm{mol}}t}\hat S'^{+}_ {\alpha}\expfunc{-\frac{i}{\hbar}\hat H_{\mathrm{mol}}t}\nonumber \\
=&\sum_{Nab}\sum_{\nu_b\nu_b}\biggr(M_{\nu_a\nu_b}\roundbracket{\lambda_{N-1,b}-\lambda_{N,a}}T^{\alpha}_{NaN-1b}\nonumber \\
&\times \vert N,a,\nu_a\rangle\langle N-1,b,\nu_b\vert\expfunc{\frac{i}{\hbar}\roundbracket{E_{N,a,\nu_a}-E_{N-1,b,\nu_b}}t} +\mathrm{h.c.}\biggr)\,,
\end{align}
and 
\begin{align}
    &B^{+I}_{\alpha}=\roundbracket{B^{-I}_{\alpha}}^{\dagger}=\expfunc{\frac{i}{\hbar}\hat H_{\mathrm{lead}}t}\hat B^{+}_ {\alpha}\expfunc{-\frac{i}{\hbar}\hat H_{\mathrm{lead}}t} \nonumber \\
    =&\sum_{\bvec k\sigma}M_{\alpha\bvec k\sigma}\hat a_{\alpha\bvec k\sigma}\expfunc{-\frac{i}{\hbar}\varepsilon_{\alpha\bvec k\sigma}t}\,.
\end{align}

A substitution of the $\hat H'^I_{\mathrm{m-l}}$ in Eq. (\ref{eq:HamMolLeadTransformedInteractionPicture}) into Eq. (\ref{eq:eombr1}) results in the following master equation,
\begin{align}\label{eq:breom-m-l}
    \left.\frac{d}{d t} \hat \rho^I\roundbracket{t}\right\rvert_{\mathrm{m-l}}
    =&
    -\frac{1}{\hbar^2}\sum_{\alpha}\int^{\infty}_0du
    \biggr(
    C_{\alpha}\roundbracket{-u}
    \hat{S}^{+I}_{\alpha}\roundbracket{t}
    \hat \rho^I\roundbracket{t}
    \hat{S}^{-I}_{\alpha}\roundbracket{t-u} \nonumber \\
    &+\bar C_{\alpha}\roundbracket{-u}
    \hat{S}^{-I}_{\alpha}\roundbracket{t}
    \hat \rho^I\roundbracket{t}
    \hat{S}^{+I}_{\alpha}\roundbracket{t-u} \nonumber \\
    &-C_{\alpha}\roundbracket{u}
    \hat{S}^{-I}_{\alpha}\roundbracket{t}
    \hat{S}^{+I}_{\alpha}\roundbracket{t-u}
    \hat \rho^I\roundbracket{t} \nonumber \\
    &-\bar C_{\alpha}\roundbracket{u}
    \hat{S}^{+I}_{\alpha}\roundbracket{t}
    \hat{S}^{-I}_{\alpha}\roundbracket{t-u}
    \hat \rho^I\roundbracket{t}
    +\mathrm{h.c.}
    \biggr),
\end{align}
where two types of correlation functions for electrode electrons are defined as
\begin{align}\label{eq:CorrlationFunc-m-l}
    C_{\alpha}\roundbracket{t-\tau}&=\mathrm{Tr}_{\mathrm{lead}}\{\hat B^{+I}_{\alpha}\roundbracket{t}\hat B^{-I}_{\alpha}\roundbracket{\tau}\} \nonumber \\
    &=\int^{+\infty}_{-\infty}d\omega J_{\alpha}\roundbracket{\omega}f_{\beta}\roundbracket{\omega,\mu_{\alpha}}\expfunc{\frac{i}{\hbar}\omega\roundbracket{t-\tau}}
\end{align}
and
\begin{align}\label{eq:CorrelationFuncBar-m-l}
    \bar C_{\alpha}\roundbracket{t-\tau}&=\mathrm{Tr}_{\mathrm{lead}}\{\hat B^{-I}_{\alpha}\roundbracket{t}\hat B^{+I}_{\alpha}\roundbracket{\tau}\} \nonumber \\
    &=\int^{+\infty}_{-\infty}d\omega J_{\alpha}\roundbracket{\omega}\roundbracket{1-f_{\beta}\roundbracket{\omega,\mu_{\alpha}}}\expfunc{-\frac{i}{\hbar}\omega\roundbracket{t-\tau}} \,.
\end{align}
The spectra density $J_{\alpha}\roundbracket{\omega}$ of the electrons in electrode $\alpha$ is specified by
\begin{align}
    J_{\alpha}\roundbracket{\omega}=\sum_{\bvec k\sigma}\vertbracket{M_{\alpha\bvec k\sigma}}^2\delta\roundbracket{\omega - \varepsilon_{\alpha\bvec k\sigma}/\hbar}
\end{align}
and the Fermi distribution function $f_{\beta}\roundbracket{\epsilon,\mu}$ is given by
\begin{align}
    f_{\beta}\roundbracket{\epsilon,\mu}=\frac{1}{1+\expfunc{\beta\roundbracket{\epsilon-\mu}}},
\end{align}
where $\mu$ is the electron chemical potential.

In order to derive the rate equations, we evaluate the diagonal matrix element with respect to electron-vibrational states $\{\ket{N,a,\nu}\}$ on both sides of Eq. (\ref{eq:breom-m-l}) and then arrive at 
\begin{align}\label{eq:eom-ml-matrixelement}
    &\left.\frac{d}{dt}\bra{N,a,\nu}\hat \rho^I\roundbracket{t}\ket{N,a,\nu}\right\rvert_{\mathrm{m-l}} \nonumber \\
    =&
    -\frac{2}{\hbar^2}\Re\sum_{\alpha}\int^{\infty}_0du
    \bra{N,a,\nu}
    C_{\alpha}\roundbracket{-u}
    \hat S'^{+I}_{\alpha}\roundbracket{t}
    \hat \rho^I\roundbracket{t}
    \hat S'^{-I}_{\alpha}\roundbracket{t-u} \nonumber \\
    &+\bar C_{\alpha}\roundbracket{-u}
    \hat S'^{-I}_{\alpha}\roundbracket{t}
    \hat \rho^I\roundbracket{t}
    \hat S'^{+I}_{\alpha}\roundbracket{t-u} \nonumber \\
    &-C_{\alpha}\roundbracket{u}
    \hat S'^{-I}_{\alpha}\roundbracket{t}
    \hat S'^{+I}_{\alpha}\roundbracket{t-u}
    \hat \rho^I\roundbracket{t} \nonumber \\
    &-\bar C_{\alpha}\roundbracket{u}
    \hat S'^{+I}_{\alpha}\roundbracket{t}
    \hat S'^{-I}_{\alpha}\roundbracket{t-u}
    \hat \rho^I\roundbracket{t}
    \ket{N,a,\nu},
\end{align}
where $\Re$ refers to the real part.

We next take the evaluation of the first term in Eq. (\ref{eq:eom-ml-matrixelement}) as an example, i.e.,
\begin{align}\label{eq:m-lME}
    &\int^{\infty}_0du\,C_{\alpha}\roundbracket{-u}\bra{N,a,\nu}\hat S'^{+I}_{\alpha}\roundbracket{t}
    \hat \rho^I\roundbracket{t}
    \hat S'^{-I}_{\alpha}\roundbracket{t-u}\ket{N,a,\nu} \nonumber \\
    =&\int^{\infty}_0du\,C_{\alpha}\roundbracket{-u}\sum_{a_1,\nu_1}\sum_{a_2,\nu_2}T^{\alpha *}_{N+1a_1,Na}T^{\alpha}_{N+1a_2,Na} \nonumber \\
    &\times M_{\nu\nu_1}\roundbracket{\lambda_{N,a}-\lambda_{N+1,a_1}}
    M_{\nu_2\nu}\roundbracket{\lambda_{N+1,a_2}-\lambda_{N,a}}
    \nonumber \\
    &\times\expfunc{\frac{i}{\hbar}\roundbracket{E_{N+1,a_2}-E_{N+1,a_1}+\roundbracket{\nu_2-\nu_1}\hbar\omega}t}\nonumber \\
    &\times
    \expfunc{-\frac{i}{\hbar}\roundbracket{E_{N+1,a_2,\nu_2}-E_{N,a,\nu}}\tau} \nonumber \\
    &\times\bra{N+1,a_1,\nu_1}\hat\rho^I\roundbracket{t}\ket{N+1,a_2,\nu_2},
\end{align}
which indicates that the evolution of $\bra{N,a,\nu}\hat \rho^I\roundbracket{t}\ket{N,a,\nu}$ (the population $P _{Na\nu}$) depends on the off-diagonal density matrix elements (coherence)\cite{Scholes2017}. 

We assume that the dynamics of the molecular system is resolved over a time step $\Delta t$ during which the reduced density matrix $\hat \rho^I\roundbracket{t}$ does not vary significantly, then the terms that satisfy $\Delta t\gg \vertbracket{E_{N+1,a_2}-E_{N+1,a_1}+\roundbracket{\nu_2-\nu_1}\hbar\omega}^{-1}$ does not contribute to the propagation of $\hat \rho^I\roundbracket{t}$ due to fast oscillations. As a result, only secular terms satisfying $E_{N+1,a_2}-E_{N+1,a_1}+(\nu_2-\nu)\hbar\omega=0$ are kept.  The consequence of the secular approximation differs for molecular systems with and without degenerate states. When there is no degeneracy in the system, the corresponding exponential factor with local time dependence can be replaced by Kronecker functions, e.g.,
\begin{align}
    \expfunc{\frac{i}{\hbar}\roundbracket{E_{N+1,a_2}-E_{N+1,a_1}+\roundbracket{\nu_2-\nu_1}\hbar\omega}t}
    \longrightarrow
    \delta_{a_1,a_2}\delta_{\nu_1,\nu_2}
\end{align}
which directly decouples the evolution of the population terms from the coherence terms of $\hat \rho^I\roundbracket{t}$. However, when the molecular system contains degenerate states, such as the system considered in this paper, terms satisfying $\nu_1=\nu_2$ and $E_{N+1,a_2}=E_{N+1,a_1}$ are still left after the secular approximation has been applied, wherein the coherence between degenerate electronic states is involved in the evolution of the populations. However, considering that the electronic dephasing is the fastest process in the system, the coherence terms can be neglected when studying steady-state transport characteristics. Consequently, we can drop all the terms containing the off-diagonal density matrix elements in Eq. (\ref{eq:m-lME}) regardless of the presence of degeneracy. As a result, the evaluation of the matrix elements gives us
\begin{align}
    &\int^{\infty}_0du\,C_{\alpha}\roundbracket{-u}\bra{N,a,\nu}\hat S'^{+I}_{\alpha}\roundbracket{t}
    \hat \rho^I\roundbracket{t}
    \hat S'^{-I}_{\alpha}\roundbracket{t-u}\ket{N,a,\nu} \nonumber \\
    =&\sum_{a_1,\nu_1}\vertbracket{T^{\alpha}_{N+1a_1,Na}}^2
    \vertbracket{M_{\nu\nu_1}\roundbracket{\lambda_{N+1,a_1}-\lambda_{N,a}}}^2
    \nonumber \\
    &\times\expfunc{-\frac{i}{\hbar}\roundbracket{E_{N+1,a_1,\nu_1}-E_{N,a,\nu}}\tau}
    P_{N+1,a_1,\nu_1},
\end{align}
where the diagonal matrix element has been replaced with the population term due to the relationship 
\begin{align}
    P_{N,a,\nu}=\bra{N,a,\nu}\hat \rho\roundbracket{t}\ket{N,a,\nu}=\bra{N,a,\nu}\hat \rho^I\roundbracket{t}\ket{N,a,\nu}\,.
\end{align}

We can use the same procedure as above to evaluate the other terms in Eq. (\ref{eq:m-lME}). Finally, we obtain the following molecule-lead rate equations
\begin{align}
    &\left.\frac{d P_{N,a,\nu}}{dt}\right\vert_{\mathrm{m-l}}\nonumber
     \\
     =&\sum_{\alpha}\sum_{N',a'}\roundbracket{\delta_{N',N+1}+\delta_{N',N-1}}\nonumber \\
    &\times\sum_{\nu'}\roundbracket{
    k^{\alpha}_{N,a,\nu\leftarrow N',a',\nu'}P_{N',a',\nu'}-k^{\alpha}_{N',a',\nu'\leftarrow N,a,\nu}P_{N,a,\nu}
    }\,.
\end{align}

The rates of electron transport and hole transport processes are thus given by
\begin{align}
\label{eq:ChargeTransferRate}
k^{\alpha}_{N+1,b,\nu'\leftarrow N,a,\nu}
=&
\gamma^{\alpha}_{N+1b\nu',Na\nu}f_{\beta}\roundbracket{\varepsilon_{N+1b\nu',Na\nu},\mu_{\alpha}}\\
\label{eq:HoleTransferRate}
k^{\alpha}_{N-1,b,\nu'\leftarrow N,a,\nu}
=&
\gamma^{\alpha}_{N-1b\nu',Na\nu}
\roundbracket{1-f_{\beta}\roundbracket{\varepsilon_{Na\nu,N-1b\nu'},\mu_{\alpha}}}
\end{align}
in which the prefactors read
\begin{align}
\gamma^{\alpha}_{N\pm 1b\nu',Na\nu}=&
\frac{2\pi}{\hbar^2}\bar J_{\alpha}\vertbracket{T^{\alpha}_{N\pm 1b,N,a}}^2\vertbracket{M_{\nu'\nu}\roundbracket{\lambda_{N\pm 1,b}-\lambda_{N,a}}}^2,
\end{align}
where $\bar J_{\alpha}$ denotes a constant spectral density of electrode $\alpha$ in the wide band limit. After evaluating $T^{\alpha}_{N\pm 1b,N,a}$ as detailed in Appendix \ref{sec:appendix1}, we can rewrite the prefactor $\gamma^{\alpha}_{N-1b\nu',Na\nu}$ as
\begin{align}\label{eq:ChargeTransferRatePrefactor}
\gamma^{\alpha}_{N\pm 1b\nu',Na\nu}=&
\Gamma_{\alpha}\nu_{N\pm 1b,Na}\vertbracket{M_{\nu'\nu}\roundbracket{\lambda_{N\pm 1,b}-\lambda_{N,a}}}^2,
\end{align}
where $\Gamma_{\alpha}=\frac{2\pi}{\hbar^2}\vertbracket{\zeta_{\alpha}}\bar J_{\alpha}$ is the characteristic rate of charge transfer transition associated with electrode $\alpha$, and the dimensionless coupling $\nu_{N\pm 1b,Na}$ of the transition $\ket{N\pm 1,b}\leftrightarrow\ket{N,a}$ has been given in \textbf{Table S1} in Ref. [\!\!\citenum{Fu2018}] for all relevant transitions considered in this paper. The meaning of $\zeta_{\alpha}$ and the evaluation of $\nu_{N\pm 1b,Na}$ are also discussed in the Appendix \ref{sec:appendix1}. 

\subsection{Vibrational Relaxation}\label{sec:vib}
The derivation of the vibrational relaxation rate equations follows the same procedure as the above section. In the interaction picture, the transformed coupling Hamiltonian $\hat H'_{\mathrm{m-th}}$ reads
\begin{align}\label{eq:m-ph-classical2-transformed-interaction}
    \hat H'^I_{\mathrm{m-th}}\roundbracket{t}= \hat S'^I_{\mathrm{m-th}}\roundbracket{t}\hat B^I_{\mathrm{m-th}}\roundbracket{t}
\end{align}
where 
\begin{align}\label{eq:m-ph-classical2-transformed-S-interaction}
    \hat S'^I_{\mathrm{m-th}}\roundbracket{t}=&\sum_{A\in\cF}\expfunc{\frac{i}{\hbar}\hat H_{\mathrm{mol}}t}\hat q\vert A\rangle \langle A\vert\,\expfunc{-\frac{i}{\hbar}\hat H_{\mathrm{mol}}t} \nonumber \\
    =&\sum_{A\in\cF}\sqrt{\frac{\hbar}{2m_{\mathrm{vib}}\omega_{\mathrm{vib}}}}\roundbracket{\hat b^{\dagger}\expfunc{i\omega_{\mathrm{vib}}t}+\hat b\,\expfunc{-i\omega_{\mathrm{vib}}t}}\vert A\rangle \langle A\vert\,,
\end{align}
and
\begin{align}\label{eq:m-ph-classical2-transformed-B-interaction}
    \hat B^I_{\mathrm{m-th}}\roundbracket{t}=&\sum_{\alpha}c_\alpha\expfunc{\frac{i}{\hbar}\hat H_{\mathrm{th}}t}\hat q_{\alpha}\,\expfunc{-\frac{i}{\hbar}\hat H_{\mathrm{th}}t} \nonumber \\
    =&\sum_{\alpha}c_\alpha\sqrt{\frac{\hbar}{2m_{\alpha}\omega_{\alpha}}}\roundbracket{\hat b^{\dagger}_{\alpha}\expfunc{i\omega_{\alpha}t}+\hat b_{\alpha}\expfunc{-i\omega_{\alpha}t}}\,.
\end{align}
For simplicity, we abbreviate $\ket{N,a}$ as $\ket{A}$ here.
We then plug Eq. (\ref{eq:m-ph-classical2-transformed-interaction}) into Eq. (\ref{eq:eombr1}) and obtain the following master equation,
\begin{align}\label{eq:breom-m-ph}
    \left.\frac{d}{d t} \hat \rho^I\roundbracket{t}\right\rvert_{\mathrm{m-th}}
    =&
    -\frac{1}{\hbar^2}\int^{\infty}_0du
    \biggr(
    C_{\mathrm{th}}\roundbracket{u}
    \hat{S}'^{I}_{\mathrm{m-th}}\roundbracket{t}
    \hat{S}'^{I}_{\mathrm{m-th}}\roundbracket{t-u}
    \hat \rho^I\roundbracket{t}
    \nonumber \\
    &+C_{\mathrm{th}}\roundbracket{-u}
    \hat \rho^I\roundbracket{t}
    \hat{S}'^{I}_{\mathrm{m-th}}\roundbracket{t-u}
    \hat{S}'^{I}_{\mathrm{m-th}}\roundbracket{t} \nonumber \\
    &-C_{\mathrm{th}}\roundbracket{-u}
    \hat{S}'^{I}_{\mathrm{m-th}}\roundbracket{t}
    \hat \rho^I\roundbracket{t}
    \hat{S}'^{I}_{\mathrm{m-th}}\roundbracket{t-u} \nonumber \\
    &-C_{\mathrm{th}}\roundbracket{u}
    \hat{S}'^{I}_{\mathrm{m-th}}\roundbracket{t-u}
    \hat \rho^I\roundbracket{t}
    \hat{S}'^{I}_{\mathrm{m-th}}\roundbracket{t},
\end{align}
where the thermal bath correlation function is defined as
\begin{align}
    C_{\mathrm{th}}\roundbracket{\tau}=&\mathrm{Tr}_{\mathrm{lead+th}}\{
    \hat B^I_{\mathrm{m-th}}\roundbracket{t}
    \hat B^I_{\mathrm{m-th}}\roundbracket{t-\tau}
    \} \nonumber \\
    =&\int^{\infty}_0d\omega\frac{\hbar}{\pi}J_{\mathrm{th}}\roundbracket{\omega}\roundbracket{n\roundbracket{\omega}\expfunc{i\omega\tau}+\roundbracket{n\roundbracket{\omega}+1}\expfunc{-i\omega\tau}}.
\end{align}
The spectral density $J_{\mathrm{th}}\roundbracket{\omega}$ of the thermal bath is defined as
\begin{align}
    J_{\mathrm{th}}\roundbracket{\omega}=\pi\sum_{\alpha}\frac{c^2_{\alpha}}{2m_{\alpha}\omega_{\alpha}}\delfunc{\omega-\omega_{\alpha}}
\end{align}
and the Bose-Einstein distribution function $n_{\beta}\roundbracket{\omega}$ is given by
\begin{align}
    n_{\beta}\roundbracket{\omega}=\frac{1}{\expfunc{\beta\hbar\omega}-1}\,.
\end{align}

A substitution of $S'^I_{\mathrm{m-th}}\roundbracket{t}$ in Eq. (\ref{eq:m-ph-classical2-transformed-S-interaction}) into Eq. (\ref{eq:breom-m-ph}) leads us to
\begin{align}\label{eq:breom-m-ph-2}
    &\left.\frac{d}{d t} \hat \rho^I\roundbracket{t}\right\rvert_{\mathrm{m-th}}\nonumber \\
    =&-\frac{1}{2m_{\mathrm{vib}}\omega_{\mathrm{vib}}}J_{\mathrm{th}}\roundbracket{\omega_{\mathrm{vib}}}\sum_{A,A'}\vert A\rangle\langle A'\vert
    \biggr[
    \roundbracket{n\roundbracket{\omega_{\mathrm{vib}}}+1} \nonumber \\
    &\times
    \biggr(
    \hat b^{\dagger}_A\hat b_A \hat\rho^I_{AA'}\roundbracket{t}
    +
    \hat\rho^I_{AA'}\roundbracket{t}\hat b^{\dagger}_{A'}\hat b_{A'} 
    -
    2\hat b_A\hat\rho^I_{AA'}\roundbracket{t}\hat b^{\dagger}_{A'}
    \biggr) \nonumber \\
    &+n\roundbracket{\omega_{\mathrm{vib}}}
    \biggr(
    \hat b_A\hat b^{\dagger}_A \hat\rho^I_{AA'}\roundbracket{t}
    +
    \hat\rho^I_{AA'}\roundbracket{t}\hat b_{A'}\hat b^{\dagger}_{A'} \nonumber \\
    &-
    2\hat b^{\dagger}_A\hat\rho^I_{AA'}\roundbracket{t}\hat b_{A'}
    \biggr)
    \biggr]\,,
\end{align}
where rotating wave approximation has been applied, and $\hat \rho_{AA'}=\langle A\vert\hat\rho\vert A'\rangle$ refers to an electronic block of the reduced density matrix.
Eq. (\ref{eq:breom-m-ph-2}) suggests that $\hat H_{\mathrm{m-th}}$ not only leads to the vibrational relaxations  
within the same electronic state manifold but also results in the dephasing of coherence between states associated with different electronic state manifold.

The vibrational relaxation rate equations $\frac{d P_{N,a,\nu}}{dt}\vert_{\mathrm{m-th}}$ is derived by taking the diagonal matrix element of Eq. (\ref{eq:breom-m-ph-2}), which gives
\begin{align}\label{eq:RateEquationVib}
    \left.\frac{dP_{N,a,\nu}}{dt}\right\vert_{\mathrm{m-th}}=\sum_{\nu'}\roundbracket{
    k^{\mathrm{vib}}_{N,a,\nu\leftarrow N,a,\nu'}P_{N,a,\nu'}
    -
    k^{\mathrm{vib}}_{N,a,\nu'\leftarrow N,a,\nu}P_{N,a,\nu}
    },
\end{align}
where the rates of vibrational relaxation is
\begin{align}\label{eq:RatesVib}
    k^{\mathrm{vib}}_{N,a,\nu\leftarrow N,a,\nu'} 
    =&
    \gamma_{\mathrm{p}}
    \biggr(
    \delta_{\nu+1,\nu'}\roundbracket{\nu+1}n\roundbracket{\omega_{\mathrm{vib}}} \nonumber \\
    &+\delta_{\nu-1,\nu'}\nu\roundbracket{n\roundbracket{\omega_{\mathrm{vib}}}+1}\biggr).
\end{align}
Here, $\gamma_{\mathrm{p}}=\frac{J_{\mathrm{th}}\roundbracket{\omega_{\mathrm{vib}}}}{m_{\mathrm{vib}}\omega_{\mathrm{vib}}}$ is the characteristics rate of vibrational relaxation. Note that the corresponding characteristic time scale $\tau_{\mathrm{p}}=\gamma_{\mathrm{p}}^{-1}$ refers to the life time of the first excited vibrational state.
\subsection{Optical Transition}\label{sec:opt}
In the interaction picture, the transformed radiative coupling Hamiltonian reads
\begin{align}\label{eq:field-transformed-interaction}
    \hat H'^I_{\mathrm{field}}
    =&-\textbf{E}(t)\cdot\textbf{d}_{01}\nonumber \\
    &\times \sum_{\nu_0,\nu_1}\biggr(M_{\nu_0\nu_1}\roundbracket{\lambda_{0,S_1}-\lambda_{0,S_0}}\ket{0,S_0,\nu_0}\bra{0,S_1,\nu_1}\nonumber \\
    &\times\expfunc{\frac{i}{\hbar}\roundbracket{E_{0,S_0,\nu_0}-E_{0,S_1,\nu_1}}}+\mathrm{h.c.}\biggr) \,.
\end{align}

Since the light-matter interaction is considered semi-classically, we derive the radiative rate equations without employing a $2$nd order expansion.

We start from $\frac{d}{dt}\hat \rho^I_{\mathrm{tot}}\roundbracket{t}\vert_{\mathrm{field}}=-\frac{i}{\hbar}[\hat H^I_{\mathrm{field}},\hat \rho^I_{\mathrm{tot}}\roundbracket{t}]$, apply the Born approximation $\hat\rho_{\mathrm{tot}}=\hat \rho\roundbracket{t}\hat\rho_{\mathrm{lead}}\hat\rho_{\mathrm{th}}$ and the trace operation $\mathrm{Tr}_{\mathrm{lead+th}}\{\cdots\}$. As a result, we have the following equation of motion of the reduced density operator $\hat{\rho}^I\roundbracket{t}$,
\begin{align}\label{eq:eom-opt}
    \left.\frac{d}{dt}\hat \rho^I\roundbracket{t}\right\vert_{\mathrm{field}}=-\frac{i}{\hbar}\roundbracket{\hat H^I_{\mathrm{field}}\hat \rho^I\roundbracket{t}-\hat \rho^I\roundbracket{t}\hat H^I_{\mathrm{field}}}.
\end{align}
By substituting Eq. (\ref{eq:field-transformed-interaction}) into Eq. (\ref{eq:eom-opt}) and taking the diagonal matrix element with respect to $\ket{0,S_0,\nu_0}$, one can derive the rate equation of $P_{0,S_0,\nu_0}$,
\begin{align}\label{eq:eom-me-1}
    &\left.\frac{d}{dt}P_{0,S_0,\nu_0}\right\vert_{\mathrm{field}}\nonumber \\
    =&\frac{i}{\hbar}\bvec E\roundbracket{t}\cdot\bvec d_{10}\nonumber \\
    &\times\sum_{\nu_1}\roundbracket{M_{\nu_0\nu_1}\biggr(\lambda_{0,S_0}-\lambda_{0,S_1}}\bra{0,S_1,\nu_1}\hat\rho^I\roundbracket{t}\ket{0,S_0,\nu_0}\nonumber \\
    &\times\expfunc{\frac{i}{\hbar}\roundbracket{E_{0,S_0,\nu_0}-E_{0,S_1,\nu_1}}t}-\mathrm{c.c.}\biggr),
\end{align}
where $\mathrm{c.c.}$ stands for complex conjugate. In order to solve the coherence term $\bra{0,S_1,\nu}\hat\rho^I\roundbracket{t}\ket{0,S_0,\nu_0}$, we turn to the following equation of motion in the Schr\"{o}dinger picture,
\begin{align}\label{eq:S0S1eom-1}
    &\frac{d}{dt}\bra{0,S_0,\nu_0}\hat\rho\roundbracket{t}\ket{0,S_1,\nu_1}\nonumber\\
    =&
    -\frac{i}{\hbar}\roundbracket{E_{0,S_0,\nu_0}-E_{0,S_1,\nu_1}-i\hbar\kappa_{0S_0\nu_0,0S_1\nu_1}}\nonumber\\
    &\times\bra{0,S_0,\nu_0}\hat\rho\roundbracket{t}\ket{0,S_1,\nu_1} +
    \frac{i}{\hbar}\bvec E\roundbracket{t}\cdot \bvec d_{10}\nonumber\\
    &\times
    \biggr(
    \sum_{\nu'}M_{\nu_0\nu'}\roundbracket{\lambda_{0,S_0}-\lambda_{0,S_1}}\bra{0,S_1,\nu'}\hat\rho\roundbracket{t}\ket{0,S_1,\nu_1}\nonumber \\
    &-\sum_{\nu}M_{\nu\nu_1}\roundbracket{\lambda_{0,S_0}-\lambda_{0,S_1}}\bra{0,S_0,\nu_0}\hat\rho\roundbracket{t}\ket{0,S_0,\nu}
    \biggr),
\end{align}
where $\kappa_{0S_0\nu_0,0S_1\nu_1}$ describes the pure dephasing of the electronic coherence which originates from the other coupling Hamiltonians, i.e., $\hat H_{\mathrm{m-l}}$, $\hat H_\mathrm{m-th}$. An explicit evaluation in Appendix \ref{sec:appendix2} shows that $\kappa_{0S_0\nu_0,0S_1\nu_1}$ is a sum over the rates of all the processes that start from $\ket{0,S_0,\nu_0}$ and $\ket{0,S_1,\nu_1}$ excluding optical transitions, e.g., charge transfer transitions and vibrational relaxations. 

When only the steady-state dynamics is concerned,  $\bra{0,S_1,\nu'}\hat\rho\roundbracket{t}\ket{0,S_1,\nu_1}$ and $\bra{0,S_0,\nu_0}\hat\rho\roundbracket{t}\ket{0,S_0,\nu}$ in  Eq.~(\ref{eq:S0S1eom-1}) are time-independent. In this case, according to the time-periodicity of $E\roundbracket{t}$, one can take a trial solution of $\bra{0,S_0,\nu_0}\hat\rho\roundbracket{t}\ket{0,S_1,\nu_1}$,
\begin{align}\label{eq:trialsolution}
    \bra{0,S_0,\nu_0}\hat\rho\roundbracket{t}\ket{0,S_1,\nu_1}=\expfunc{i\omega t}\rho_{0S_0\nu_0,0S_1\nu_1}\roundbracket{\omega}.
\end{align}
$\rho_{0S_0\nu_0,0S_1\nu_1}\roundbracket{\omega}$ is then solved as
\begin{align}\label{eq:freqsolution}
    &\rho_{0S_0\nu_0,0S_1\nu_1}\roundbracket{\omega}\nonumber \\
    &=\frac{1}{\hbar}\bvec E_0\cdot\bvec d_{10}
    \frac{1}{\omega -\roundbracket{E_{0,S_1,\nu_1}-E_{0,S_0,\nu_0}}/\hbar-i\kappa_{0S_0\nu,0S_1\nu_1}}\nonumber\\
    &\times
    \biggr(
    \sum_{\nu'}M_{\nu_0\nu'}\roundbracket{\lambda_{0,S_0}-\lambda_{0,S_1}}\bra{0,S_1,\nu'}\hat\rho\roundbracket{t}\ket{0,S_1,\nu_1}\nonumber \\
    &-\sum_{\nu}M_{\nu\nu_1}\roundbracket{\lambda_{0,S_0}-\lambda_{0,S_1}}\bra{0,S_0,\nu_0}\hat\rho\roundbracket{t}\ket{0,S_0,\nu}
    \biggr),
\end{align}
where we have dropped the fast oscillating terms containing $\expfunc{\pm 2i\omega}$ (rotating wave approximation). After substituting the solution Eq. (\ref{eq:trialsolution}) and (\ref{eq:freqsolution}) into Eq. (\ref{eq:eom-me-1}), we obtain the rate equation of $P_{0,S_0,\nu_0}$ in the following,
\begin{align}\label{eq:eom-0S0-me}
    &\left.\frac{d}{dt}P_{0,S_0,\nu_0}\right\vert_{\mathrm{field}}\nonumber \\
    =&
    \frac{i}{\hbar^2}\vertbracket{\bvec E_0\cdot\bvec d_{10}}^2
    \biggr[
    \sum_{\nu_1}M_{\nu_1\nu_0}\roundbracket{\lambda_{0,S_1}-\lambda_{0,S_0}}\nonumber \\
    &\times\frac{1}{\omega -\roundbracket{E_{0,S_1,\nu_1}-E_{0,S_0,\nu_0}}/\hbar-i\kappa_{0S_0\nu,0S_1\nu_1}}\nonumber \\
    &\times
    \sum_{\nu}\biggr(
    M_{\nu_0\nu}\roundbracket{\lambda_{0,S_0}-\lambda_{0,S_1}}\bra{0,S_1,\nu}\hat\rho\roundbracket{t}\ket{0,S_1,\nu_1}\nonumber\\
    &-M_{\nu\nu_1}\roundbracket{\lambda_{0,S_0}-\lambda_{0,S_1}}\bra{0,S_0,\nu_0}\hat\rho\roundbracket{t}\ket{0,S_0,\nu}
    \biggr)
    +\mathrm{c.c.}
    \biggr]\,.
\end{align}
The coherence terms $\bra{0,S_a,\nu}\hat\rho\roundbracket{t}\ket{0,S_a,\nu'}$ in Eq. (\ref{eq:eom-0S0-me}) can be dropped by transforming Eq. (\ref{eq:eom-0S0-me}) back to the interaction picture and applying the secular approximation. We finally obtain the radiative rate equation of $\ket{0,S_0,\nu_0}$, i.e.,
\begin{align}
    &\left.\frac{dP_{0,S_0,\nu_0}}{dt}\right\vert_{\mathrm{field}}\nonumber\\
    =&\sum_{\nu_1}\roundbracket{k^{\mathrm{field}}_{0,S_0,\nu_0\leftarrow 0,S_1,\nu_1}P_{0,S_1,\nu_1}-k^{\mathrm{field}}_{0,S_1,\nu_1\leftarrow 0,S_0,\nu_0}P_{0,S_0,\nu_0}},
\end{align}
with the rates of stimulated optical transitions $k_{0,S_0,\nu_1\leftrightarrow 0,S_1,\nu_1}^{\mathrm{field}}$ given by
\begin{widetext}
\begin{align}\label{eq:rate-eom-opt-0S0}
    k_{0,S_0,\nu_1\leftrightarrow 0,S_1,\nu_1}^{\mathrm{field}}
    =\frac{2}{\hbar^2}
    \vertbracket{\bvec E_0\cdot\bvec d_{01}}^2
    \vertbracket{M_{\nu_0\nu_1}\roundbracket{\lambda_{0,S_1}-\lambda_{0,S_0}}}^2  \frac{\vertbracket{\Re\kappa_{0S_0\nu_0,S_1\nu_1}}}{\squarebracket{\omega-\roundbracket{E_{0,S_1,\nu_1}-E_{0,S_0,\nu_0}}/\hbar+\Im\kappa_{0S_0\nu_0,S_1\nu_1}}^2+\vertbracket{\Re\kappa_{0S_0\nu_0,S_1\nu_1}}^2}\,,
\end{align}
where $\Im$ refers to the imaginary part.

The rate equation of the other singlet state $\ket{0,S_1,\nu_1}$ could be obtained similarly as
\begin{align}\label{eq:rate-eom-opt-0S1}
    \left.\frac{dP_{0,S_1,\nu_1}}{dt}\right\vert_{\mathrm{field}}
    =\sum_{\nu_0}\roundbracket{k^{\mathrm{field}}_{0,S_1,\nu_1\leftarrow 0,S_0,\nu_0}P_{0,S_0,\nu_0}-k^{\mathrm{field}}_{0,S_0,\nu_0\leftarrow 0,S_1,\nu_1}P_{0,S_1,\nu_1}}\,.
\end{align}
\end{widetext}
\subsection{Pauli Master Equation}\label{sec:pme}
The final form of the Pauli master equations  is achieved by phenomenologically introducing the rates of singlet-triplet transitions and the rate of spontaneous emission. The singlet-triplet transitions include the processes of intersystem crossing $\ket{0,S_1,\nu}\rightarrow\ket{0,T^m_1,\nu'}$ and phosphorescence $\ket{0,T^m_1,\nu}\rightarrow\ket{0,S_0,\nu'}$. We collect all the rates discussed above and arrive at the following form of Pauli master equations,
\begin{widetext}
\begin{align}\label{eq:PauliQME}
    \frac{d P_{N,a,\nu}}{dt}
    =&\sum_{N',a',\nu'}\roundbracket{\delta_{N',N+1}+\delta_{N',N-1}} \sum_{\alpha}\roundbracket{k^{\alpha}_{N,a,\nu\leftarrow N',a',\nu'}P_{N',a',\nu'}-k^{\alpha}_{N',a',\nu'\leftarrow N,a,\nu}P_{N,a,\nu}}
    +\sum_{\nu'}\roundbracket{
    k^{\mathrm{th}}_{N,a,\nu\leftarrow N,a,\nu'}P_{N,a,\nu'}
    -
    k^{\mathrm{th}}_{N,a,\nu'\leftarrow N,a,\nu}P_{N,a,\nu} 
    } \nonumber \\
    &+
    \delta_{N,0}\delta_{a,S_0}\sum_{\nu'}\roundbracket{k^{\mathrm{field}}_{N,a,\nu\leftarrow 0,S_1,\nu'}P_{0,S_1,\nu'}-k^{\mathrm{field}}_{0,S_1,\nu'\leftarrow 0,S_0,\nu}P_{N,a,\nu}} +
    \delta_{N,0}\delta_{a,S_1}\sum_{\nu'}\roundbracket{k^{\mathrm{field}}_{N,a,\nu\leftarrow 0,S_0,\nu'}P_{0,S_0,\nu'}-k^{\mathrm{field}}_{0,S_0,\nu'\leftarrow N,a,\nu}P_{N,a,\nu}} \nonumber \\
    &+\sum_{m=0,\pm 1}\sum_{\nu'}\biggr( \delta_{N,0}\delta_{a,T^m_1}\roundbracket{
    k^{\mathrm{SOC}}_{N,a,\nu\leftarrow 0,S_1,\nu'}P_{0,S_1,\nu'} 
    -
    k^{\mathrm{SOC}}_{0,S_0,\nu'\leftarrow N,a,\nu}P_{N,a,\nu}} 
    -\delta_{N,0}\delta_{a,S_1}k^{\mathrm{SOC}}_{0,T^m_1,\nu'\leftarrow N,a,\nu}P_{N,a,\nu} \nonumber \\ &+\delta_{N,0}\delta_{a,S_0}k^{\mathrm{SOC}}_{N,a,\nu\leftarrow0,T^m_1,\nu' }P_{0,T^m_1,\nu'}
    \biggr)
    +\delta_{N,0}\delta_{a,S_0}\sum_{\nu'}k^{\mathrm{spon}}_{N,a,
    \nu\leftarrow 0,S_1,\nu'}P_{0,S_1,\nu'}-\delta_{N,0}\delta_{a,S_1}\sum_{\nu'}k^{\mathrm{spon}}_{0,S_0,\nu'\leftarrow N,a,\nu}P_{N,a,\nu} 
    .
\end{align}
\end{widetext}
where $k^{\mathrm{SOC}}_{N,a,\nu\leftarrow N',a',\nu' }$ refers to the rate due to SOC and $k^{\mathrm{spon}}_{0,S_0,\nu\leftarrow 0,S_1,\nu'}$ refers to the rate of spontaneous emission.

\subsection{Current Formula}\label{sec:current}
The electric current through the molecular junction is defined by the flow of electron through electrode $\alpha$, i.e., 
\begin{align}
    I_{\alpha}\roundbracket{t}
    =-\vertbracket{e}\frac{d}{d t} \langle \hat  N_{\alpha}\rangle
    =-\vertbracket{e}\frac{i}{\hbar}\mathrm{Tr}\{\left[\hat N_{\mathrm{lead},\alpha}\,,\hat H_{\mathrm{tot}}\right]\hat\rho_{\mathrm{tot}}\roundbracket{t}\},
\end{align}
where the trace $\mathrm{Tr}\{\cdots\}$ includes the average over the electrons in the electrodes, the thermal bath modes and the electron-vibrational states of the molecular system. Following the same procedure for deriving the molecule-lead rate equations, we obtain the following current formula,
\begin{align}\label{eq:current1}
    I_{\alpha}\roundbracket{t}=
    \vertbracket{e}
    \sum_{N,a,b}\sum_{v,v'}
    \biggr(
    k^{\alpha}_{N+1,b,v'\leftarrow N,a,v}
    -  
    k^{\alpha}_{N-1,b,v'\leftarrow N,a,v}
    \biggr)
    P_{N,a,v}\,.
\end{align}
Since we study the steady-state current, the population $P_{N,a,v}$ takes the stationary solution of the Pauli master equation.

The current formula Eq.\!\!\!\!\!  (\ref{eq:current1}) suggests that the current through the molecular junction is determined by both the rates of the charge transfer transitions and the steady-state population of the involved states, wherein the population is determined by the rates of relevant electronic transitions. It is noted that the source-drain bias $V_{\mathrm{SD}}$ and the gate voltage $V_{\mathrm{G}}$ control the rates of charge transfer transitions in Eqs. (\ref{eq:ChargeTransferRate}) and (\ref{eq:HoleTransferRate}) by virtue of the determination of the Fermi distribution function. 
In the low temperature limit, the Fermi distribution function becomes a step function. In this case, the rates of transitions between a neutral state $\ket{0,a,\nu_a}$ and a charged state $\ket{N,b,\nu_b}$ ($N=\pm 1$) can be clearly expressed as a function of $V_{\mathrm{SD}}$ and $V_{\mathrm{G}}$,
\begin{align}\label{eq:ChargeTransferLowT}
    k^{\alpha}_{N,b,\nu_b\leftarrow 0,a,\nu_a}
=&
\gamma^{\alpha}_{Nb\nu_b,0a\nu_a}\theta\roundbracket{E_{0,a,\nu_a}-\tilde{E}^{\alpha}_{N,b,\nu_b}\roundbracket{V_{\mathrm{G}},V_{\mathrm{\mathrm{SD}}}}} 
\end{align}
and
\begin{align}\label{eq:HoleTransferLowT}
k^{\alpha}_{0,a,\nu_a\leftarrow N,b,\nu_b}
=&
\gamma^{\alpha}_{0a\nu_a,Nb\nu_b}
\theta\roundbracket{\tilde{E}^{\alpha}_{N,b,\nu_b}\roundbracket{V_{\mathrm{G}},V_{\mathrm{V_{\mathrm{SD}}}}}-E_{0,a,\nu_a}}\,.
\end{align}
The step function $\theta\roundbracket{x}$ takes a value of $1$ for $x>0$ and a value of $0$ otherwise. The \textit{renormalized state energy} associated with electrode $\alpha$ is defined by
\begin{align}\label{eq:ReChgEn}
    \tilde{E}^{\alpha}_{N,a,\nu}\roundbracket{V_{\mathrm{G}},V_{\mathrm{SD}}} = E_{N,a,\nu}-N(\mu_0+\zeta_{\alpha} eV_{\mathrm{SD}}/2)-NeV_{\mathrm{G}},
\end{align} 
where $\zeta_{\mathrm{S}}=1$ and $\zeta_{\mathrm{D}}=-1$. Eq. (\ref{eq:ReChgEn}) combines the energy level of $\ket{N,a,\nu}$ with the Fermi energy on electrode $\alpha$, i.e., $\mu_0+\zeta_{\alpha} eV_{\mathrm{SD}}/2$, and the energy shift caused by gate voltage, i.e., $-NeV_{\mathrm{G}}$. Eqs. (\ref{eq:ChargeTransferLowT}) and (\ref{eq:HoleTransferLowT}) clearly reveal that the charge transfer transitions $\ket{0,a,\nu_a}\leftrightarrow\ket{N,b,\nu_b}$ ($N=\pm 1$) are explicitly determined by the energy level alignments between $E_{0,a,\nu}$ and $ \tilde{E}^{\alpha}_{N,a,\nu}\roundbracket{V_{\mathrm{G}},V_{\mathrm{SD}}}$, which provides us with a handy tool for identifying the dominant transport channels.

\section{Parameters}
\label{Parameters}

\begin{figure}
\center{\includegraphics[width=1.0\linewidth]{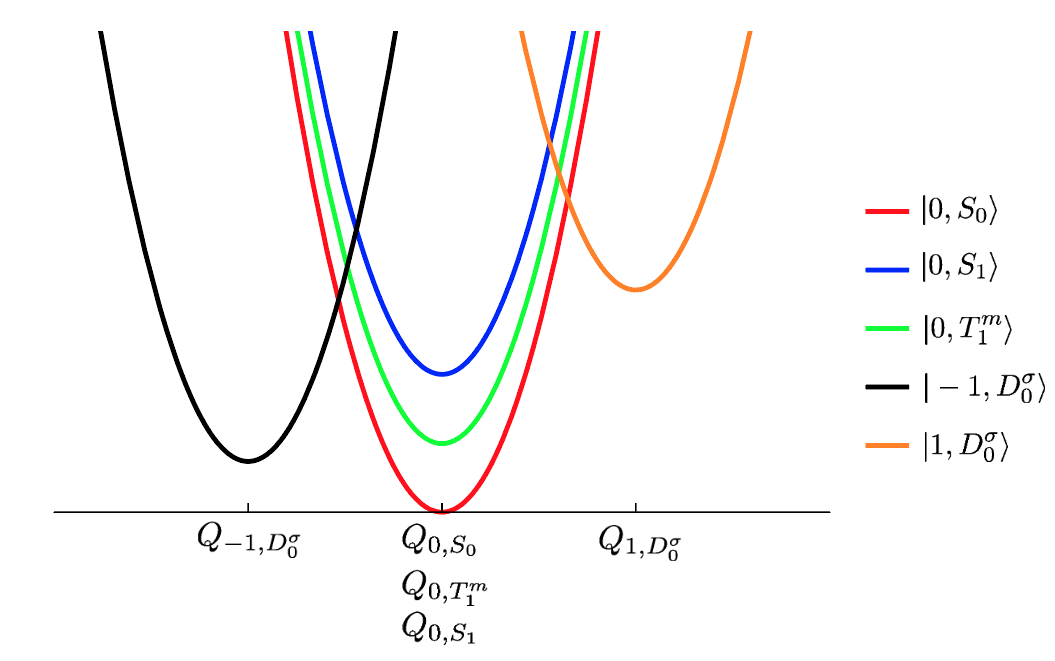}}
\caption{The potential energy surfaces of relevant electronic states is plotted according to $V_{N,a}=E_{N,a}+\frac{1}{4}\roundbracket{Q-Q_{N,a}}^2$, where $Q_{N,a}=2\lambda_{N,a}$ is the dimensionless equilibrium coordinate of $\ket{N,a}$.}
\label{fig:PES}
\end{figure}
We compute the electron transport characteristics in the succeeding section using the system parameters compiled in Table \ref{table:paras}. As shown in Figure \ref{fig:PES}, we consider a model system with a specific configuration of PESs. In this configuration, the electronic transitions between neutral states are not coupled to the reaction coordinate, whereas the PESs of cation states and anion states are displaced oppositely with respect to the PESs of neutral states by the same amount. The displacement between the PESs is characterized by the e-p coupling parameter $\lambda$. 
The energies of the involved many-electron states $\ket{N,a}\in\cF$, which corresponds to the local minimum of the associated PESs as shown in Figure \ref{fig:PES}, are determined by the \textit{ab initio} calculations of zinc phthalocyanine at the level of TDDFT/B3LYP/6-311g(d,p) using Gaussian 16.\cite{g16} The charged state energies are corrected for the image charge effect, see \textbf{Supporting Information} of Ref.~ [\!\!\citenum{Fu2018}] for details. A value of $\mu_0=-5.3$ eV is adopted for the chemical potential of Au(111) facet throughout all the computed transport characteristics. A crossover from weak to strong e-p couplings will be explored using $\lambda=0.25,0.5,1.0,2.0$. In order to comply with the weak molecule-lead coupling limit, we adopt a value of $0.0001$ eV for $\hbar\Gamma$ with symmetric molecule-lead couplings $\Gamma_{\mathrm{S}}=\Gamma_{\mathrm{D}}=\Gamma$. The energy of vibrational quanta is chosen as $0.2$ eV and $0.02$ eV. Moreover, we investigate electron transport characteristics in the low temperature regime  ($\beta=0.05\hbar\omega_{\mathrm{vib}}$)  in order to resolve the Frank-Condon allowed charge transfer transitions in the stability diagram and the current-voltage characteristics. 

Since there is no displacement between electronic manifold $\ket{0,S_0}$ and $\ket{0,S_1}$, a value of $10^{12}$ $\mathrm{s}^{-1}$ is chosen for $k^{\mathrm{field}}_{0,S_0,\nu\leftrightarrow 0,S_1,\nu}$ such that $k^{\mathrm{field}}\Gamma^{-1}\gtrsim 1$, where the rate of optical transition is  hereinafter referred to as $k^{\mathrm{field}}$. Assuming the optical excitation is in resonance with the electronic transition $\ket{0,S_0,\nu}\leftrightarrow\ket{0,S_1,\nu}$, the respective electric field strength takes the value of $100$ $\mathrm{kV}/\mathrm{cm}$ if the transition dipole moment is $5$ Debye. A spontaneous emission rate of $10^8$ $\mathrm{s}^{-1}$ estimated through $4\alpha n\omega_{10}^3\vert\bvec d_{01}\vert^2/(3c^2)$ is also introduced phenomenologically, where $\alpha$ is the fine structure constant, $\omega_{10}=(E_{0,S_1}-E_{0,S_0})/\hbar$, $n$ is the refractive index and $c$ is the speed of light in vacuum. The rates of singlet-triplet transitions are then chosen for typical values, i.e., $10^6$ $\mathrm{s}^{-1}$ for intersystem crossing processes $\ket{0,S_1}\rightarrow\ket{0,T^m_1}$ and $10^3$ $\mathrm{s}^{-1}$ for phosphorescent processes $\ket{0,T^m_1}\rightarrow\ket{0,S_0}$. 

\begin{table*}[tbh!]
\centering
\begin{ruledtabular}
\begin{tabular}{c c c c c} 
 \hline
 Parameter && Value && Meaning \\
 \hline
 \hline
 $E_{0,S_0}$&& $0$ eV  && Energy level of the singlet ground state $\ket{0,S_0}$\\
 $E_{0,S_1}$ && $2.01$ eV && Energy level of the singlet excited state $\ket{0,S_1}$\\
 $E_{0,T^m_1}$ && $1.00$ eV && Energy level of the degenerate triplet states $\ket{0,T^m_1}$\\
 $E_{-1,D^{\sigma}_0}$ && $6.045$ eV  && Energy level of the cation doublet states $\ket{-1,D^{\sigma}_0}$, image charge correction included.\\
 $E_{+1,D^{\sigma}_0}$ && $-2.06$ eV && Energy level of the anion doublet states $\ket{+1,D^{\sigma}_0}$, image charge correction included. \\
 $\hbar\omega_{\mathrm{vib}}$ && $0.2$, $0.02$ eV && Energy of the intramolecular vibrational mode\\
 $\lambda$ && $0$,$0.25$,$0.5$,$1.0$,$2.0$ && Electron-Phonon coupling \\
 $\mu_0$ && $-5.3$ eV && Fermi energy in the electrode at zero source-drain bias voltage\\
 $\hbar\Gamma$ && $10^{-4}$ eV && $\Gamma_{\mathrm{S}}=\Gamma_{\mathrm{D}}=\Gamma$ in a symmetric molecular junction\\
 $k_BT$ && 0.05$\hbar\omega_{\mathrm{vib}}$ && Thermal energy at temperate T\\
 $k_{0,T_1\leftarrow 0,S_1}$ && $10^{6}$ $\mathrm{s}^{-1}$ && Intersystem crossing rate\\
 $k_{0,S_0\leftarrow 0,T_1}$ && $10^{3}$ $\mathrm{s}^{-1}$ && Phosphorescence rate\\
 $k^{\mathrm{spon}}_{0,S_0\leftarrow 0,S_1}$ && $10^8$ $\mathrm{s}^{-1}$ && Spontaneous emission rate\\
 \hline
\end{tabular}
\end{ruledtabular}
\caption{Parameters adopted in the computations. Details are discussed in main text.}
\label{table:paras}
\end{table*}

\section{Results and Discussion}
\label{Results}

\begin{figure*}
\center{\includegraphics[width=1.0\linewidth]{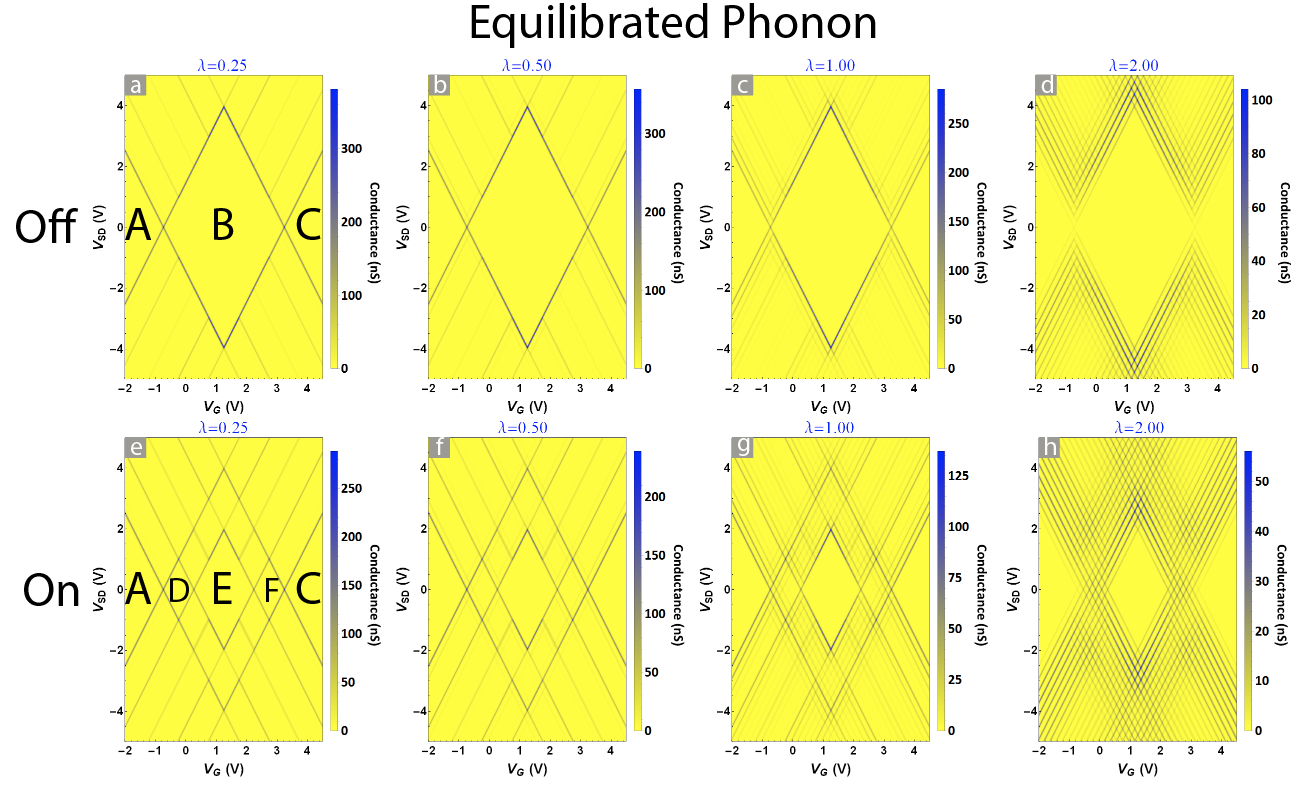}}
\caption{The charge stability diagrams are computed for vibrational frequency $\hbar\omega_{\mathrm{vib}}=0.2$ eV and the equilibrated phonon regime. The field-off charge stability diagrams are presented for (a) $\lambda=0.25$, (b) $\lambda=0.50$ , (c) $\lambda=1.0$  and (d) $\lambda=2.0$. The field-on charge stability diagrams are also presented for  (e) $\lambda=0.25$, (f) $\lambda=0.50$ , (g) $\lambda=1.0$  and (h) $\lambda=2.0$.}
\label{fig:StabilityDiagrameqhw0.2}
\end{figure*}
\begin{figure*}
\center{\includegraphics[width=1.0\linewidth]{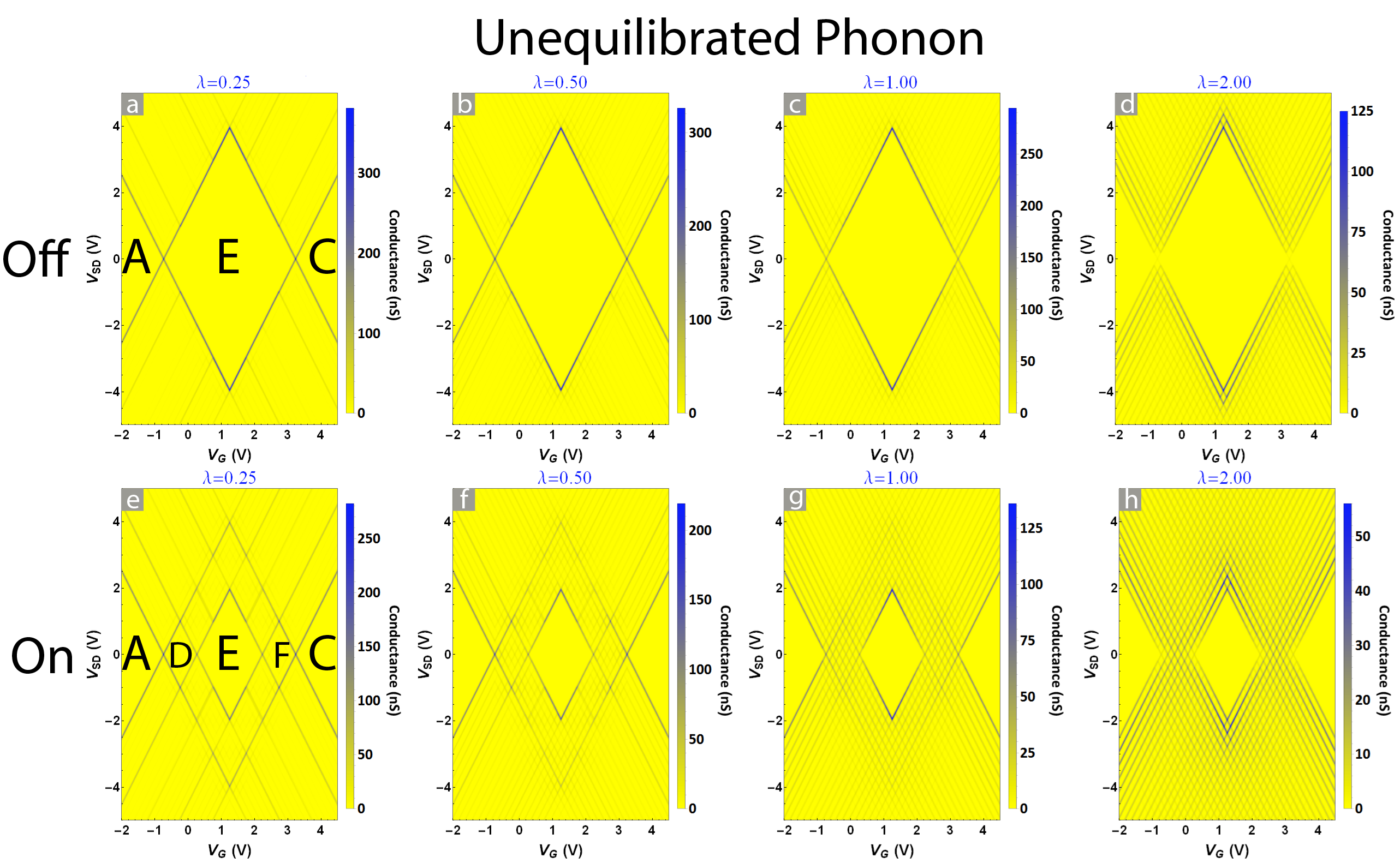}}
\caption{The charge stability diagrams are computed for vibrational frequency $\hbar\omega_{\mathrm{vib}}=0.2$ eV and the unequilibrated phonon regime. The field-off charge stability diagrams are presented for (a) $\lambda=0.25$,  (b) $\lambda=0.50$, (c) $\lambda=1.0$ and (d) $\lambda=2.0$. The field-on charge stability diagrams are also presented for (e) $\lambda=0.25$, (f) $\lambda=0.50$, (g) $\lambda=1.0$ and (h) $\lambda=2.0$.}
\label{fig:StabilityDiagramUneqhw0.2}
\end{figure*}

In this section, we investigate the influence of e-p coupling and vibrational relaxation on the transport characteristics under an optical field. Light-driven electron transport is explored from weak to strong e-p coupling regimes. For simplicity, the effect of vibrational relaxation is discussed in two limits: the \textit{equilibrated phonon} and the \textit{unequilibrated phonon}. We analyze the transport characteristics from the following aspects. In section \ref{sec:ChargeStabilityDiagram}, we study the charge stability diagram (the conductance spectra) of the irradiated molecular junction in a broad range of source-drain bias voltage and gate voltage. Charge stability diagrams have been extensively studied in nanoscale electron transport and they can provide rich information about electronic structures. Moreover, considering that low-bias current-voltage characteristics is accessible in most experiments, we focus on low-bias transport characteristics in section \ref{sec:CurrentVsGate} and \ref{sec:CurrentVsBias}. In section \ref{sec:CurrentVsGate}, we control the gate voltage, compute the low-bias current at $V_{\mathrm{SD}}=0.1$ V, and compare the situations of high- and low-frequency vibrational modes in a range of gate voltage $V_{\mathrm{G}}$ that covers all relevant energy level alignment schemes between the charged states $\ket{-1,D^{\alpha}_0}$ and the neutral states. In section \ref{sec:CurrentVsBias}, focusing on four representative schemes of energy level alignments, we compute the current-voltage characteristics and derive the analytical current formula. This offers a quantitative point of view on the light-driven transport when the charge transfer transitions are coupled to intramolecular vibrations.
\subsection{Equilibrated and Unequilibrated Phonon regimes}\label{sec:PhononRegimes}
These two extreme limits of vibrational relaxation, i.e., $\gamma_{\mathrm{p}}\gg\Gamma$ (equilibrated phonon regime) and $\gamma_{\mathrm{p}}\ll\Gamma$ (unequilibrated phonon regime), offer us a clear picture to understand the role of the vibrational relaxation on the light-driven transport characteristics. 

In the equilibrated phonon regime, the vibrational relaxation is much faster than all the other transitions about the molecule. In other words, the vibrational state distribution of each electronic state manifold instantaneously relax into its thermal equilibrium upon any electronic transition. As a result, the vibrational relaxation part of the Pauli master equation Eq. (\ref{eq:PauliQME}), i.e., $\frac{d P_{N,a,\nu}}{dt}\vert_{\mathrm{m-th}}$, reduces into
\begin{align}\label{eq:rateeq-eqphonon}
    \left.\frac{d P_{N,a,\nu}}{dt}\right\vert_{\mathrm{m-th}}=-\lim_{\gamma_{\mathrm{p}}\rightarrow\infty}\gamma_{\mathrm{p}}\roundbracket{P_{N,a,\nu}-P^{\mathrm{eq}}_{\nu}\sum_{\nu'=0}P_{N,a,\nu'}},
\end{align}
where $P^{\mathrm{eq}}_{\nu}=\expfunc{\nu\beta\hbar\omega}/\sum_{\nu'=0}^{\infty}\expfunc{\nu'\beta\hbar\omega}$ describes the equilibrium vibrational distribution. The role of Eq. (\ref{eq:rateeq-eqphonon}) on the Pauli master equation Eq. (\ref{eq:PauliQME}) is to force the vibrational state distribution to be kept at the thermal equilibrium distribution at all times. This matter of fact enables us to derive analytial solutions for transport characteristics in section \ref{sec:CurrentVsBias}. 

In the unequilibrated phonon regime, on the contrary, we assume that the vibrational relaxation is slower than any other processes about the molecule, which simply modifies the Pauli master equation Eq. (\ref{eq:PauliQME}) by dropping the terms belonging to $\frac{d P_{Na\nu}}{dt}\vert_{\mathrm{m-th}}$. In other words, the non-equilibrium vibrational excitations brought by other electronic transitions are completely preserved during the time scale of the relevant processes. The role of the non-equilibrium vibrational excitations are then investigated in this regime.

\subsection{Charge Stability Diagram: Effect of e-p Coupling on Photoinduced Coulomb Diamond}\label{sec:ChargeStabilityDiagram}
We explore the influence of e-p coupling ($\lambda$) on the charge stability diagram, i.e., a plot of differential conductance $dI/dV$ versus $V_{\mathrm{SD}}$ and $V_{\mathrm{G}}$, in the equilibrated phonon regime 
and the unequilibrated phonon regime 
\cite{Koch2005}. We discuss the equilibrated phonon regime in details. For the unequilibrated phonon regime, we only discuss the difference from the equilibrated phonon regime.

We first investigate the equilibrated phonon regime, which is characterized by the equilibrium vibrational distribution. 
As presented in Figure \ref{fig:StabilityDiagrameqhw0.2}, we compute the conductance spectra for $\hbar\omega_{\mathrm{vib}}=0.2$ eV and $\lambda=0.25, 0.5, 1.0, 2.0$ under a field-off condition $k^{\mathrm{field}}=0$ and a field-on condition $k^{\mathrm{field}}=10^{12}$ $\mathrm{s}^{-1}$.


When the radiation is off, the conventional Coulomb blockade diamonds A, B and C, respectively corresponding to $\ket{-1,D^{\pm 1/2}_0}$, $\ket{0,S_0}$ and $\ket{1,D^{\pm 1/2}_0}$, remain intact for e-p couplings from $\lambda=0.25$ to $\lambda=1.0$, as shown in Figure \ref{fig:StabilityDiagrameqhw0.2}a-c. The diamond B, corresponding to $\ket{0,S_0}$, exhibits the width $E_{-1,D^{\sigma}_0}+E_{1,D^{\sigma}_0}-2E_{0,S_0}$ and the height-to-width ratio $2:1$. From Figure \ref{fig:StabilityDiagrameqhw0.2}a to Figure \ref{fig:StabilityDiagrameqhw0.2}c, as the e-p coupling $\lambda$ increases, conductance lines spaced by $\hbar\omega_{\mathrm{vib}}$ along $V_{\mathrm{G}}$ axis and $2\hbar\omega_{\mathrm{vib}}$ along $V_{\mathrm{SD}}$ axis become pronounced. The equally spaced conductance lines arise from the step-wise activation of charge transport channels. In Figure \ref{fig:StabilityDiagrameqhw0.2}d, when $\lambda$ becomes $2.0$, we observe a clear Frack-Condon blockade pattern, which has been reported both theoretically\cite{Koch2005} and experimentally\cite{Burzuri2014}. The two intersections in between the three Coulomb diamonds break because the low-bias conduction is suppressed by the Frank-Condon factor ($M_{\nu_1\nu_2}\roundbracket{\lambda}$ in Eqs. (\ref{eq:ChargeTransferRate}) and (\ref{eq:HoleTransferRate})), i.e., the transitions between low-lying vibrational states  decrease exponentially in the regime of strong e-p coupling ($\lambda>1$).

When the radiation is on, the anomalous Coulomb blockade pattern is clearly identified in the weak e-p coupling regime ($\lambda<1.0$) from Figure \ref{fig:StabilityDiagrameqhw0.2}e to Figure \ref{fig:StabilityDiagrameqhw0.2}g. In the zero e-p coupling limit, the widths of the diamonds D, E and F are $E_{0,T^m_1}-E_{0,S_0}$, $E_{-1,D^{\sigma}_0}+E_{1,D^{\sigma}_0}-2E_{0,T^m_1}$ and $E_{0,T^m_1}-E_{0,S_0}$, respectively. As the e-p coupling increases, the diamonds D and F (corresponding to partial charged states) shrink by steps of $\Delta V_{\mathrm{G}}=\hbar\omega_{\mathrm{vib}}/\vertbracket{e}$ ,while the diamond E (corresponding to the triply degenerate states $\ket{0,T^m_1}$) remains invariant. The conductance lines in the diamonds D and F result from the additional transport channels activated by e-p coupling, while the robustness of the diamond E with respect to the e-p coupling is attributed to the origin of the diamond E, i.e., the presence of the triplet states. 

In the limit of unequilibrated phonon regime, the field-off and field-on charge stability diagrams (conductance spectra) are computed for $\lambda=0.25,0.5,1.0,2.0$, see Figure \ref{fig:StabilityDiagramUneqhw0.2}. A comparison between Figure \ref{fig:StabilityDiagrameqhw0.2} and Figure \ref{fig:StabilityDiagramUneqhw0.2} shows that the non-equilibrium vibrational exicitations result in more reduction in the sizes of diamonds D and F at the same e-p coupling. At a  low bias, the equilibrated and unequilibrated phonon regimes exhibit the same trend in response to e-p coupling and optical excitation.

\subsection{Current vs Gate Voltage: Role of Vibrational Frequency and Vibrational Relaxation}\label{sec:CurrentVsGate}
\begin{figure*}
\center{\includegraphics[width=1.0\linewidth]{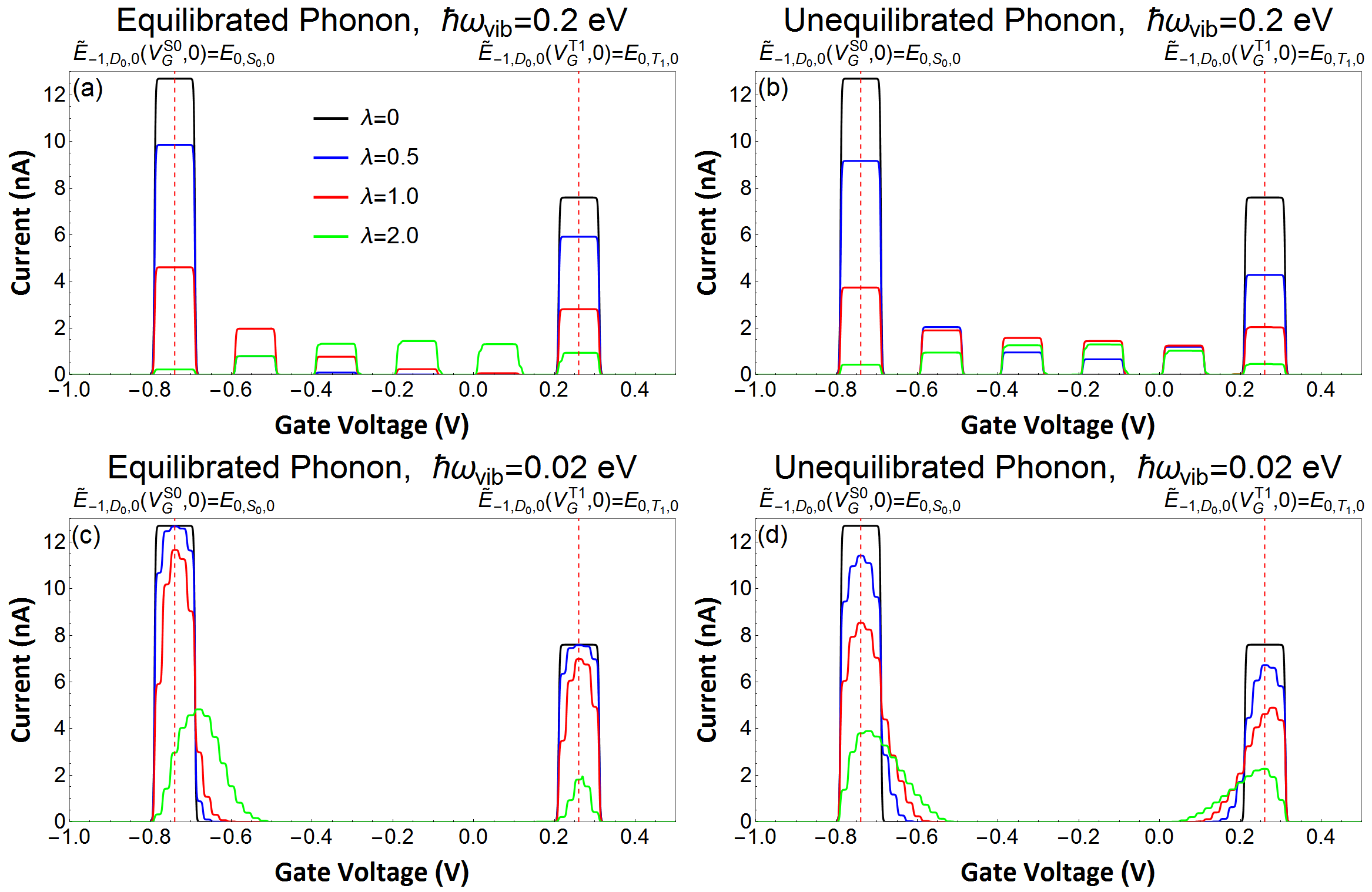}}
\caption{Field-on current at $V_{\mathrm{SD}}=0.1$ V is plotted versus $V_{\mathrm{G}}$ for $\lambda=$ $0$, $0.5$, $1.0$, $2.0$. The effect of the vibrational relaxation and the vibrational frequency is investigated for (a) high-frequency intramolecular vibration ($\hbar\omega_{\mathrm{vib}}=0.2$ eV) in the equilibrated phonon regime ($\gamma_p\gg\Gamma$), (b) high-frequency intramolecular vibration ($\hbar\omega_{\mathrm{vib}}=0.2$ eV) in the unequilibrated phonon regime ($\gamma_p\ll\Gamma$), (c) low-frequency intramolecular vibration ($\hbar\omega_{\mathrm{vib}}=0.02$ eV) in the equilibrated phonon regime ($\gamma_p\gg\Gamma$) and low-frequency intramolecular vibration ($\hbar\omega_{\mathrm{vib}}=0.02$ eV) in the unequilibrated phonon regime ($\gamma_p\ll\Gamma$).}
\label{fig:IVGVd01}
\end{figure*}

In most cases, the transport characteristics of a molecular junction is only available at a low bias voltage due to the instability caused by the high-bias electric field. Therefore, in this section, we concentrate on the field-on current at a low bias, i.e., $V_{\mathrm{SD}}=0.1$ V, and explore the current response via changing the gate voltage $V_{\mathrm{G}}$. Similar to the previous section, we carry out the calculation for both the equilibrated phonon regime and the unequilibrated phonon regime. Furthermore, in additional to the situation with high frequency mode $\hbar\omega_{\mathrm{vib}}=0.2$ eV,  we also consider a situation in which a low frequency vibrational mode ($\hbar\omega_{\mathrm{vib}}=0.02$ eV) dominates. The results are presented in Figure \ref{fig:IVGVd01}. In view of the symmetry in the charge stability diagrams, we only present the current within a range of gate voltages that covers all possible energy level alignment schemes between $\tilde{E}^{\alpha}_{-1,D^{\sigma}_0,0}\roundbracket{V_{\mathrm{G}},0}$ and neutral states. For simplicity, we hereafter denote the gate voltage corresponding to $\tilde{E}^{\alpha}_{-1,D^{\sigma}_0,0}\roundbracket{V_{\mathrm{G}},0}=E_{0,S_0,0}$ as $V^{\mathrm{S0}}_{\mathrm{G}}$ and the gate voltage corresponding to $\tilde{E}^{\alpha}_{-1,D^{\sigma}_0,0}\roundbracket{V_{\mathrm{G}},0}=E_{0,T^m_1,0}$ as $V^{\mathrm{T1}}_{\mathrm{G}}$.

Generally, from Figure \ref{fig:IVGVd01}a to \ref{fig:IVGVd01}d, we identify the plateaus of current centered at the gate voltages that correspond to the energy level alignments $\tilde{E}^{\alpha}_{-1,D^{\sigma}_0,0}\roundbracket{V_{\mathrm{G}},0}=E_{0,S_0,0}+n\hbar\omega_{\mathrm{vib}}$ ($n$ is non-negative integer) and $\tilde{E}^{\alpha}_{-1,D^{\sigma}_0,0}\roundbracket{V_{\mathrm{G}},0}=E_{0,T^m_1,0}$. 
At zero e-p coupling, the electronic transitions $\ket{-1,D^{\sigma}_0,\nu}\leftrightarrow\ket{0,S_0,\nu'}$ and $\ket{-1,D^{\sigma}_0,\nu}\leftrightarrow\ket{0,T^m_1,\nu'}$ are allowed for $\nu=\nu'$, so we observe the current plateaus only at $V^{\mathrm{S0}}_{\mathrm{G}}$ and $V^{\mathrm{T1}}_{\mathrm{G}}$. As the e-p coupling increases, the current at these two plateaus decrease monotonically and additional photoinduced current plateaus show up between $V^{\mathrm{S0}}_{\mathrm{G}}$ and $V^{\mathrm{T1}}_{\mathrm{G}}$, because the e-p coupling suppresses the diagonal vibrational transitions, i.e., the electronic transitions with $\nu=\nu'$, and invokes the off-diagonal vibrational transitions, i.e., the electronic transitions with $\nu\neq\nu'$.


Considering a high-frequency mode, e.g., $\hbar\omega_{\mathrm{vib}}=0.2$ eV, we observe four isolated current plateaus  between $V^{\mathrm{S0}}_{\mathrm{G}}$ and $V^{\mathrm{T1}}_{\mathrm{G}}$ in Figure \ref{fig:IVGVd01}a and \ref{fig:IVGVd01}b. The current plateaus are isolated from each other because $\hbar\omega_{\mathrm{vib}}/\vertbracket{e}>V_{\mathrm{SD}}$.
A comparison between Figure \ref{fig:IVGVd01}a and \ref{fig:IVGVd01}b demonstrates that, at the same e-p coupling, the non-equilibrium vibrational population in the unequilibrated phonon regime leads to more pronounced current plateaus within $[V^{\mathrm{S0}}_{\mathrm{G}},V^{\mathrm{T1}}_{\mathrm{G}}]$.


In order to find out the origin of the e-p coupling induced current plateaus within $[V^{\mathrm{S0}}_{\mathrm{G}},V^{\mathrm{T1}}_{\mathrm{G}}]$, we consider a low-frequency mode, i.e., $\hbar\omega_{\mathrm{vib}}=0.02$ eV. In this case, the e-p coupling induced current plateaus are observed only in the vicinity of $V^{\mathrm{S0}}_{\mathrm{G}}$ and $V^{\mathrm{T1}}_{\mathrm{G}}$, because $\hbar\omega_{\mathrm{vib}}\ll \vertbracket{V^{\mathrm{T1}}_{\mathrm{G}}-V^{\mathrm{S0}}_{\mathrm{G}}}$.  As a result, the role of the e-p coupling behaves like a broadening of the current plateaus at $V^{\mathrm{S0}}_{\mathrm{G}}$ and $V^{\mathrm{T1}}_\mathrm{G}$. In both extreme limits of vibrational relaxation, the additional Franck-Condon allowed transport channels is activated when $\tilde{E}^{\alpha}_{-1,D^{\sigma}_0,0}$ is aligned with vibrational excitations of the singlet ground state. This broadens the current plateau at $V^{\mathrm{S0}}_{\mathrm{G}}$ toward higher gate voltages. In contrast, the e-p coupling induced current plateaus due to $\ket{-1,D^{\sigma}_0,\nu}\leftrightarrow\ket{0,T^m_1,\nu'}$, i.e., the broadening of the current plateau at $V^{\mathrm{T1}}_\mathrm{G}$, requires the nonequilibrium vibrational population and shows up only in the unequilibrated phonon regime.

\subsection{Current vs Source-Drain Voltage: Energy Level Alignment and Analytic Analysis}\label{sec:CurrentVsBias}

\begin{figure*}
\center{\includegraphics[width=1.0\linewidth]{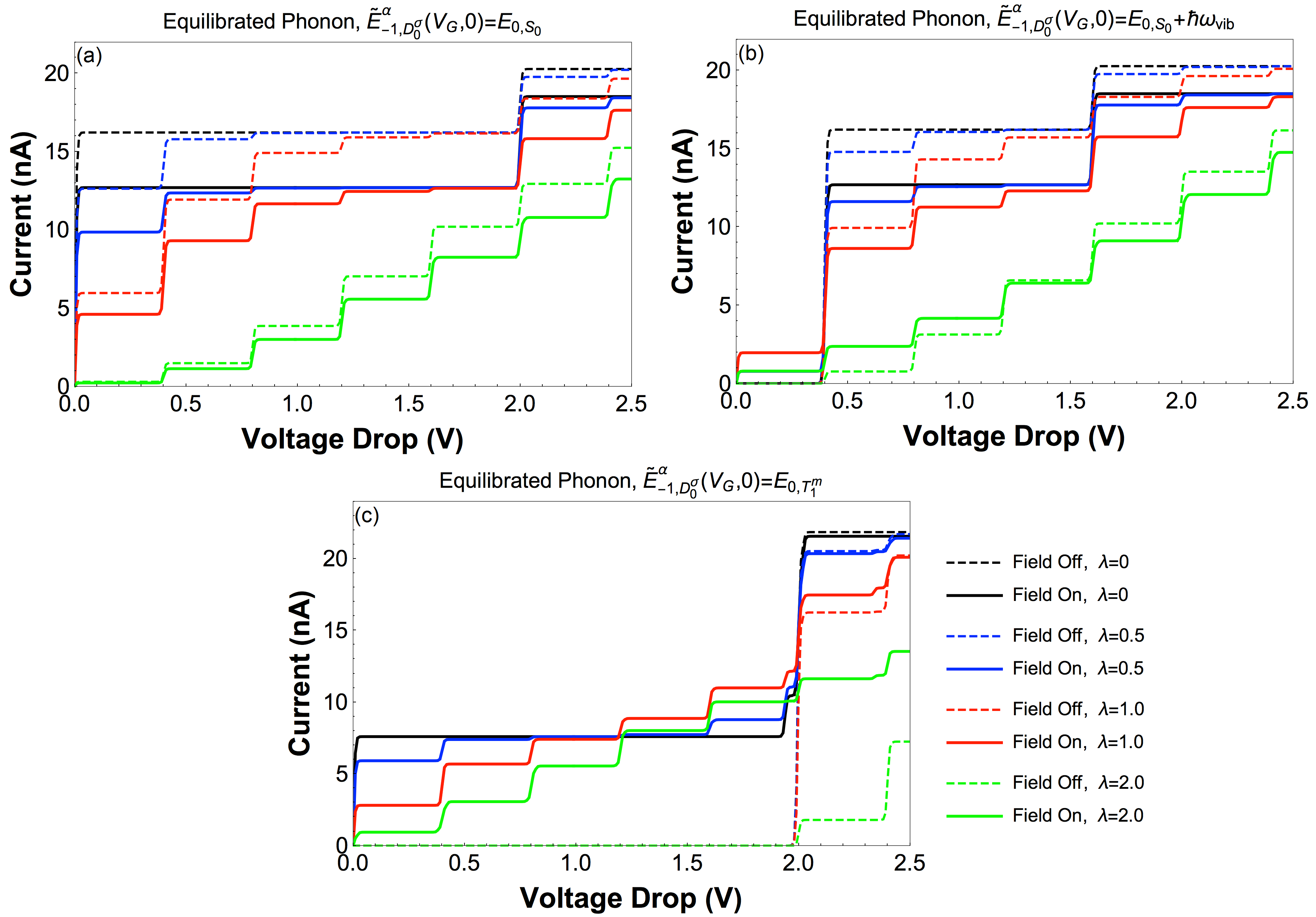}}
\caption{Current-voltage characteristics computed for equilibrated phonon regime, vibrational frequency $\hbar\omega_{\mathrm{vib}}=0.2$ eV, field-off and field-on condition, and e-p couplings of $0,0.5,1.0,2.0$ are presented for situations specified by (a) $\tilde{E}^{\alpha}_{-1,D^{\sigma}_0,0}\roundbracket{V_{\mathrm{G}},0}=E_{0,S_0,0}$, (b) $\tilde{E}^{\alpha}_{-1,D^{\sigma}_0,0}\roundbracket{V_{\mathrm{G}},0}=E_{0,S_0,0}+\hbar\omega_{\mathrm{vib}}$, (c) $\tilde{E}^{\alpha}_{-1,D^{\sigma}_0,0}\roundbracket{V_{\mathrm{G}},0}=E_{0,T^m_1,0}$ and (d) $\tilde{E}^{\alpha}_{-1,D^{\sigma}_0,0}\roundbracket{V_{\mathrm{G}},0}=E_{0,T^m_1,0}+\hbar\omega_{\mathrm{vib}}$.}
\label{fig:IV0S0epfield}
\end{figure*}
\begin{figure*}
\center{\includegraphics[width=1.0\linewidth]{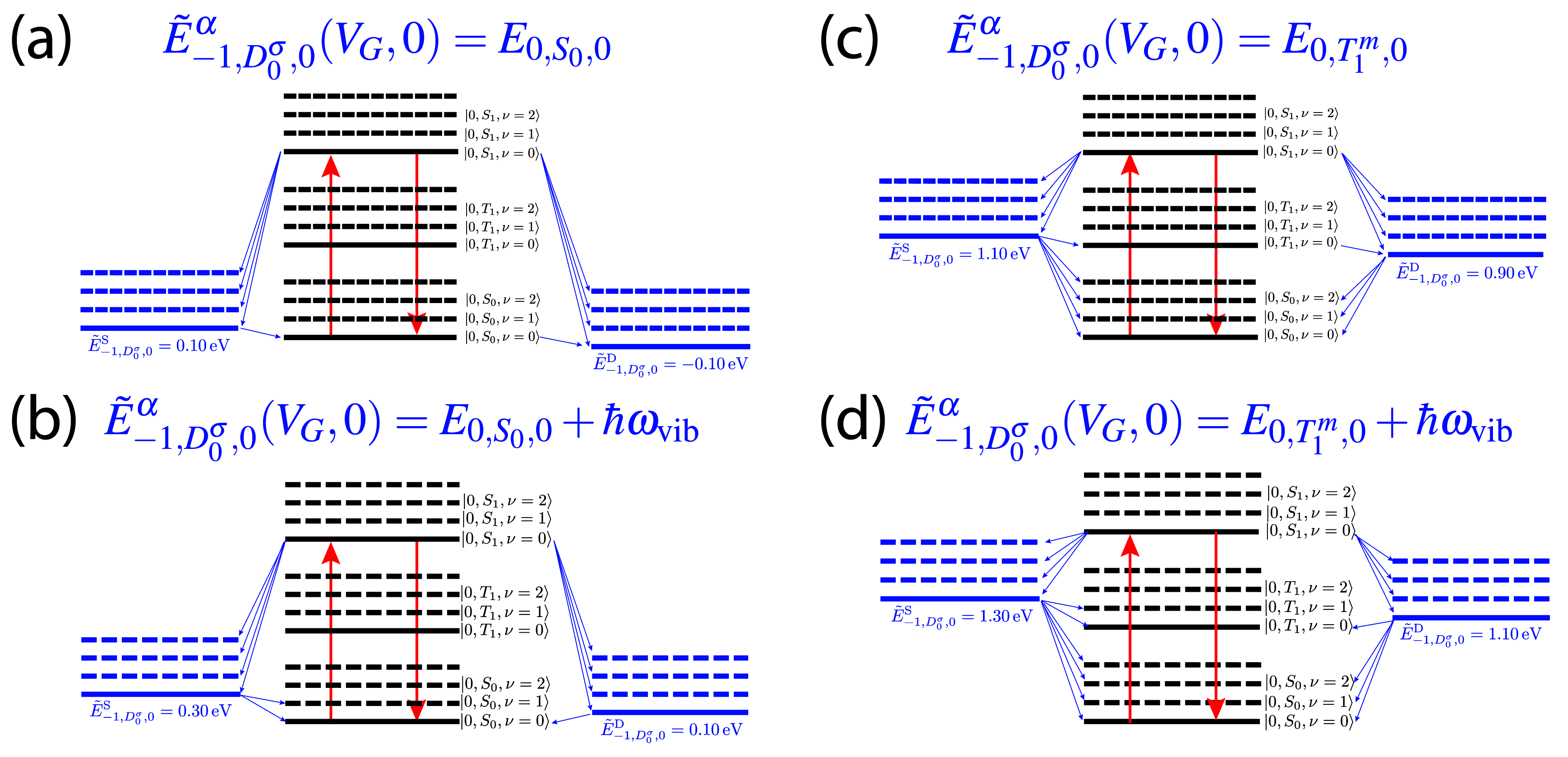}}
\caption{The diagrams of energy level alignments corresponding to four situations specified by (a) $\tilde{E}^{\alpha}_{-1,D^{\sigma}_0,0}\roundbracket{V_{\mathrm{G}},0}=E_{0,S_0,0}$, (b) $\tilde{E}^{\alpha}_{-1,D^{\sigma}_0,0}\roundbracket{V_{\mathrm{G}},0}=E_{0,S_0,0}+\hbar\omega_{\mathrm{vib}}$, (c) $\tilde{E}^{\alpha}_{-1,D^{\sigma}_0,0}\roundbracket{V_{\mathrm{G}},0}=E_{0,T^m_1,0}$ and (d) $\tilde{E}^{\alpha}_{-1,D^{\sigma}_0,0}\roundbracket{V_{\mathrm{G}},0}=E_{0,T^m_1,0}+\hbar\omega_{\mathrm{vib}}$. The source-drain bias is set to $V_{\mathrm{SD}}=0.2$ V through (a) to (d).}
\label{fig:EnergyLevelAlignment}
\end{figure*}

In our previous work\cite{Fu2018},  we obtained the analytical solutions to the current-voltage characteristics without considering e-p coupling in the cases of several specific energy level alignments. These analytical solutions provide an in-depth and quantitative understanding on the transport characteristics of an irradiated molecular junction.

When the e-p coupling is significant, it is challenging to derive general analytical solutions, because the Pauli master equations are constructed within an enormous Fock space spanned by the electron-vibrational states. In order to analyze the effect of e-p coupling, for simplicity, we focus on the equilibrated phonon regime, which allows us to work on a smaller Fock space that is spanned by the electronic states. The equilibrated phonon regime enables the analytical solutions for two reasons. First, this regime excludes the vibrational relaxations, thus allowing us to consider only the transitions between different electronic manifolds, i.e., $\ket{N,a}\rightarrow\ket{N',a'}$. Second, the equilibrium vibrational state distribution on each electronic state manifold is dominated by the vibrational ground state, because $\hbar\omega_{\mathrm{vib}}\gg k_BT$ is valid here.


Now we reconstruct the rate equations built upon the Fock space $\{\ket{N,a}\}$ in the equilibrated phonon regime. Considering that $\hbar\omega_{\mathrm{vib}}\gg k_BT$, one can equate the population of an electronic state $\ket{N,a}$ to the population of its associated lowest lying vibrational state $\ket{N,a,0}$, i.e., $P_{N,a}=\sum_{\nu}P_{N,a,\nu}=P_{N,a,0}$. 
Therefore, the effective rate of an electronic transition $\ket{N,a}\rightarrow\ket{N',a'}$ is the sum of the rates of all outgoing processes from $\ket{N,a,0}$, i.e.,
\begin{align}
    k^{\roundbracket{\mathrm{eff}}}_{N',a'\leftarrow N,a}=\sum_{\nu'}k_{N',a',\nu'\leftarrow N,a,0}\,.
\end{align}
In this way, the effective rate of the charge transfer transitions can be expressed as 
\begin{align}\label{eq:ChargeTranferRateEffective}
    k^{\alpha\roundbracket{\mathrm{eff}}}_{N',a'\leftarrow N,a}=\Lambda^{\alpha}_{N',a'\leftarrow N,a}\nu_{N'a',Na}\Gamma,
\end{align}
where 
\begin{align}
    \Lambda^{\alpha}_{N',a'\leftarrow N,a}=\sum_{\nu'}\vertbracket{M_{\nu'0}}^2\theta\roundbracket{\tilde{E}^{\alpha}_{N,a,0}\roundbracket{V_{\mathrm{G}},V_{\mathrm{SD}}}-\tilde{E}^{\alpha}_{N',a',\nu'}\roundbracket{V_{\mathrm{G}},V_{\mathrm{SD}}}}
\end{align}
refers the ratio of the effective rate $k^{\alpha\roundbracket{\mathrm{eff}}}_{N',a'\leftarrow N,a}$ to its zero e-p coupling limit.
Note that $\Lambda^{\alpha}_{N',a'\leftarrow N,a}$ could be alternatively expressed as
\begin{align}\label{eq:DefLambda}
    \Lambda^{\alpha}_{N',a'\leftarrow N,a}=\sum^{\nu_m}_{\nu=0}\vertbracket{M_{\nu0}\roundbracket{\lambda_{\mathrm{N,a}}-\lambda_{\mathrm{N',a'}}}}^2,
\end{align}
where 
\begin{align}\label{eq:FC0q}
    \vertbracket{M_{0\nu}\roundbracket{\lambda}}^2=\vertbracket{M_{\nu0}\roundbracket{\lambda}}^2=\frac{1}{\nu !}\lambda^{2\nu}\expfunc{-\lambda^2}\,.
\end{align}
and 
\begin{align}
    \nu_m\roundbracket{V_{\mathrm{G}},V_{\mathrm{SD}}}=\lfloor \frac{\tilde{E}^{\alpha}_{N,a,0}\roundbracket{V_{\mathrm{G}},V_{\mathrm{SD}}}-\tilde{E}^{\alpha}_{N',a',0}\roundbracket{V_{\mathrm{G}},V_{\mathrm{SD}}}}{\hbar\omega_{\mathrm{vib}}}\rfloor
\end{align}
with the floor function $\lfloor x\rfloor$ returning the largest integer less than $x$.  Eq.~(\ref{eq:FC0q}) indicates that $\Lambda^{\alpha}_{N',a'\leftarrow N,a}$ ranges from 0 to 1.0. Besides, $\Lambda^{\alpha}_{N',a'\leftarrow N,a}$ reaches its upper limit when
$\tilde{E}^{\alpha}_{Na0}\roundbracket{V_{\mathrm{G}},V_{\mathrm{SD}}}\gg\tilde{E}^{\alpha}_{N'a'0}\roundbracket{V_{\mathrm{G}},V_{\mathrm{SD}}}$ due to the relationship
\begin{align}
    \sum^{\nu=+\infty}_{\nu=0}\vertbracket{M_{\nu0}\roundbracket{\lambda}}^2=1.
\end{align}

In order to better understand the effect of energy level alignment on light-driven electron transport when coupled to vibrations, we analyze the analytical solutions of the current-voltage characteristics and the corresponding numerical calculations in the four representative cases specified by the following energy alignments: (1) $\tilde{E}^{\alpha}_{-1,D^{\sigma}_0,0}\roundbracket{V_{\mathrm{G}},0}=E_{0,S_0,0}$, (2) $E_{0,S_0,0}<\tilde{E}^{\alpha}_{-1,D^{\sigma}_0,0}\roundbracket{V_{\mathrm{G}},0}<E_{0,T^m_1,0}$, (3)  $\tilde{E}^{\alpha}_{-1,D^{\sigma}_0,0}\roundbracket{V_{\mathrm{G}},0}=E_{0,T^m_1,0}$ and (4) $\tilde{E}^{\alpha}_{-1,D^{\sigma}_0,0}\roundbracket{V_{\mathrm{G}},0}>E_{0,T^m_1,0}$. 
\subsubsection{$\tilde{E}^{\alpha}_{-1,D^{\sigma}_0,0}\roundbracket{V_{\mathrm{G}},0}=E_{0,S_0,0}$}\label{sec:situation1}
In this situation, the relevant electronic state manifolds are $\ket{-1,D^{\uparrow}}$, $\ket{-1,D^{\downarrow}}$, $\ket{0,S_0}$ and $\ket{0,S_1}$. When the e-p coupling is zero, the energy level alignment $\tilde{E}^{\alpha}_{-1,D^{\sigma}_0,0}\roundbracket{V_{\mathrm{G}},0}=E_{0,S_0,0}$ leads to a field-off electric current with a zero onset voltage, similar to the resonant tunneling described by the Landauer theory. Compared to the field-off transport characteristics, the field-on electron transport in this case exhibits a suppression in the current, hereinafter referred to as \textit{photoresistance}.  

We now move on to the cases with e-p couplings. One can construct the Pauli master equations based on the electronic states using the effective rates given in Eq. (\ref{eq:ChargeTranferRateEffective}). The rate equations in a matrix form is 
\begin{align}
    \frac{d \bvec P}{dt}=\bvec W\cdot\bvec P,
\end{align}
where the occupation probabilities in the vector form are
\begin{align}
    \bvec P =\roundbracket{P_{-1,D^{\uparrow}_0}\,,P_{-1,D^{\downarrow}_0}\,,P_{0,S_0}\,,P_{0,S_1}}^\mathrm{T}.
\end{align}
Moreover, based on the energy level alignment in Figure \ref{fig:EnergyLevelAlignment}a, the rate matrix $\bvec W$ is constructed as
\begin{widetext}
\begin{align}
    \bvec W 
    =
    \left(\begin{array}{cccc}
         -\Lambda^{\mathrm{S}}_0\Gamma& 0 & \Lambda^{\mathrm{D}}_0\Gamma&  0.5(\Lambda^{\mathrm{S}}_1+\Lambda^{\mathrm{D}}_1)\Gamma\\
         0& -\Lambda^{\mathrm{S}}_0\Gamma& \Lambda^{\mathrm{D}}_0\Gamma &  0.5(\Lambda^{\mathrm{S}}_1+\Lambda^{\mathrm{D}}_1)\Gamma\\
         \Lambda^{\mathrm{S}}_0\Gamma& \Lambda^{\mathrm{S}}_0\Gamma& -k^{\mathrm{field}}-2\Lambda^{\mathrm{D}}_0\Gamma& k^{\mathrm{field}} \\
         0&0 &k^{\mathrm{field}} &-k^{\mathrm{field}}-(\Lambda^{\mathrm{S}}_1+\Lambda^{\mathrm{D}}_1)\Gamma
    \end{array}
    \right)\,.
\end{align}
\end{widetext}
For simplicity, we hereinafter denote $\Lambda^{\alpha}_{0,S_0\leftrightarrow -1,D^{\uparrow/\downarrow}_0}$ as $\Lambda^{\alpha}_0$, $\Lambda^{\alpha}_{0,S_1\leftrightarrow -1,D^{\uparrow/\downarrow}_0}$ as $\Lambda^{\alpha}_1$ and neglect the insignificant processes, i.e., the singlet-triplet transitions and the spontaneous emission.

By solving the steady-state dynamics from $\bvec W\cdot\bvec P=0$, we obtain an analytical expression of the stationary current as
\begin{align}\label{eq:CurrentExprCase1v1}
    I_{\mathrm{S}}=-I_{\mathrm{D}}=\vertbracket{e}\Gamma\Lambda^{\mathrm{S}}_0\frac{\roundbracket{2\Lambda^{\mathrm{D}}_0+\Lambda^{\mathrm{D}}_1}k^{\mathrm{field}}+2\Lambda^{\mathrm{D}}_0\Lambda_1\Gamma}{\roundbracket{2\Lambda_0+\Lambda_1}k^{\mathrm{field}}+\roundbracket{2\Lambda^{\mathrm{D}}_0+\Lambda^{\mathrm{S}}_0}\Lambda_1\Gamma},
\end{align}
where we employ the abbreviations $\Lambda_{0/1}=\Lambda^{\mathrm{S}}_{0/1}+\Lambda^{\mathrm{D}}_{0/1}$. Since $I_{\mathrm{S}}+I_{\mathrm{D}}=0$ is valid in stationary transport, we only refer to $I_{\mathrm{S}}$ below. It is worth noting that, when $\lambda\rightarrow 0$, the analytical expression Eq. (\ref{eq:CurrentExprCase1v1}) is reduced to $\vertbracket{e}\Gamma\roundbracket{3k^{\mathrm{field}}+4\Gamma}/\roundbracket{6k^{\mathrm{field}}+6\Gamma}$, which agrees with our previous work\cite{Fu2018}.

Since $\Lambda^{\mathrm{S}}_0=\Lambda^{\mathrm{D}}_0=0.5\Lambda_0$ in the situation $\tilde{E}^{\alpha}_{-1,D^{\sigma}_0,0}\roundbracket{V_{\mathrm{G}},0}=E_{0,S_0,0}$, the analytical expression of electric current, Eq. (\ref{eq:CurrentExprCase1v1}), could be further reduced to
\begin{align}\label{eq:CurrentExprCase1v2}
    I_{\mathrm{S}}=&\vertbracket{e}\Gamma\frac{\Lambda_0}{2}\frac{\roundbracket{\Lambda_0+\Lambda^{\mathrm{D}}_1}k^{\mathrm{field}}+\Lambda_0\Lambda_1\Gamma}{\roundbracket{2\Lambda_0+\Lambda_1}k^{\mathrm{field}}+1.5\Lambda_0\Lambda_1\Gamma} \nonumber \\
    =& \vertbracket{e}\Gamma\frac{\Lambda_0}{2}\roundbracket{\frac{2}{3}-\frac{1}{3}\frac{\roundbracket{\Lambda_0+2\Lambda^{\mathrm{S}}_1-\Lambda^{\mathrm{D}}_1}k^{\mathrm{field}}}{\roundbracket{2\Lambda_0+\Lambda_1}k^{\mathrm{field}}+1.5\Lambda_0\Lambda_1\Gamma}},
\end{align}
which clearly reveals that the field-off current (Recall $k^{\mathrm{field}}=0$) is 
\begin{align}\label{eq:CurrentExprCase1Off}
    I^{\mathrm{off}}_S=\vertbracket{e}\Gamma\frac{\Lambda_0}{3},
\end{align}
and the amount of current suppression led by optical excitation is 
\begin{align}
    \Delta I=I^{\mathrm{off}}_S-I_S=\vertbracket{e}\Gamma\frac{\Lambda_0}{6}\frac{\roundbracket{\Lambda_0+2\Lambda^{\mathrm{S}}_1-\Lambda^{\mathrm{D}}_1}k^{\mathrm{field}}}{\roundbracket{2\Lambda_0+\Lambda_1}k^{\mathrm{field}}+1.5\Lambda_0\Lambda_1\Gamma}\,.
\end{align}

In Figure \ref{fig:IV0S0epfield}a, the photoresistivity is clearly identified in all explored e-p coupled regimes.
Obviously, the current decreases as the e-p coupling increases, which is also clearly described in Eq. (\ref{eq:CurrentExprCase1v1}).

Furthermore, the stepwise dependence of $\Lambda^{\alpha}_{0/1}$ on $V_{\mathrm{SD}}$ and $V_{\mathrm{G}}$ results in the steplike current-voltage characteristics with width of $\Delta V_{\mathrm{SD}}=2\hbar\omega_{\mathrm{vib}}/\vertbracket{e}$ and the equally spaced conductance lines in the charge stability diagrams. In other words, from the aspects of the energy level alignment, the current changes 
whenever $\tilde{E}^{\alpha}_{\pm 1,D^{\sigma}_0,0}\roundbracket{V_{\mathrm{G}},V_{\mathrm{SD}}}$ goes across an electron-vibrational state $\ket{0,S_0,\nu}$. Therefore, a change of $\hbar\omega_{\mathrm{vib}}$ in the energy level alignment corresponds to 
either a $\Delta V_{\mathrm{SD}}=2\hbar\omega_{\mathrm{vib}}/\vertbracket{e}$ or a $\Delta V_{\mathrm{G}}=\hbar\omega_{\mathrm{vib}}/\vertbracket{e}$.

\subsubsection{$E_{0,S_0,0}<\tilde{E}^{\alpha}_{-1,D^{\sigma}_0,0}\roundbracket{V_{\mathrm{G}},0}<E_{0,T^m_1,0}$}\label{sec:situation2}
In this case, we focus on the diamond D and study the current-voltage characteristics for gate voltages $V_{\mathrm{G}}\in[V^{\mathrm{S0}}_{\mathrm{G}},V^{\mathrm{T1}}_{\mathrm{G}}]$ and source-drain bias voltages $V_{\mathrm{SD}}\in[0,2\roundbracket{\tilde{E}^{\alpha}_{-1,D^{\sigma}_0,0}\roundbracket{V_{\mathrm{G}},0}-E_{0,S_0,0}}/\vertbracket{e}]$. At zero e-p coupling, the field-off and field-on conditions result in the same blockade of electron transport but differs in the origin of the blockade. The diamond D is associated with $\ket{0,S_0}$ when optical field is off, but it is related to a partial charged state when the radiation is on\cite{Fu2018}. The blockade under optical excitation is referred to as \textit{anomalous Coulomb blockade} in Ref. [\!\!\citenum{Fu2018}].

In contrast with the situation in section \ref{sec:situation1} where the e-p coupling suppresses the current, 
in this case, the e-p coupling lifts the anomalous Coulomb blockade, resulting in the photoconductivity within the diamond D. The IV curves in Figure \ref{fig:IV0S0epfield}b are computed for a representative energy level alignment $\tilde{E}^{\alpha}_{-1,D^{\sigma}_0,0}\roundbracket{V_{\mathrm{G}},0}=E_{0,S_0,0}+\hbar\omega$ which is illustrated in Figure \ref{fig:EnergyLevelAlignment}b. In the bias voltage range $[0,2\hbar\omega_{\mathrm{vib}}/\vertbracket{e}]$, i.e., within the diamond D, Figure \ref{fig:IV0S0epfield}b shows a current step under optical excitation when $\lambda>0$. The photoinduced current within $V_{\mathrm{SD}}\in[0,2\hbar\omega_{\mathrm{vib}}/\vertbracket{e}]$ does not monotonically change with $\lambda$, but first increases and then decreases. This dependence of photoinduced current on $\lambda$ can be illustrated quantitatively by the analytical solutions.

In order to derive the analytical solutions, we also choose the effective electronic states $\ket{-1,D^{\uparrow}}$, $\ket{-1,D^{\downarrow}}$, $\ket{0,S_0}$ and $\ket{0,S_1}$. According to the energy level alignment scheme in Figure \ref{fig:EnergyLevelAlignment}b, the rate matrix $\bvec W$ is constructed as
\begin{widetext}
\begin{align}
    \bvec W=\left(
    \begin{array}{cccc}
        -\roundbracket{\Lambda^{\mathrm{S}}_0+\Lambda^{\mathrm{D}}_0} & 0 &  0 & 0.5\roundbracket{\Lambda^{\mathrm{S}}_1+\Lambda^{\mathrm{D}}_1}  \\
        0 & -\roundbracket{\Lambda^{\mathrm{S}}_0+\Lambda^{\mathrm{D}}_0} & 0  & 0.5\roundbracket{\Lambda^{\mathrm{S}}_1+\Lambda^{\mathrm{D}}_1}  \\
        \Lambda^{\mathrm{S}}_0+\Lambda^{\mathrm{D}}_0 & 0 & -k^{\mathrm{field}}  & k^{\mathrm{field}}  \\
        0 & 0 & k^{\mathrm{field}}  & -k^{\mathrm{field}}-\roundbracket{\Lambda^{\mathrm{S}}_1+\Lambda^{\mathrm{D}}_1}
    \end{array}
    \right)\,.
\end{align}
\end{widetext}
The stationary solution then gives the populations as
\begin{align}
    \bvec P =\frac{1}{\roundbracket{2\Lambda_0+\Lambda_1}k^{\mathrm{field}}+\Lambda_0\Lambda_1\Gamma}\left(
    \begin{array}{c}
        0.5\Lambda_1k^{\mathrm{field}} \\
        0.5\Lambda_1k^{\mathrm{field}} \\
        \Lambda_0 k^{\mathrm{field}}+\Lambda_0\Lambda_1\Gamma \\
        \Lambda_0 k^{\mathrm{field}}
    \end{array}
    \right),
\end{align}
and the electric current within the bias voltage range $[0,2\hbar\omega_{\mathrm{vib}}/\vertbracket{e}]$ as
\begin{align}\label{eq:CurrentExprCase2v1}
    I_{\mathrm{S}}=-I_{\mathrm{D}}=\vertbracket{e}\Gamma k^{\mathrm{field}}\frac{\Lambda^{\mathrm{S}}_0\Lambda^{\mathrm{D}}_1-\Lambda^{\mathrm{D}}_0\Lambda^{\mathrm{S}}_1}{\roundbracket{2\Lambda_0+\Lambda_1}k^{\mathrm{field}}+\Lambda_0\Lambda_1\Gamma}\,.
\end{align}
Similar to the previous section, the zero e-p coupling limit of Eq. (\ref{eq:CurrentExprCase2v1}) reduces to our analytical results in Ref. [\!\!\citenum{Fu2018}], i.e., $I_{\mathrm{S}}=-I_{\mathrm{D}}\rightarrow 0$ and the net charge on the molecule  $\vertbracket{Q}=2\vertbracket{e}P_{-1,D^{\sigma}_0}\rightarrow\frac{\vertbracket{e}k^{\mathrm{field}}}{3k^{\mathrm{field}}+\Gamma}$ . 

Eq. (\ref{eq:CurrentExprCase2v1}) reveals that \textit{the photoinduced current can originate from the asymmetry in $\Lambda^{\mathrm{S}}_{0/1}$ and $\Lambda^{\mathrm{D}}_{0/1}$ in the case of $E_{0,S_0,0}<\tilde{E}^{\alpha}_{-1,D^{\sigma}_0,0}\roundbracket{V_{\mathrm{G}},0}<E_{0,T^m_1,0}$.} Under the situation specified in Figure \ref{fig:EnergyLevelAlignment}b, this asymmetry is immediately achieved when a finite $V_{\mathrm{SD}}$ is applied. 

It is worthy to point out that the asymmetry in Eq. (\ref{eq:CurrentExprCase2v1}) resembles the Eq. (16) of Ref. [\!\!\citenum{Galperin2005}] which attributes the observation of photocurrent to the asymmetric molecule-lead coupling, i.e., $\Gamma_{\mathrm{S}}\neq\Gamma_{\mathrm{D}}$. The Eq. (16) of Ref. [\!\!\citenum{Galperin2005}] is derived under an off-resonant tunnelling situation using a HOMO-LUMO model in the single particle picture. The situation of off-resonant tunneling is similar to the energy level alignment considered in this section, especially if we consider that $\Lambda^{\mathrm{\alpha}}_{0}$ and $\Lambda^{\mathrm{\alpha}}_{1}$ are respectively the analogs of the  coupling between HOMO and electrode $\alpha$, and the coupling between LUMO and electrode $\alpha$. However, it should be clarified that the asymmetry in Eq. (\ref{eq:CurrentExprCase2v1}) results from the e-p coupling and does not rely on the symmetry of the molecular system, while the photocurrent predicted in Ref. [\!\!\citenum{Galperin2005}] requires an intrinsic asymmetry of the molecule.


For the transport characteristics in the case  $V_{\mathrm{SD}}>2\hbar\omega_{\mathrm{vib}}/\vertbracket{e}$ (out of the diamond D), as shown in Figure \ref{fig:IV0S0epfield}, the IV curves for $\lambda=2.0$ exhibit a transition from the photoconductive behavior to the photoresistive behavior at the fourth current step, in contrast with the IV curves for $\lambda=0.5$ and $1.0$ that exhibit the same photoresistive behavior as in section \ref{sec:situation1}.  
The current-voltage characteristics for $\lambda=2.0$ thus deserves further discussion.
When $\lambda=2.0$,
the asymmetry between $\Lambda^{\mathrm{S}}_{0/1}$ and $\Lambda^{\mathrm{D}}_{0/1}$ dominates the transport characteristics within $V_{\mathrm{SD}}\in[2,6]\hbar\omega_{\mathrm{vib}}/\vertbracket{e}$ and results in the photoconductive behavior within this bias range. However, as the $V_{\mathrm{SD}}$ increases, the asymmetry gradually disappears due to the activation of more transport channels, which leads to the photoresistive behavior when $V_{\mathrm{SD}}>6\hbar\omega_{\mathrm{vib}}/\vertbracket{e}$.


Apart from the representative situation given in Figure \ref{fig:EnergyLevelAlignment}b, other situations corresponding to $\tilde{E}^{\alpha}_{-1,D^{\sigma}_0,0}\roundbracket{V_{\mathrm{G}},0})=E_{0,S_0}+n \hbar\omega_{\mathrm{vib}}$ ($n>1$) also deserve further exploration. We highlight these situations because their zero onset voltage of the photocurrent is of experimental interest. The current-voltage characteristics under these situations share a number of key features with the one given in Figure \ref{fig:EnergyLevelAlignment}b, i.e., (1) e-p coupling induced photocurrent when $V_{\mathrm{SD}}\in[0,2n\hbar\omega_{\mathrm{vib}}/\vertbracket{e}]$, (2) a photoresistive current-voltage characteristics for $\lambda\leq 1$ and a crossover from photoconductive behavior to photoresistive behavior for $\lambda=2.0$ when $V_{\mathrm{SD}}>2n\hbar\omega_{\mathrm{vib}}/\vertbracket{e}$.



In order to gain more insight about the photocurrent under the above situations, we investigate the first current step with the help of Eq.~(\ref{eq:CurrentExprCase2v1}). Note that Eq. (\ref{eq:CurrentExprCase2v1}) encodes the dependence of $V_{\mathrm{G}}$ and $V_{\mathrm{SD}}$ in $\Lambda^{\alpha}_{0/1}$ and generally applies to $V_{\mathrm{G}}\in[V^{\mathrm{S0}}_\mathrm{G},V^{\mathrm{T1}}_\mathrm{G}]$ and $V_{\mathrm{SD}}\in[0,2\mathrm{min}(\vertbracket{V_{\mathrm{G}}-V^{\mathrm{S0}}_\mathrm{G}},\vertbracket{V_{\mathrm{G}}-V^{\mathrm{T1}}_\mathrm{G}})]$. If we only focus on the gate voltages close to $V^{\mathrm{S0}}_{\mathrm{G}}$ such that  $(E_{0,S_1,0}-\tilde{E}^{\alpha}_{-1,D^{\sigma}_0,0}\roundbracket{V_{\mathrm{G}},V_{\mathrm{SD}}})\gg \hbar\omega_{\mathrm{vib}}$ and  $\Lambda^{\alpha}_1\rightarrow 1$, Eq. (\ref{eq:CurrentExprCase2v1}) is simplified to
\begin{align}\label{eq:CurrentExprCase2v2}
    I_{\mathrm{S}}\sim&\vertbracket{e}\Gamma k^{\mathrm{field}}\frac{\Lambda^{\mathrm{S}}_0-\Lambda^{\mathrm{D}}_0}{\roundbracket{2\Lambda_0+\Lambda_1}k^{\mathrm{field}}+\Lambda_0\Lambda_1\Gamma}\,.
\end{align}
Since we are interested in the first current step, i.e., $V_{\mathrm{SD}}\in[0,2\hbar\omega_{\mathrm{vib}}/\vertbracket{e}]$, Eq. (\ref{eq:CurrentExprCase2v2}) turns to
\begin{align}\label{eq:CurrentExprCase2v3}
    I_{\mathrm{S}}
    =&\vertbracket{e}\Gamma k^{\mathrm{field}}\frac{\vertbracket{M_{0n}\roundbracket{\lambda}}^2}{\roundbracket{2\Lambda_0+\Lambda_1}k^{\mathrm{field}}+\Lambda_0\Lambda_1\Gamma}\,.
\end{align}
where $n$, determined by $V_{\mathrm{G}}$, labels the vibrational excitation associated with $\ket{0,S_0}$ that aligns with $\tilde{E}^{\alpha}_{-1,D^{\sigma}_0,0}\roundbracket{V_{\mathrm{G}},0})$.

\begin{figure*}
\center{\includegraphics[width=1.0\linewidth]{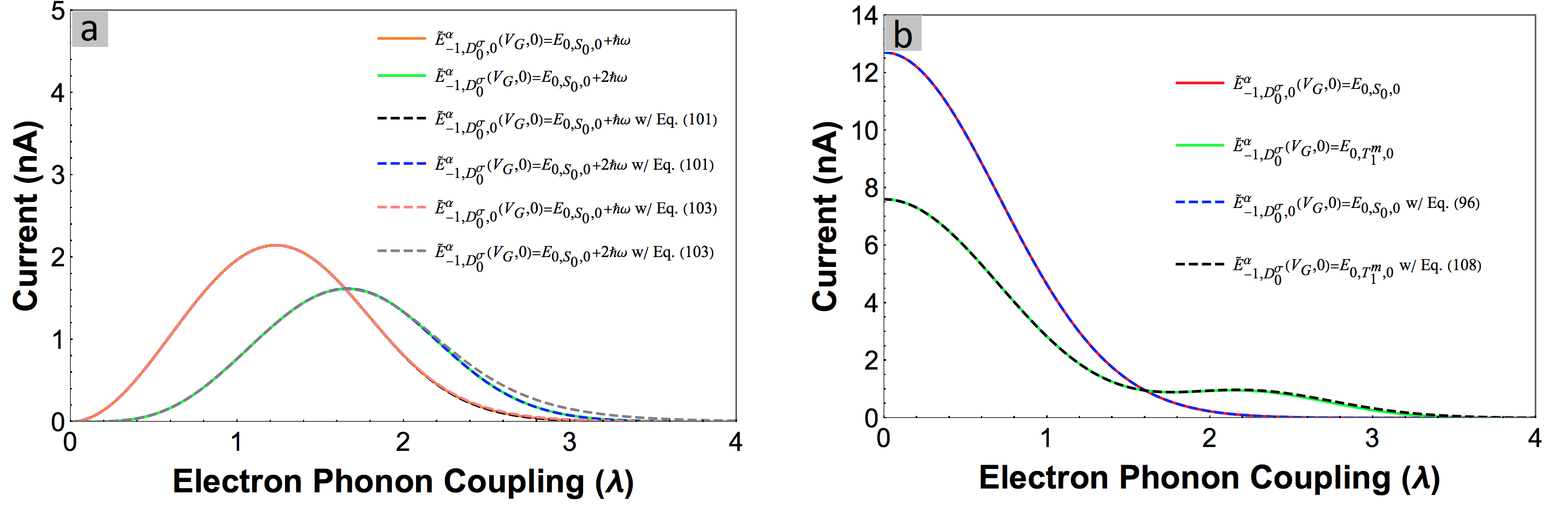}}
\caption{The current is plotted versus the e-p coupling $\lambda$ for four different energy level alignment schemes, i.e., (a) $\tilde{E}^{\alpha}_{-1,D^{\sigma}_0,0}\roundbracket{V_{\mathrm{G}},0}=E_{0,S_0,0}+\hbar\omega_{\mathrm{vib}}$ and $\tilde{E}^{\alpha}_{-1,D^{\sigma}_0,0}\roundbracket{V_{\mathrm{G}},0}=E_{0,S_0,0}+2\hbar\omega_{\mathrm{vib}}$, (b)  $\tilde{E}^{\alpha}_{-1,D^{\sigma}_0,0}\roundbracket{V_{\mathrm{G}},0}=E_{0,S_0,0}$ and $\tilde{E}^{\alpha}_{-1,D^{\sigma}_0,0}\roundbracket{V_{\mathrm{G}},0}=E_{0,T^m_1,0}$. In (a), the current is computed by the numerical solution of the Pauli master equation, the exact analytic solution in Eq. (\ref{eq:CurrentExprCase2v1}) and the approximate analytic solution in Eq. (\ref{eq:CurrentExprCase2v3}). In (b), the current is computed by the numerical solution of the Pauli master equation, the analytic solution in Eq. (\ref{eq:CurrentExprCase1v2}) for $\tilde{E}^{\alpha}_{-1,D^{\sigma}_0,0}\roundbracket{V_{\mathrm{G}},0}=E_{0,S_0,0}$ and the analytic solution in Eq. (\ref{eq:CurrentExpr3v2}) for $\tilde{E}^{\alpha}_{-1,D^{\sigma}_0,0}\roundbracket{V_{\mathrm{G}},0}=E_{0,T^m_1,0}$. }
\label{fig:IepEnergyAlign}
\end{figure*}

According to Eq. (\ref{eq:CurrentExprCase2v3}), the photocurrent can be analyzed  by the dependence of $\vertbracket{M_{0n}\roundbracket{\lambda}}^2$ on $\lambda$ and $n$, respectively. On one hand, when $n$ is fixed and $\lambda$ is varied, i.e., we are dealing with a specific energy level alignment $\tilde{E}^{\alpha}_{-1,D^{\sigma}_0,0}\roundbracket{V_{\mathrm{G}},0})=E_{0,S_0,0}+n \hbar\omega_{\mathrm{vib}}$, the photoinduced current approximately maximizes at $\lambda=\sqrt{n}$, which is consistent with the behavior of the first current step in Figure \ref{fig:IV0S0epfield}b.
Figure \ref{fig:IepEnergyAlign}a presents the photocurrent as a function of $\lambda$ for situations of $n=1$ and $n=2$. In Figure \ref{fig:IepEnergyAlign}a, the approximate analytical solution in Eq. (\ref{eq:CurrentExprCase2v3}) agrees well with the numerical calculation and the exact analytical solution in Eq. (\ref{eq:CurrentExprCase2v1}), where a slight deviation is observed under the situation of $n=2$ when $\lambda$ exceeds $2.0$.
It should be pointed out that the peak of the $I-\lambda$ curve in Figure \ref{fig:IepEnergyAlign}a slightly deviates from $\sqrt{n}$ due to the monotonous decline of the denominator of Eq. (\ref{eq:CurrentExprCase2v2}) with respect to $\lambda$.  On the other hand, when $\lambda$ is fixed and $n$ is varied, the photoinduced current peaks at $n\sim\lambda^2$. This explains why the most pronounced conductance lines within the diamond D in Figure \ref{fig:StabilityDiagrameqhw0.2} shift towards larger gate voltage as the electron phonon coupling increases.


\subsubsection{$\tilde{E}^{\alpha}_{-1,D^{\sigma}_0,0}\roundbracket{V_{\mathrm{G}},0}=E_{0,T^m_0,0}$}\label{sec:situation3}
Under this situation, we restrict ourselves to the bias range $V_{\mathrm{SD}}\in[0,2\roundbracket{E_{0,T^m_1,0}-E_{0,S_0,0}}/\vertbracket{e}]$. In this bias range, the field-on transport characteristics is activated when $V_{\mathrm{SD}}>0$, while the field-off transport characteristics in this range is completely blocked. The photoinduced current is attributed to the energy level alignment between $\tilde{E}^{\alpha}_{-1,D^{\sigma}_0,0}\roundbracket{V_{\mathrm{G}},0}$ and $E_{0,T^m_0,0}$. 

The influence of the e-p coupling on the photoinduced transport characteristics is presented in Figure \ref{fig:IV0S0epfield}c. Among all the explored values of $\lambda$, with the IV curves at $\lambda=0$ as a reference, the e-p coupling suppresses the first three photoinduced currents steps but enhances the subsequent two current steps. These observations are also quantitatively explained by the analytical solution.

Next, we derive the analytical current expression using the effective electronic states $\ket{-1,D^{\uparrow}}$, $\ket{-1,D^{\downarrow}}$, $\ket{0,S_0}$, $\ket{0,S_1}$, $\ket{0,T^{+1}_0}$, $\ket{0,T^0_0}$ and $\ket{0,T^{-1}_0}$. The vector form of the occupation probabilities is
\begin{align}
    \bvec P =\roundbracket{P_{-1,D^{\uparrow}_0}\,,P_{-1,D^{\downarrow}_0}\,,P_{0,S_0}\,,P_{0,S_1}\,,P_{0,T^{+1}_0}\,,P_{0,T^0_0}\,,P_{0,T^{-1}_0}}^\mathrm{T}\,.
\end{align}

According to the energy level alignment illustrated in Figure \ref{fig:EnergyLevelAlignment}c, the rate matrix $\bvec W$ is constructed as
\begin{widetext}
\begin{align}
    \bvec W=\left(
    \begin{array}{ccccccc}
       -\roundbracket{\Lambda^{\mathrm{S}}_0+\Lambda^{\mathrm{D}}_0+1.5\Lambda^{\mathrm{S}}_T}\Gamma  & 0 & 0  &  0.5\roundbracket{\Lambda^{\mathrm{S}}_1+\Lambda^{\mathrm{D}}_1}\Gamma & \Lambda^{\mathrm{D}}_T\Gamma  & 0.5\Lambda^{\mathrm{D}}_T\Gamma  & 0 \\
       0 & -\roundbracket{\Lambda^{\mathrm{S}}_0+\Lambda^{\mathrm{D}}_0+1.5\Lambda^{\mathrm{S}}_T}\Gamma & 0  & 0.5\roundbracket{\Lambda^{\mathrm{S}}_1+\Lambda^{\mathrm{D}}_1}\Gamma  & 0  & 0.5\Lambda^{\mathrm{D}}_T  &  \Lambda^{\mathrm{D}}_T\Gamma \\
       \roundbracket{\Lambda^{\mathrm{S}}_0+\Lambda^{\mathrm{D}}_0}\Gamma  & \roundbracket{\Lambda^{\mathrm{S}}_0+\Lambda^{\mathrm{D}}_0}\Gamma & -k^{\mathrm{field}}  & k^{\mathrm{field}}  & 0  &  0 & 0 \\
        0 & 0 & k^{\mathrm{field}}  & -k^{\mathrm{field}}-0.5\roundbracket{\Lambda^{\mathrm{S}}_1+\Lambda^{\mathrm{D}}_1}\Gamma  & 0  & 0  & 0 \\
        \Lambda^{\mathrm{S}}_T\Gamma & 0 & 0  & 0  & -\Lambda^{\mathrm{D}}_T\Gamma  &  0 &  0\\
        0.5\Lambda^{\mathrm{S}}_T\Gamma & 0.5\Lambda^{\mathrm{S}}_T\Gamma & 0  & 0  & 0  &  -\Lambda^{\mathrm{D}}_T\Gamma & 0 \\
        0 & \Lambda^{\mathrm{S}}_T\Gamma & 0  & 0  &  0 & 0  &  -\Lambda^{\mathrm{D}}_T\Gamma
    \end{array}
    \right)\,,
\end{align}
\end{widetext}
where we further abbreviate $\Lambda^{\alpha}_{-1,D^{\sigma}_0\leftarrow 0,T^m_1}$ as $\Lambda^{\alpha}_T$ for simplicity.

The stationary current corresponding to the situation in Figure \ref{fig:EnergyLevelAlignment}c is then solved as
\begin{align}\label{eq:CurrentExpr3v1}
    I_{\mathrm{S}}=\frac{\vertbracket{e}\Gamma k^{\mathrm{field}}\Lambda^{\mathrm{D}}_T\roundbracket{3\Lambda_1\Lambda^{\mathrm{S}}_T+2\roundbracket{\Lambda^{\mathrm{S}}_0\Lambda^{\mathrm{D}}_1-\Lambda^{\mathrm{D}}_0\Lambda^{\mathrm{S}}_1}}}{\roundbracket{\roundbracket{4\Lambda_0-\Lambda_1}\Lambda^{\mathrm{D}}_T+3\Lambda_1\Lambda_T}k^{\mathrm{field}}+2\Lambda_0\Lambda_1\Lambda^{\mathrm{D}}_T\Gamma}\,,
\end{align}
whose zero e-p coupling limit also agrees with the results in Ref. [\!\!\citenum{Fu2018}], i.e., 
\begin{align}\label{eq:CurrentExpr3ZeroEp}
    \lim_{\lambda\rightarrow 0}I_{\mathrm{S}}=3\vertbracket{e}\Gamma k^{\mathrm{field}}/\roundbracket{4\Gamma+9k^{\mathrm{field}}}\,.
\end{align}

Since the energy level alignment diagram in Figure \ref{fig:EnergyLevelAlignment}c indicates  $\Lambda^{\mathrm{S}}_T=\Lambda^{\mathrm{D}}_T=0.5\Lambda_T$, we can further simplify Eq. (\ref{eq:CurrentExpr3v1}) into
\begin{align}\label{eq:CurrentExpr3v2}
    I_{\mathrm{S}}=I_{\mathrm{T}}+I_{\mathrm{as}}
\end{align}
with
\begin{align}\label{eq:CurrentExpr3v2-1}
    I_{\mathrm{T}}=\vertbracket{e}\Gamma k^{\mathrm{field}}\frac{3\Lambda_1\Lambda_T/4}{\roundbracket{2\Lambda_0+2.5\Lambda_1}k^{\mathrm{field}}+\Lambda_0\Lambda_1\Gamma}
\end{align}
and
\begin{align}\label{eq:CurrentExpr3v2-2}
    I_{\mathrm{as}}=\vertbracket{e}\Gamma k^{\mathrm{field}}\frac{\Lambda^{\mathrm{S}}_0\Lambda^{\mathrm{D}}_1-\Lambda^{\mathrm{D}}_0\Lambda^{\mathrm{S}}_1}{\roundbracket{2\Lambda_0+2.5\Lambda_1}k^{\mathrm{field}}+\Lambda_0\Lambda_1\Gamma}\,,
\end{align}
where $I_{\mathrm{T}}$ and $I_{\mathrm{as}}$ reveal two origins of the photoinduced current under this situation. First, the component $I_{\mathrm{T}}$ originates from the energy level alignment between $\tilde{E}^{\alpha}_{-1,D^{\sigma}_0,0}\roundbracket{V_{\mathrm{G}},0}$ and the triplet states, so it 
exactly reduces to Eq. (\ref{eq:CurrentExpr3ZeroEp}) as $\lambda\rightarrow 0$ and also contributes to the suppression of the first three photocurrent steps in Figure \ref{fig:IV0S0epfield}c. 
Second, the component $I_{\mathrm{as}}$ originates from the asymmetry in $\Lambda^{\mathrm{S/D}}_{0/1}$. $I_{\mathrm{as}}$ resembles Eq. (\ref{eq:CurrentExprCase2v1}) which describes the photocurrent in section \ref{sec:situation2} 
and vanishes at $\lambda=0$. It is worth noting that $I_{\mathrm{as}}$ contributes to the enhancement of 4th and 5th photoinduced current steps in Figure \ref{fig:IV0S0epfield}c. 

Figure \ref{fig:IepEnergyAlign}b offers more insights about $I_{\mathrm{T}}$ and $I_{\mathrm{as}}$, where the first photoinduced current step in this case, together with first field-on current step in section \ref{sec:situation1} is plotted as a function of the e-p coupling $\lambda$. 
In Figure \ref{fig:IepEnergyAlign}b, the numerical calculations agrees well with the analytical solutions in Eq. (\ref{eq:CurrentExprCase1v1}) and (\ref{eq:CurrentExpr3v2}). The $I-\lambda$ calculated at $V_{\mathrm{SD}}=0.2$ V and $V_{\mathrm{G}}=V^{\mathrm{T1}}_{\mathrm{G}}$ first displays an exponential decay due to $I_{\mathrm{T}}$ and then exhibits a maximum led by $I_{\mathrm{as}}$ nearby $\lambda=2.0$, while on the contrary the situation of $V_{\mathrm{G}}=V^{\mathrm{S0}}_{\mathrm{G}}$ shows a monotonous decrease in the field-on current with respect to $\lambda$. 

Furthermore, the decomposition of the photocurrent into $I_{\mathrm{T}}$ and $I_{\mathrm{as}}$ also explains the role of the e-p coupling on the photoinduced current as shown in Figure \ref{fig:IV0S0epfield}c. Within $V_{\mathrm{SD}}\in[0,6\hbar\omega_{\mathrm{vib}}/\vertbracket{e}]$, i.e., during the first three current steps, $I_{\mathrm{T}}$ dominates the photoinduced transport characteristics. In this bias range, $I_{\mathrm{T}}$ is responsible for the e-p coupling induced suppression in the photocurrent, 
while $I_{\mathrm{as}}$ offers negligible contribution because the asymmetry between $\Lambda^{\mathrm{S}}_{0/1}$ and $\Lambda^{\mathrm{D}}_{0/1}$ is insignificant.
The energy level alignment in Figure \ref{fig:EnergyLevelAlignment}c clearly reveals that the extent of the asymmetry required by $I_{\mathrm{as}}$ grows with $V_{\mathrm{SD}}$. Meanwhile, as $V_{\mathrm{SD}}$ increases, the e-p coupling induced photocurrent suppression due to $I_{\mathrm{T}}$ is also lifted and will be completely eliminated when more transport channels are activated. Henceforth, the gradual elimination of the photocurrent current suppression given by $I_{\mathrm{T}}$, together with the increase in $I_{\mathrm{as}}$, finally results in the e-p coupling enhanced photoinduced current at the 4th and 5th current step in Figure \ref{fig:IV0S0epfield}c.


\subsubsection{$\tilde{E}^{\alpha}_{-1,D^{\sigma}_0,0}\roundbracket{V_{\mathrm{G}},0}>E_{0,T^m_0,0}$}\label{sec:situation4}
Our discussion for this situation focuses on the bias range $V_{\mathrm{SD}}\in[0,2(\tilde{E}^{\alpha}_{-1,D^{\sigma}_0,0}\roundbracket{V_{\mathrm{G}},0}-E_{0,T^m_0,0})/\vertbracket{e}]$, beyond which the transport characteristics is governed by Eq. (\ref{eq:CurrentExpr3v1}). The ranges of $V_{\mathrm{G}}$ and $V_{\mathrm{SD}}$ we specified correspond to the diamond E, where the electron tranport is blocked under both field-off and field-on conditions. 

Unlike the photoinduced current described in section \ref{sec:situation2}, the e-p coupling does not lift the blockade under this situation, which has been clearly revealed in the charge stability diagrams. According to the representative energy level alignment diagram in Figure \ref{fig:EnergyLevelAlignment}d, the transport characteristics in this case is completely blocked by the triplet states. Figure \ref{fig:EnergyLevelAlignment}d shows that there is no outgoing electronic transitions from triplet states, therefore the triplet states accumulate all the population of the molecular system in the steady state. The analytical solution under this situation is straightforward and obvious, i.e., $P_{0,T^{+1}_1}=P_{0,T^0_1}=P_{0,T^{-1}_1}=1/3$. Note that the relevant singlet-triplet transitions is insignificant and negligible because the optical excitations dominate in this situation. 

\section{Conclusion and Perspective}\label{Conculsion}

In conclusion, we have examined the influence of e-p coupling and vibrational relaxation on electron transport through an irradiated molecular junction. Moreover, we have shown that the roles of the triplet states and the energy level alignment between charge states and neutral states are crucial for the transport characteristics.
Our main findings are summarized as follows:

(1) In the charge stability diagram, the width of the middle diamond (corresponding to the triplet states) is robust to the e-p coupling.  In contrast, the two side diamonds (corresponding to the partial charged states) gradually shrink with the increasing e-p coupling due to the activation of transport channels induced by
vibrational transitions. In addition, the effect of the e-p coupling is more pronounced in the unequilibrated phonon regime than in the equilibrated phonon regime because of the vibrational transitions caused by the nonequilibrium vibratinal population.

(2) We have demonstrated a new type of photoconduction mechanism due to the e-p coupling. Our previous study shows that, in the absence of the e-p coupling, the energy level alignment between the charge states and the triplet states uniquely leads to the photocurrent. In this study, in the presence of the e-p coupling, the asymmetry in the rates of charge transfer transitions also results in the photocurrent when the renormalized state energy of a charge state falls between $E_{0,S_0,0}$ and $E_{0,T^m_1,0}$. 
This photoconduction mechanism is also significant when the triplet states come into play. 

(3) 
Our study can be used as a guide for the selection of photoconductive molecules. When a high-frequency vibrational mode is coupled to the charge transfer transition, the observation of photocurrent requires 
$\tilde{E}^{\alpha}_{\pm 1,D^{\sigma}_0,0}\sim E_{0,T^m_1,0},E_{0,S_0,n} (n\geq 1)$, i.e., the renormalized state energy of charged states approximately aligns with the triplet states $\ket{0,T^m_1}$ or the excited vibrational states of $\ket{0,S_0}$. When a low-frequency vibrational mode dominates, the observation of photocurrent requires $\tilde{E}^{\alpha}_{\pm 1,D^{\sigma}_0,0}\gtrsim E_{0,S_0,0}$ or  $\tilde{E}^{\alpha}_{\pm 1,D^{\sigma}_0,0}\lesssim E_{0,T^m_1,0}$, whereas the required proximity of the alignment is determined by the e-p coupling.

(4) Compared with the photoinduced current in the zero e-p coupling limit, the presence of the e-p coupling can either enhance or suppress the photocurrent.  In the case of $\tilde{E}^{\alpha}_{\pm 1,D^{\sigma}_0,0}\roundbracket{V_{\mathrm{G}},0}=E_{0,T^m_1,0}$, the influence of the e-p coupling on the photocurrent changes from suppression to enhancement as the source-drain bias increases.

Although we have conducted a comprehensive study on the effect of e-p couplings and vibrational relaxation on an irradiated molecular junction, several issues deserve further discussion.
First, the master equation approach cannot include the level broadening due to the molecule-lead coupling and thus fails to describe the tunneling current within the blockade region. Our future work will address this issue through two routes,
i.e., the master equation approaches that go beyond 2nd order expansion\cite{Wegesijs2010} and the Hubbard operator NEGF approach\cite{Galperin2008} that intrinsically considers the level broadening at the lowest order expansion. 
Second, the energy level alignment of molecular systems can be adjusted not only by a gate electrode but also by the functionalization of molecules (e.g., changing the substituent group\cite{Li2013}), electrochemical gating\cite{Bai2019} and molecular orbital gating\cite{Hines2010,Choi2010}.
Finally, we hope that our work can motivate further studies on irradiated molecular junctions and promote the development of molecular electronics.

\begin{acknowledgments}
This research was supported by Academia Sinica and the Ministry of Science and Technology of Taiwan (MOST 106-2113-M-001-036-MY3). This work used the Extreme Science and Engineering Discovery Environment (XSEDE), which is supported by National Science Foundation grant number ACI-1548562.
\end{acknowledgments}

\appendix

\section{The Evaluation of $V_{\alpha\bvec k\sigma,NaN-1b}$}\label{sec:appendix1}
For the purpose of evaluating the scattering amplitude $V^*_{\alpha\bvec k\sigma,NaN-1b}$ ($V_{\alpha\bvec k\sigma,NaN-1b}$), we recast the molecule-lead coupling Hamiltonian in Eq. (\ref{eq:mol-lead}) in terms of single particle basis as,
\begin{align}\label{eq:m-lv2}
    H_{\mathrm{m-l}}
    =
    \sum_{n,\alpha\bvec k \sigma}\roundbracket{t^*_{n\alpha\bvec k\sigma}\hat a^{\dagger}_{\alpha\bvec k\sigma}\hat d_{n\sigma}+t_{n\alpha,\bvec k\sigma}\hat d^{\dagger}_{n\sigma}\hat a_{\alpha\bvec k\sigma}},
\end{align}
where $\hat d_{n\sigma}$($\hat d_{n\sigma}^{\dagger}$) is the annihilation(creation) operator of an electron with spin $\sigma$ on the single particle level $\ket{n}$, and $t^*_{n\alpha\bvec k\sigma}$ ($t_{n\alpha\bvec k\sigma}$) is the scattering amplitude that describes the hopping of an electron with momentum $\bvec k$ and spin $\sigma$ from $\ket{n}$ (electrode $\alpha$) to the electrode $\alpha$ ($\ket{n}$). In principle, the single particle basis could be any complete orthonormal basis, hereafter we choose the molecular orbitals for convenience.

We next denote a many-body electronic state $\ket{N,a}$ as $\ket{A}$ and insert the relationship $\hat 1=\sum_A\ket{A}\bra{A}$ into Eq. (\ref{eq:m-lv2}), which results in
\begin{align}\label{eq:m-lv3}
    H_{\mathrm{m-l}}&=\sum_{n,\alpha\bvec k \sigma}\sum_{A,B}\roundbracket{t^*_{n\alpha\bvec k\sigma}\hat a^{\dagger}_{\alpha\bvec k\sigma}\ket{B}\bra{B}\hat d_{n\sigma}\ket{A}\bra{A}
    +
    \mathrm{h.c.}}\,.
\end{align}
By comparing Eq. (\ref{eq:m-lv3}) with Eq. (\ref{eq:mol-lead}), we obtain the identities
\begin{align}
\label{eq:tranfercoupling1}
    V^*_{\alpha\bvec k\sigma,AB}
    &=\sum_n t^*_{n\alpha\bvec k\sigma}\bra{B}\hat d_{n\sigma}\ket{A} \\
\label{eq:tranfercoupling2}
    V_{\alpha\bvec k\sigma,AB}
    &=\sum_n t_{n\alpha\bvec k\sigma}\bra{A}\hat d^{\dagger}_{n\sigma}\ket{B}
\end{align}
which relates the $V_{\alpha\bvec k\sigma,AB}$($V^*_{\alpha\bvec k\sigma,AB}$) in the picture of molecular many-electron states with the $t_{n\alpha\bvec k\sigma}$($t^*_{n\alpha\bvec k\sigma}$) in the single particle basis.

Similar to the factorization in Eq. (\ref{eq:factorization1}), we assume that $t_{n\alpha\bvec k\sigma}$($t^*_{n\alpha\bvec k\sigma}$) could factorize as\cite{Peskin2017}
\begin{align}
\label{eq:decompose1}
    t_{n\alpha\bvec k\sigma}=\zeta_{n\alpha}M_{\alpha\bvec k\sigma} \\
\label{eq:decompose2}
    t^*_{n\alpha\bvec k\sigma}=\zeta^*_{n\alpha}M^*_{\alpha\bvec k\sigma} \,.
\end{align}
where $\zeta_{n\alpha}$($\zeta^*_{n\alpha}$) characterizes the overlap between the molecular orbital $\ket{n}$ and the electrons in the electrode $\alpha$.

Plugging Eq. (\ref{eq:decompose1}) and (\ref{eq:decompose2}) into Eq. (\ref{eq:tranfercoupling1}) and (\ref{eq:tranfercoupling2}), we arrive at the definitions of $T_{\alpha,AB}$ and $T^*_{\alpha,AB}$ as
\begin{align}
\label{eq:transfercouplingmanybodystates1}
    T^*_{\alpha,AB}=&\sum_n \zeta^*_{n\alpha}\bra{B}\hat d_{n\sigma}\ket{A} \\
\label{eq:transfercouplingmanybodystates2}
    T_{\alpha,AB}=&\sum_n \zeta_{n\alpha}\bra{A}\hat d^{\dagger}_{n\sigma}\ket{B}\,.
\end{align}
Therefore, it is obvious that $T_{\alpha,AB}$ and $T^*_{\alpha,AB}$ survive only when $\ket{A}$ differs from $\ket{B}$ by one electron.



In order to explicitly evaluate $T_{\alpha,AB}$ and $T^*_{\alpha,AB}$, we approximate the many-electron states considered in the main text in terms of the Kohn-Sham orbitals,
\begin{align}
    \ket{0,S_0}&=\hat c^{\dagger}_{\mathrm{H}\uparrow}\hat c^{\dagger}_{\mathrm{H}\downarrow}\ket{\chi} \nonumber \\
    \ket{0,S_1}&=\frac{1}{\sqrt{2}}\roundbracket{\hat c^{\dagger}_{\mathrm{H}\uparrow}\hat c^{\dagger}_{\mathrm{L}\downarrow}-\hat c^{\dagger}_{\mathrm{H}\downarrow}\hat c^{\dagger}_{\mathrm{L}\uparrow}} \ket{\chi} \nonumber \\
    \ket{0,T_1^0}&=\frac{1}{\sqrt{2}}\roundbracket{\hat c^{\dagger}_{\mathrm{H}\uparrow}\hat c^{\dagger}_{\mathrm{L}\downarrow}+\hat c^{\dagger}_{\mathrm{H}\downarrow}\hat c^{\dagger}_{\mathrm{L}\uparrow}}\ket{\chi} \nonumber \\
    \ket{0,T_1^{+1}}&=\hat c^{\dagger}_{\mathrm{H}\uparrow}\hat c^{\dagger}_{\mathrm{L}\uparrow}\ket{\chi} \nonumber \\
    \ket{0,T_1^{-1}}&=\hat c^{\dagger}_{\mathrm{H}\downarrow}\hat c^{\dagger}_{\mathrm{L}\downarrow}\ket{\chi} \nonumber \\
    \ket{-1,D_0^{+\frac{1}{2}}}&=\hat c^{\dagger}_{\mathrm{H}\uparrow}\ket{\chi} \nonumber \\
    \ket{-1,D_0^{-\frac{1}{2}}}&=\hat c^{\dagger}_{\mathrm{H}\downarrow}\ket{\chi} \nonumber \\
    \ket{1,D_0^{+\frac{1}{2}}}&=\hat c^{\dagger}_{\mathrm{H}\uparrow}\hat c^{\dagger}_{\mathrm{H}\downarrow}\hat c^{\dagger}_{\mathrm{L}\uparrow}\ket{\chi} \nonumber \\
    \ket{1,D_0^{-\frac{1}{2}}}&=\hat c^{\dagger}_{\mathrm{H}\downarrow}\hat c^{\dagger}_{\mathrm{H}\downarrow}\hat c^{\dagger}_{\mathrm{L}\downarrow}\ket{\chi} 
\end{align}
where $\hat c^{\dagger}_{n\sigma}$ and $\hat c_{n\sigma}$ are respectively the creation operator and the annihilation operator of an electron on Kohn-Sham orbital $\ket{n}$ with spin $\sigma$ ,  H refers to HOMO, L refers to LUMO, and $\ket{\chi}$ is a reference state with empty HOMO and LUMO. Here we assume that all the many-electron states could be constructed from the same set of Kohn-Sham orbitals.

Note that, in $T_{\alpha,N\pm 1b,N a}$, only up to one term survives in the summation over orbital index $n$, i.e. either $\zeta_{1\alpha}$ or $\zeta_{2\alpha}$. We suppose that $\zeta_{1\alpha}=\zeta_{2\alpha}=\zeta_{\alpha}$, and then reorganize $\frac{2\pi}{\hbar^2}\vertbracket{T_{\alpha,N\pm 1b,N a}}^2\bar J_{\alpha}$ into $\Gamma_{\alpha}\nu_{N\pm 1b,Na}$, where the characteristic rate of charge transfer transition $\Gamma_{\alpha}$ is defined as 
\begin{align}
    \Gamma_{\alpha}=\frac{2\pi}{\hbar^2}\vertbracket{\zeta_{\alpha}}^2\bar{J}_{\alpha}
\end{align} 
and the dimensionless coupling coefficient $\nu_{N\pm 1b,Na}$ of the transition $\ket{N\pm 1,b}\leftrightarrow\ket{N,a}$ has been given in Table S1 of Ref. [\!\!\citenum{Fu2018}].

\section{Derivation of $\kappa_{0S_0\nu_0,0S_1\nu_1}$ in Eq. (\ref{eq:S0S1eom-1})}\label{sec:appendix2}
According to Eq. (\ref{eq:eom-3terms}) and (\ref{eq:S0S1eom-1}), $\kappa_{0S_0\nu_0,0S_1\nu_1}$ is defined as
\begin{align}\label{eq:eom-dephasing}
    &\left.\frac{d \rho_{0S_0\nu_0,0S_1\nu_1}}{dt}\right\vert_{\mathrm{m-l}}
    +\left.\frac{d \rho_{0S_0\nu_0,0S_1\nu_1}}{dt}\right\vert_{\mathrm{m-th}}
    +\left.\frac{d \rho_{0S_0\nu_0,0S_1\nu_1}}{dt}\right\vert_{\mathrm{SOC}} \nonumber\\
    =&-\kappa_{0S_0\nu_0,0S_1\nu_1}\rho_{0S_0\nu_0,0S_1\nu_1}\,,
\end{align}
where $\rho_{0S_0\nu_0,0S_1\nu_1}=\bra{0,S_0,\nu_0}\hat\rho\roundbracket{t}\ket{0,S_1,\nu_1}$.

In order to derive $\kappa_{0S_0\nu_0,0S_1\nu_1}$, we follow the same procedure as we used to derive the rate equation. Since the SOC-induced transitions are significantly slower than the other processes, we restrict ourselves to contributions from $\hat H_{\mathrm{m-l}}$ and $\hat H_\mathrm{m-th}$, which are hereafter referred to as $\kappa^{\mathrm{m-l}}_{0S_0\nu_0,0S_1\nu_1}$ and $\kappa^{\mathrm{m-th}}_{0S_0\nu_0,0S_1\nu_1}$ respectively, i.e.,
\begin{align}
    \kappa_{0S_0\nu_0,0S_1\nu_1}=\kappa^{\mathrm{m-l}}_{0S_0\nu_0,0S_1\nu_1}+\kappa^{\mathrm{m-th}}_{0S_0\nu_0,0S_1\nu_1}\,.
\end{align}

The derivation of $\kappa_{0S_0\nu_0,0S_1\nu_1}$ starts from taking the off-diagonal matrix element of Eq. (\ref{eq:breom-m-l}). Unlike the derivation of charge transfer rate equation, not all the terms in Eq. (\ref{eq:breom-m-l}) contribute to $\kappa^{\mathrm{m-th}}_{0S_0\nu_0,0S_1\nu_1}$. In the following, we take the derivation for the 1st and 4th terms in Eq. (\ref{eq:breom-m-l}) as examples, one of which contributes to $\kappa_{0S_0\nu_0,0S_1\nu_1}$ and the other does not. 

The matrix element of the $1$st term in Eq. (\ref{eq:breom-m-l}) with respect to $\ket{N,a,\nu_a}$ and $\ket{M,b,\nu_b}$ is  derived as
\begin{align}\label{eq:breom-OffME-ml-1}
    &\sum_{\alpha}\int^{\infty}_0du\,\,C_{\alpha}\roundbracket{-u}
    \bra{N,a,\nu_a}
    \hat{S}^{+I}_{\alpha}\roundbracket{t}
    \hat \rho^I\roundbracket{t}
    \hat{S}^{-I}_{\alpha}\roundbracket{t-u}
    \ket{M,b,\nu_b} \nonumber \\
    =&\sum_{\alpha}\int^{\infty}_0du\,\,C_{\alpha}\roundbracket{-u}\sum_{b_1,b_2}\sum_{\nu_1,\nu_2}
    T^{\alpha}_{Na,N-1b_1}T^{*\alpha}_{Mb,M-1b_2}\nonumber \\ 
    &\times M_{\nu_a\nu_{1}}\roundbracket{\lambda_{N-1,b_1}-\lambda_{N,a}}M_{\nu_{2}\nu_b}\roundbracket{\lambda_{M,b}-\lambda_{M-1,b_2}}\nonumber\\
    &\times\expfunc{\frac{i}{\hbar}\roundbracket{E_{M,b,\nu_b}-E_{M-1,b_2,\nu_{2}}}u} \nonumber \\
    &\times
    \expfunc{\frac{i}{\hbar}\roundbracket{E_{N,a,\nu_a}-E_{N-1,b_1,\nu_{1}}-E_{M,b,\nu_b}+E_{M-1,b_2,\nu_{2}}}t} \nonumber \\
    &\times \bra{N-1,b_1,\nu_{1}}\hat\rho^I\roundbracket{t}\ket{M-1,b_2,\nu_{2}}
    \,.
\end{align}
Similar to the Section \ref{sec:QME-m-l}, the secular approximation leads to
\begin{align}
    &\expfunc{\frac{i}{\hbar}\roundbracket{E_{N,a,\nu_a}-E_{N-1,b_1,\nu_{1}}-E_{M,b,\nu_b}+E_{M-1,b_2,\nu_{2}}}t}\nonumber\\
    \longrightarrow&\delta_{N,M}\delta_{a,b}\delta_{\nu_a,\nu_b}\delta_{b_1,b_2}\delta_{\nu_1,\nu_2}\,,
\end{align}
which means that Eq. (\ref{eq:breom-OffME-ml-1}) does not contribute to $\kappa^{\mathrm{m-l}}_{0S_0\nu_0,0S_1\nu_1}$. Note that we here neglect the possibility of $E_{N,a,\nu_a}-E_{M,b,\nu_b}=E_{M-1,b_2,\nu_{2}}-E_{N-1,b_1,\nu_{1}}$ when $E_{N,a,\nu_a}\neq E_{M,b,\nu_b}$, which is a reasonable assumption when considering the electronic structure of a molecular system.

We next derive the matrix element of the $4$th term in Eq. (\ref{eq:breom-m-l}), which gives
\begin{align}
    &\sum_{\alpha}\int^{\infty}_0du\,\,\bar C_{\alpha}\roundbracket{u}
    \bra{N,a,\nu_a}
    \hat{S}^{+I}_{\alpha}\roundbracket{t}
    \hat{S}^{-I}_{\alpha}\roundbracket{t-u}
    \hat \rho^I\roundbracket{t}
    \ket{M,b,\nu_b}\nonumber \\
    =&\sum_{\alpha}\int^{\infty}_0du\,\,\bar C_{\alpha}\roundbracket{u}
    \sum_{b_1,b_2}\sum_{\nu_1,\nu_2}T^{\alpha}_{Na,N-1b_2}T^{*\alpha}_{Nb_1,N-1b_2}\nonumber\\
    &\times M_{\nu_a\nu_{2}}\roundbracket{\lambda_{N-1,b_2}-\lambda_{N,a}}M_{\nu_{2}\nu_{1}}\roundbracket{\lambda_{N,b_1}-\lambda_{N-1,b_2}}\nonumber\\
    &\times\expfunc{\frac{i}{\hbar}\roundbracket{E_{N,b_1,\nu_{1}}-E_{N-1,b_2,\nu_{2}}}u} \expfunc{\frac{i}{\hbar}\roundbracket{E_{N,a,\nu_a}-E_{N,b_1,\nu_{1}}}t}\nonumber \\
    &\times \bra{N,b_1,\nu_{1}}\hat\rho^I\roundbracket{t}\ket{M,b,\nu_b}\,.
\end{align}

With the help of the secular approximation, we have $\expfunc{\frac{i}{\hbar}\roundbracket{E_{N,a,\nu_a}-E_{N,b_1,\nu_{1}}}t}\rightarrow\delta_{a,b_1}\delta_{\nu_a,\nu_1}$
, which leads to
\begin{align}
     &\sum_{\alpha}\int^{\infty}_0du\,\,\bar C_{\alpha}\roundbracket{u}
    \bra{N,a,\nu_a}
    \hat{S}^{+I}_{\alpha}\roundbracket{t}
    \hat{S}^{-I}_{\alpha}\roundbracket{t-u}
    \hat \rho^I\roundbracket{t}
    \ket{M,b,\nu_b}\nonumber \\
    =&\hbar^2\kappa^{\mathrm{m-l}(4)}_{Na\nu_a,Mb\nu_b}\bra{N,a,\nu_{a}}\hat\rho^I\roundbracket{t}\ket{M,b,\nu_b}\,,
\end{align}
where $\kappa^{\mathrm{m-l}(4)}_{Na\nu_a,Mb\nu_b}$ refers to the contribution from the 4th term in Eq. (\ref{eq:breom-m-l}). In order to find out the physical meaning of $\kappa^{\mathrm{m-l}(4)}_{Na\nu_a,Mb\nu_b}$, we apply the wide band limit and the formula
\begin{align}
    \int^{\infty}_0dt\expfunc{i\omega t}=\pi\delta\roundbracket{\omega}+i\mathrm{\cal P}\frac{1}{\omega}\,,
\end{align}
where $\cal P$ refers to the Cauchy principal value. As a result, we obtain
\begin{align}
    &\kappa^{\mathrm{m-l}(4)}_{Na\nu_a,Mb\nu_b}\nonumber \\
    =&\frac{1}{\hbar^2}\sum_{\alpha}\int^{\infty}_0du\,\,\bar C_{\alpha}\roundbracket{u}
    \sum_{b_2,\nu_2}
    \vertbracket{T^{\alpha}_{Na,N-1b_2}}^2
    \vertbracket{M_{\nu_a\nu_{2}}\roundbracket{\lambda_{N-1,b_2}-\lambda_{N,a}}}^2\nonumber\\
    &\times \expfunc{\frac{i}{\hbar}\roundbracket{E_{N,a,\nu_{a}}-E_{N-1,b_2,\nu_{2}}}u} 
    \bra{N,a,\nu_{a}}\hat\rho^I\roundbracket{t}\ket{M,b,\nu_b}\nonumber \\
    =&\frac{1}{\hbar^2}
    \sum_{\alpha}\int^{\infty}_0du\,\,
    \int^{\infty}_{\infty}d\omega\,\,\bar J_{\alpha}(1-f_{\beta}\roundbracket{\omega,\mu_{\alpha}}
    \sum_{b_2,\nu_2}
    \vertbracket{T^{\alpha}_{Na,N-1b_2}}^2 \nonumber\\
    &\times \vertbracket{M_{\nu_a\nu_{2}}\roundbracket{\lambda_{N-1,b_2}-\lambda_{N,a}}}^2 \expfunc{\frac{i}{\hbar}\roundbracket{E_{N,a,\nu_{a}}-E_{N-1,b_2,\nu_{2}}-\omega}u}\nonumber \\
    =&\sum_{\alpha}\sum_{b_2,\nu_2}
    \roundbracket{\frac{1}{2}k^{\alpha}_{N-1,b_2,\nu_2\leftarrow N,a,\nu_a}+i\Delta^{\alpha}_{N-1,b_2,\nu_2\leftarrow N,a,\nu_a}}\,,
\end{align}
where 
\begin{align}
    &\Delta^{\alpha}_{N-1,b_2,\nu_2\leftarrow N,a,\nu_a}\nonumber\\
    =&\frac{1}{\hbar^2}\mathrm{\cal P}\int^{\infty}_{\infty}d\omega\,\,
    \frac{1}{E_{N,a,\nu_{a}}-E_{N-1,b_2,\nu_{2}}-\omega}\bar J_{\alpha}(1-f_{\beta}\roundbracket{\omega,\mu_{\alpha}}\nonumber \\
    &\times\vertbracket{T^{\alpha}_{Na,N-1b_2}}^2
    \vertbracket{M_{\nu_a\nu_{2}}\roundbracket{\lambda_{N-1,b_2}-\lambda_{N,a}}}^2
\end{align}
is an energy shift caused by the charge transfer transition $\ket{N,a,\nu_a}\rightarrow \ket{N-1,b_2,\nu_2}$. 

We can similarly evaluate the other terms in Eq. (\ref{eq:breom-m-l}) and then obtain
\begin{align}
    &\kappa^{\mathrm{m-l}}_{0S_0\nu_0,0S_1\nu_1}\nonumber\\
    =&\frac{1}{2}\sum_{\alpha,a,\nu}\biggr(
    k^{\alpha}_{-1,a,\nu\leftarrow 0,S_0,\nu_0}+
    k^{\alpha}_{1,a,\nu\leftarrow 0,S_0,\nu_0}+
    k^{\alpha}_{-1,a,\nu\leftarrow 0,S_1,\nu_1}\nonumber\\
    &+k^{\alpha}_{1,a,\nu\leftarrow 0,S_1,\nu_1}
    \biggr)+i\sum_{\alpha,a,\nu}\biggr(
    \Delta^{\alpha}_{-1,a,\nu\leftarrow 0,S_0,\nu_0}+
    \Delta^{\alpha}_{1,a,\nu\leftarrow 0,S_0,\nu_0}\nonumber\\
    &+
    \Delta^{\alpha}_{-1,a,\nu\leftarrow 0,S_1,\nu_1}+
    \Delta^{\alpha}_{1,a,\nu\leftarrow 0,S_1,\nu_1}
    \biggr)\,.
\end{align}

\begin{widetext}
Next, we move forward onto the derivation of $\kappa^{\mathrm{m-th}}_{0S_0\nu_0,0S_1\nu_1}$, which starts from taking the off-diagonal matrix element of Eq. (\ref{eq:breom-m-ph-2}). We then obtain
\begin{align}
    &\frac{d}{dt}\bra{N,a,\nu_a}\hat \rho^I\roundbracket{t}\ket{M,b,\nu_b}\nonumber\\
    =&-\frac{1}{2m_{\mathrm{vib}}\omega_{\mathrm{vib}}}J_{\mathrm{th}}\roundbracket{\omega_{\mathrm{vib}}}
    \biggr[
    \roundbracket{n\roundbracket{\omega_{\mathrm{vib}}}+1}\roundbracket{(\nu_a+\nu_b)\bra{N,a,\nu_a}\hat \rho^I\roundbracket{t}\ket{M,b,\nu_b}-2\sqrt{\nu_a+1}\sqrt{\nu_b+1}\bra{N,a,\nu_a+1}\hat \rho^I\roundbracket{t}\ket{M,b,\nu_b+1}}\nonumber \\
    &+n\roundbracket{\omega_{\mathrm{vib}}}\roundbracket{(\nu_a+\nu_b+2)\bra{N,a,\nu_a}\hat \rho^I\roundbracket{t}\ket{M,b,\nu_b}
    -2\sqrt{\nu_a}\sqrt{\nu_b}\bra{N,a,\nu_a-1}\hat \rho^I\roundbracket{t}\ket{M,b,\nu_b-1}
    }\nonumber \\
    \approx&-\frac{1}{2m_{\mathrm{vib}}\omega_{\mathrm{vib}}}J_{\mathrm{th}}\roundbracket{\omega_{\mathrm{vib}}}
    \biggr[
    \roundbracket{n\roundbracket{\omega_{\mathrm{vib}}}+1}(\nu_a+\nu_b)\bra{N,a,\nu_a}\hat \rho^I\roundbracket{t}\ket{M,b,\nu_b}+n\roundbracket{\omega_{\mathrm{vib}}}(\nu_a+\nu_b+2)\bra{N,a,\nu_a}\hat \rho^I\roundbracket{t}\ket{M,b,\nu_b} \nonumber \\
    =&\kappa^{\mathrm{m-th}}_{Na\nu_a,Mb\nu_b}\bra{N,a,\nu_a}\hat \rho^I\roundbracket{t}\ket{M,b,\nu_b}\,,
\end{align}
where the terms $\bra{N,a,\nu_a-1}\hat \rho^I\roundbracket{t}\ket{M,b,\nu_b-1}$ and $\bra{N,a,\nu_a+1}\hat \rho^I\roundbracket{t}\ket{M,b,\nu_b+1}$ are dropped because they introduce higher order terms of the coupling Hamiltonian $\hat H_{\mathrm{m-th}}$, which is reasonable in the limit of weak system-bath coupling.
According to Eq. (\ref{eq:RatesVib}), we can express $\kappa^{\mathrm{m-th}}_{0S_0\nu_a,0S_1\nu_b}$ as
\begin{align}
    \kappa^{\mathrm{m-th}}_{0S_0\nu_a,0S_1\nu_b}=\frac{1}{2}\roundbracket{
    k^{\mathrm{vib}}_{0,S_0,\nu_a+1\leftarrow0,S_0,\nu_a}
    +
    k^{\mathrm{vib}}_{0,S_0,\nu_a-1\leftarrow0,S_0,\nu_a}
    +
    k^{\mathrm{vib}}_{0,S_1,\nu_b+1\leftarrow0,S_1,\nu_b}
    +
    k^{\mathrm{vib}}_{0,S_1,\nu_b-1\leftarrow0,S_1,\nu_b}
    }\,.
\end{align}
\end{widetext}

\bibliography{PET}

\end{document}